\documentclass[a4paper,11pt]{article}
\usepackage{jinstpub} 
\usepackage{lineno}

\usepackage[utf8]{inputenc}
\usepackage{graphicx}
\usepackage{siunitx}
\usepackage{url}
\usepackage[export]{adjustbox}

\title{Instrument design and performance of the first seven stations of RNO-G}

\collaboration{\includegraphics[height=19mm]{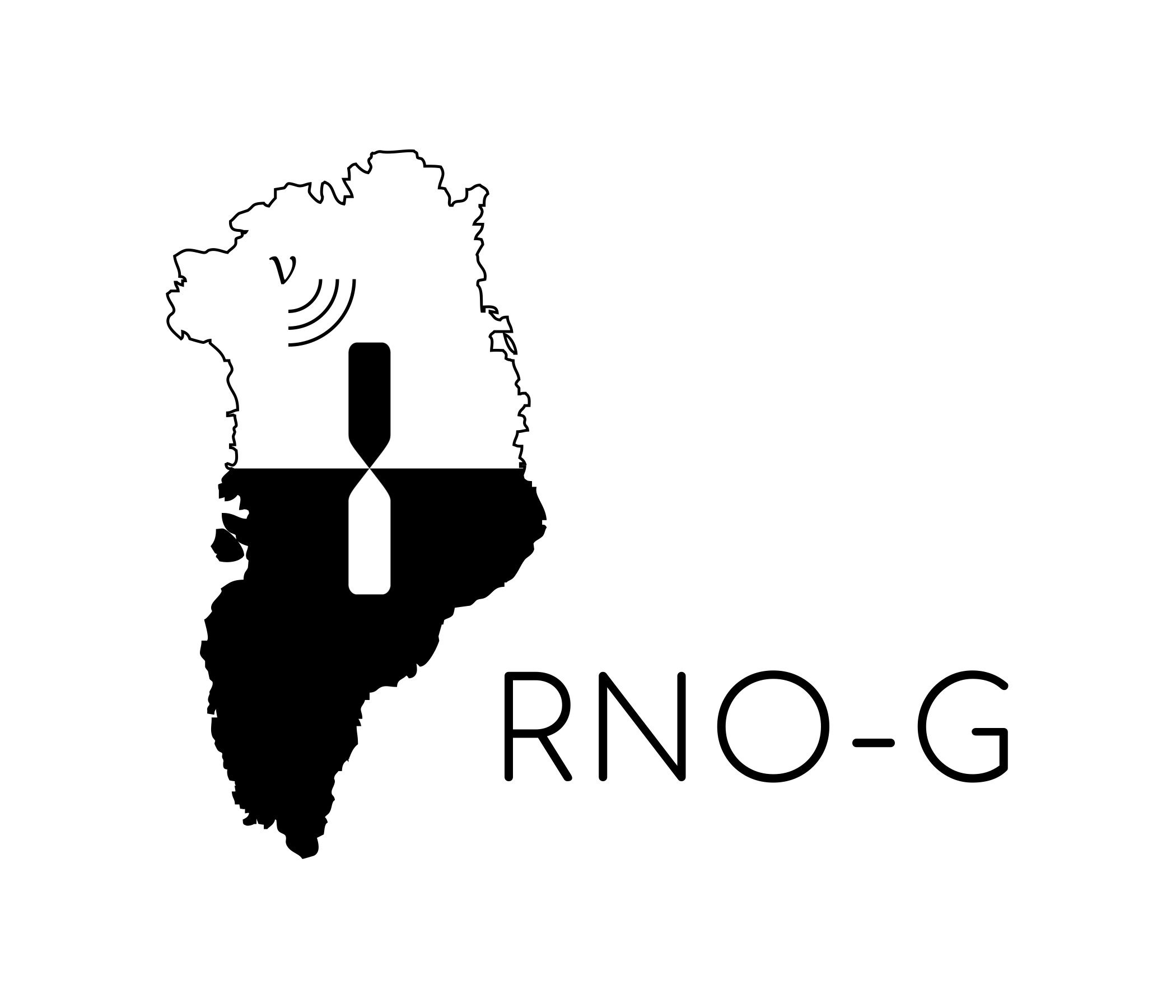}\\[6pt]
RNO-G Collaboration}

\author[1]{S.~Agarwal}
\author[2]{J.~A.~Aguilar}
\author[3]{N.~Alden}
\author[1]{S.~Ali}
\author[4]{P.~Allison}
\author[5]{M.~Betts}
\author[1]{D.~Besson}
\author[6]{A.~Bishop}
\author[8]{O.~Botner}
\author[9]{S.~Bouma}
\author[10,11]{S.~Buitink}
\author[2]{R.~Camphyn}
\author[9]{M.~Cataldo}
\author[2]{S.~Chiche}
\author[12]{B.~A.~Clark}
\author[8]{A.~Coleman}
\author[1]{K.~Couberly}
\author[7]{S.~de~Kockere}
\author[7]{K.~D.~de~Vries}
\author[3]{C.~Deaconu}
\author[8]{C.~Glaser}
\author[8]{T.~Gl{\"u}senkamp}
\author[8]{A.~Hallgren}
\author[13]{S.~Hallmann}
\author[14]{J.~C.~Hanson}
\author[5]{B.~Hendricks}
\author[13,9]{J.~Henrichs}
\author[8]{N.~Heyer}
\author[1]{C.~Hornhuber}
\author[4]{K.~Hughes}
\author[13]{T.~Karg}
\author[6]{A.~Karle}
\author[6]{J.~L.~Kelley}
\author[21]{C.~Kerr}
\author[9]{C.~Klein}
\author[2,7]{M.~Korntheuer}
\author[13,15]{M.~Kowalski}
\author[16]{I.~Kravchenko}
\author[5]{R.~Krebs}
\author[9]{R.~Lahmann}
\author[7]{U.~Latif}
\author[9]{P.~Laub}
\author[16]{C.-H. Liu}
\author[17]{M.~J.~Marsee}
\author[13,9]{Z.~S.~Meyers}
\author[1]{M.~Mikhailova}
\author[11]{K.~Mulrey}
\author[5,6]{M.~Muzio}
\author[13,9]{A.~Nelles}
\author[18]{A.~Novikov}
\author[1]{A.~Nozdrina}
\author[3]{E.~Oberla}
\author[19]{B.~Oeyen}
\author[21]{S.~Polfrey}
\author[18]{N.~Punsuebsay}
\author[13,9]{L.~Pyras}
\author[8]{M.~Ravn}
\author[9]{M.~Reichert}
\author[21]{J.~Rix}
\author[19]{D.~Ryckbosch}
\author[2]{F.~Schl{\"u}ter}
\author[7,20]{O.~Scholten}
\author[18]{D.~Seckel}
\author[1]{M.~F.~H.~Seikh}
\author[3]{D.~Smith}
\author[7]{J.~Stoffels}
\author[9]{K.~Terveer}
\author[2]{S.~Toscano}
\author[6]{D.~Tosi}
\author[5]{J.~Tutt}
\author[7,10]{D.~J.~Van Den Broeck}
\author[7]{N.~van Eijndhoven}
\author[3]{A.~G.~Vieregg}
\author[12]{A.~Vijai}
\author[3]{C.~Welling}
\author[17]{D.~R.~Williams}
\author[3]{P.~Windischhofer}
\author[21]{J.~Veale}
\author[5]{S.~Wissel}
\author[1]{R.~Young}
\author[9]{A.~Zink}
\affiliation[1]{ University of Kansas, Dept.\ of Physics and Astronomy, Lawrence, KS 66045, USA }
\affiliation[2]{Universit\'e Libre de Bruxelles, Science Faculty CP230, B-1050 Brussels, Belgium }
\affiliation[3]{ Dept.\ of Physics, Enrico Fermi Inst., Kavli Inst.\ for Cosmological Physics, University of Chicago, Chicago, IL 60637, USA }
\affiliation[4]{Dept.\ of Physics, Center for Cosmology and AstroParticle Physics, Ohio State University, Columbus, OH 43210, USA }
\affiliation[5]{Dept.\ of Physics, Dept.\ of Astronomy \& Astrophysics, Center for Multimessenger Astrophysics, Institute of Gravitation and the Cosmos, Pennsylvania State University, University Park, PA 16802, USA }
\affiliation[6]{ Wisconsin IceCube Particle Astrophysics Center (WIPAC) and Dept.\ of Physics, University of Wisconsin-Madison, Madison, WI 53703,  USA }
\affiliation[7]{ Vrije Universiteit Brussel, Dienst ELEM, B-1050 Brussels, Belgium }
\affiliation[8]{ Uppsala University, Dept.\ of Physics and Astronomy, Uppsala, SE-752 37, Sweden }
\affiliation[9]{ Erlangen Centre for Astroparticle Physics (ECAP), Friedrich-Alexander-University Erlangen-N\"urnberg, 91058 Erlangen, Germany }
\affiliation[10]{ Vrije Universiteit Brussel, Astrophysical Institute, Pleinlaan 2, 1050 Brussels, Belgium }
\affiliation[11]{ Dept.\ of Astrophysics/IMAPP, Radboud University, PO Box 9010, 6500 GL, The Netherlands }
\affiliation[12]{ Department of Physics, University of Maryland, College Park, MD 20742, USA }
\affiliation[13]{ Deutsches Elektronen-Synchrotron DESY, Platanenallee 6, 15738 Zeuthen, Germany }
\affiliation[14]{ Whittier College, Whittier, CA 90602, USA }
\affiliation[15]{ Institut f\"ur Physik, Humboldt-Universit\"at zu Berlin, 12489 Berlin, Germany }
\affiliation[16]{ Dept.\ of Physics and Astronomy, Univ.\ of Nebraska-Lincoln, NE, 68588, USA }
\affiliation[17]{ Dept.\ of Physics and Astronomy, University of Alabama, Tuscaloosa, AL 35487, USA }
\affiliation[18]{ Dept.\ of Physics and Astronomy, University of Delaware, Newark, DE 19716, USA }
\affiliation[19]{ Ghent University, Dept.\ of Physics and Astronomy, B-9000 Gent, Belgium }
\affiliation[20]{ Kapteyn Institute, University of Groningen, PO Box 800, 9700 AV, The Netherlands }
\affiliation[21]{British Antarctic Survey (BAS), Cambridge
CB3 0ET, United Kingdom}

\emailAdd{authors@rno-g.org}
\emailAdd{ejo@uchicago.edu}

\abstract{The Radio Neutrino Observatory in Greenland (RNO-G) is the first in-ice radio array in the northern hemisphere for the detection of ultra-high energy neutrinos via the coherent radio emission from neutrino-induced particle cascades within the ice. The array is currently in phased construction near Summit Station on the Greenland ice sheet, with 7~stations deployed during the first two boreal summer field seasons of 2021 and 2022. In this paper, we describe the installation and system design of these initial RNO-G stations, and discuss the performance of the array as of summer 2024.}


\arxivnumber{1234.56789} 

\begin{document}
\maketitle

\section{Introduction}

Ultra-high energy (UHE) neutrinos, above 10~PeV, can be detected in dense radio-frequency (RF) transparent media through the Askaryan effect~\cite{Askaryan:1961pfb}.
Cold glacial ice, available in the polar regions with the necessary abundant volumes, has been measured to have exceptional RF properties~\cite{Barwick_2005} and continues to be the best detector medium for measuring these UHE particles nearly 40~years from  the inception of the experimental concept~\cite{MARKOV1986242}. Experimental efforts towards large-scale in-ice radio detectors have so far been focused in Antarctica, through the RICE, ARA, and ARIANNA projects~\cite{Kravchenko:2011im,Allison:2011wk,Anker:2019rzo}.
Complementary radio-detection efforts, which use long-duration balloon payloads overflying Antarctica, were done in a series of four flights by the ANITA project~\cite{GORHAM200910} and will be advanced by the upcoming PUEO mission~\cite{PUEO}. There has yet to be a detection of UHE neutrinos with these radio instruments, though the continued technology advancement, simulation maturity, and field deployment experiences have provided a valuable knowledge base to scale up the experimental design to the requisite detector volumes.

The Greenlandic ice sheet was first proposed as a suitable
location for a UHE neutrino radio detector with measurements 
of the RF attenuation length near Summit Station in 
2013~\cite{Avva:2014ena} and then a conceptual
experimental design known as the Greenland Neutrino Observatory (GNO)
that presented early field testing results~\cite{Avva:2016ggs}.  
 With GNO on indefinite hiatus, a new experimental effort to build a 
large-scale radio array reemerged in the form of the 
Radio Neutrino Observatory Greenland (RNO-G),  
which was first described in a 2021 whitepaper~\cite{RNO-G:2020rmc}.
RNO-G followed the initial RF attenuation length measurements
with a detailed study in 2021, confirming the suitability of the 
Greenland ice sheet as a radio-detection volume for UHE 
neutrinos with a measured attenuation length of 
$\sim$1~km at 200~MHz~\cite{Aguilar:2022kgi}.

Neutrino-induced particle cascades in dense media, such as ice,
emit coherent broadband radio waves via the Askaryan effect, which
describes the charge-excess nature
of such cascades~\cite{Askaryan:1961pfb}. 
The high-energy particle shower
develops a negative charge excess, which coherently emits
radio over a broadband frequency spectrum, 
and for high-energy extended
tracks, the radio power is beamed in a relatively
narrow angular range around a conical section
defined by the Cherenkov angle. This emission mechanism has been 
confirmed in a set of accelerator based measurements 
using a variety of dense media, including ice~\cite{ANITA:2006nif,T-510:2015pyu,Bechtol:2021tyd}.
Thus, the neutrino signature
at the radio receiver, within the RNO-G band of $\sim$100-700~MHz,
is a nanosecond-scale electric field impulse with Cherenkov-like
characteristics.  The radio-wave polarization of the electric field
points radially inwards towards the shower axis. 

The science case of RNO-G, namely the discovery and measurement of neutrinos above 10~PeV in energy, has been extensively discussed in recent comprehensive reviews, i.e.~\cite{Ackermann:2022rqc, MammenAbraham:2022xoc, Ackermann:2019ows,Ackermann:2019cxh}. A conclusive picture emerges that predicts a potentially detectable flux of neutrinos, which is a result of the interaction of ultra-high energy cosmic rays and various photon backgrounds in the universe. RNO-G aims to improve existing limits \cite{ANITA:2019wyx,ARA:2019wcf,Anker:2019rzo,IceCube:2018fhm,PierreAuger:2015aqe} on the neutrino flux, if not discover the flux of EeV neutrinos. 

 This article reports on the design and performance of the first seven stations of RNO-G, which is planned to consist of 35 stations when the first phase of construction is complete. Each RNO-G station comprises 24-antenna receivers installed down to a depth of 100~m. The RNO-G work effort began in earnest in mid-2019, when design and construction of 10~full RNO-G stations of hardware components was initiated, followed by field deployments to Greenland in 2021 and 2022 during which time the initial seven stations were installed and commissioned.
 After these first field seasons, and affected by the global chip shortage, RNO-G paused deployment in 2023 in order to evaluate the instrument performance
 and design aspects before pushing forward to complete the array. 
 This article will describe these in detail, 
 with the experimental overview including drilling in \autoref{sec:experiment},
 the instrument design in \autoref{sec:instrument},
 and a review of the initial experiment performance in \autoref{sec:performance}.

\section{Experiment and Deployment Overview}
\label{sec:experiment}

RNO-G is a radio array composed of independent stations, which are designed to autonomously detect and record transient signals. The array is currently under construction adjacent to Summit Station near the top of the Greenlandic ice sheet. Summit Station is operated by the National Science Foundation (NSF) and provides the deployment base-camp and communications hub for the experiment.

The RNO-G design requires a scalable and robust design approach such that the stations can be efficiently deployed and can operate within a small power envelope to run from renewable-energy power sources. 
The experiment is enabled by a wide range of custom-designed low-power radio-frequency (RF), digital, and power electronics that were developed at an expeditious pace to meet the deployment goals for the 2021 and 2022 field seasons. 
Additionally, a new drill developed by British Antarctic Survey (BAS) is used to make the \SI{100}{m} boreholes for the stations. In this section, we discuss the overall array plan, 
the RNO-G station layout, and the drilling aspects of the project.

\subsection{Array layout}
\begin{figure}
\centering
\includegraphics[width=0.64\textwidth]{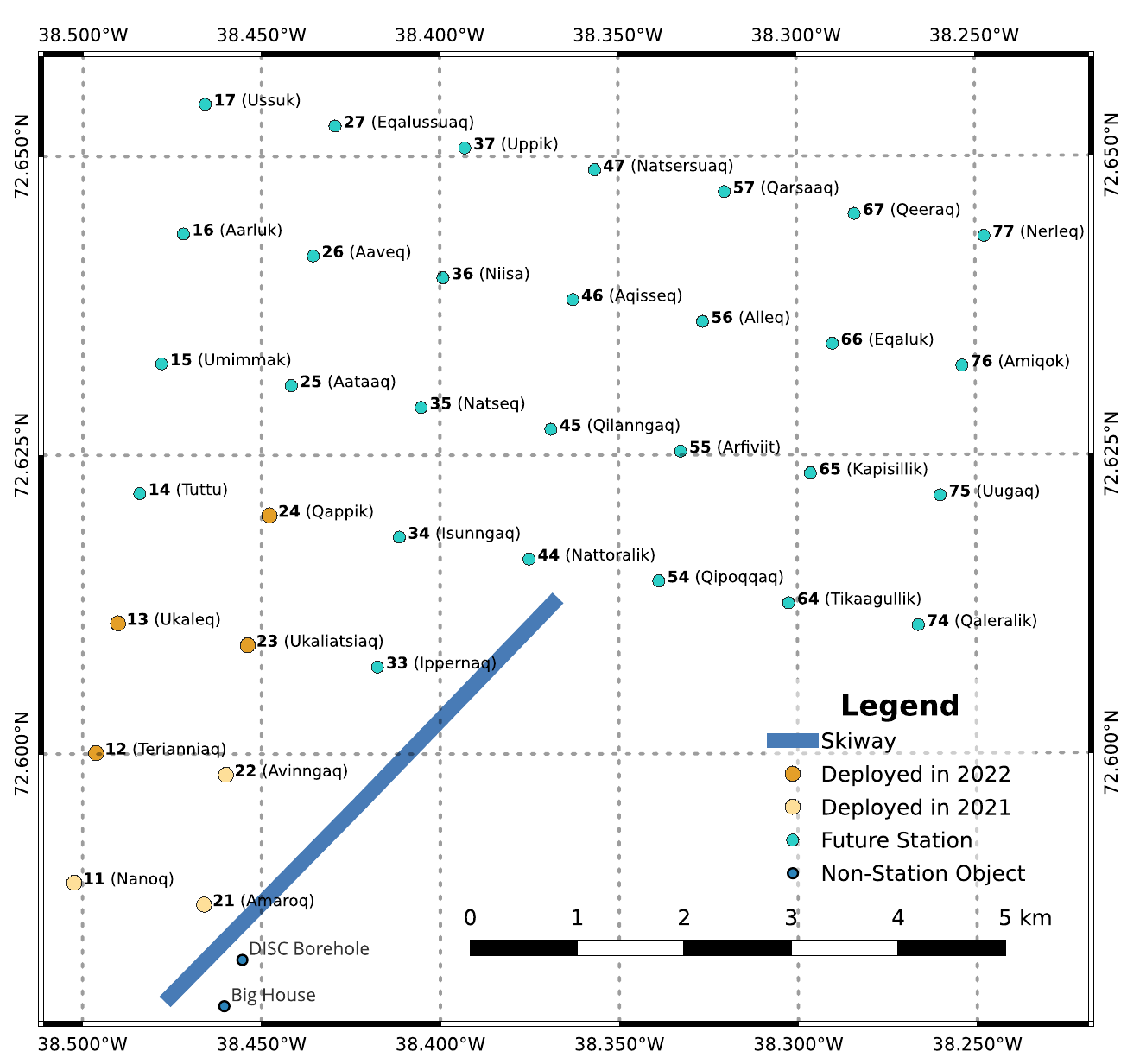}
\includegraphics[width=0.35\textwidth]{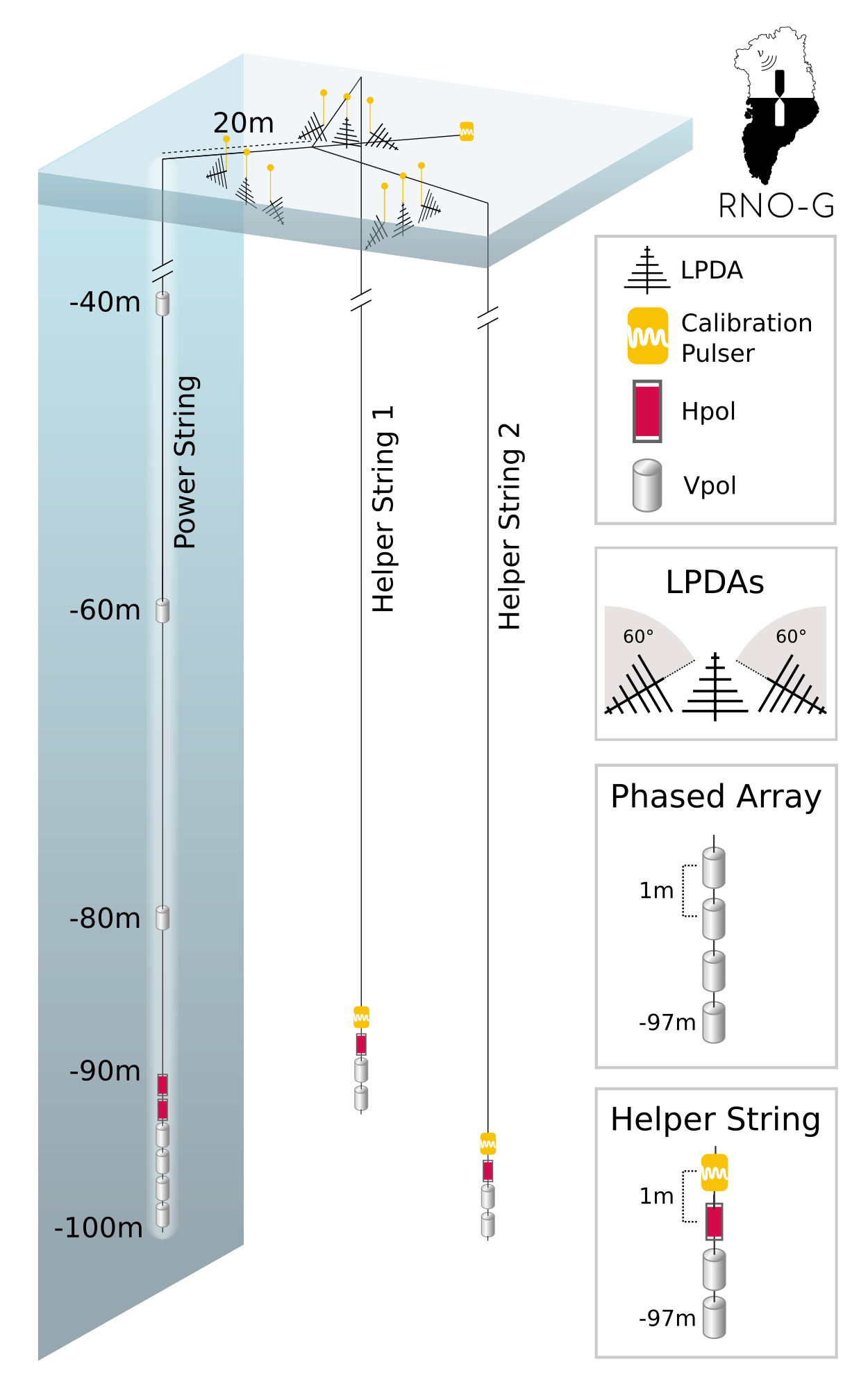}
\caption{Left: Planned and existing layout of the 35 station array of RNO-G. The first seven stations, the subject of this paper, were deployed in 2021 and 2022. The main building of Summit Station is labeled \textit{Big House}. Right: The station construction as installed for the first seven stations of RNO-G, showing the three instrumented boreholes down to \SI{100}{m}, as well as the antennas in shallow trenches near the surface.}
\label{fig:array_and_station_layout}
\end{figure}

The array layout as planned for the first 35 stations is shown in \autoref{fig:array_and_station_layout}. The square grid has an inter-station spacing of \SI{1.25}{km}, which was chosen so that the effective neutrino-detection volume of each station is independent in the lower energy range ($<$300~PeV) and leads to inter-station coincident measurements at the 10\% level for neutrino interactions at the EeV energy scale, based on simulation studies in the design phase of the project~\cite{RNO-G:2020rmc}. This is driven mostly by the attenuation length \cite{Aguilar:2022kgi,Aguilar:2022ijn} and the refractive index profile of the ice \cite{Deaconu:2018bkf,Aguilar:2023udv}.

The stations are denoted by a grid numbering scheme (XY) and, colloquially, by fauna endemic to Greenland, using the local language \emph{kalaallisut}. The first seven installed stations of RNO-G (RNO-G-7) are \emph{Nanoq} (Polar Bear, 11), \emph{Terianniaq} (Arctic Fox, 12), \emph{Ukaleq} (Arctic Hare, 13), \emph{Amaroq} (Arctic Wolf, 21), \emph{Avinngaq} (Arctic Lemming, 22), \emph{Ukaliatsiaq} (Stoat, 23), and \emph{Qappik} (Wolverine, 24). The location of the array with respect to Summit Camp proper and the skiway is restricted by the requirements of aviation safety, the clean air and snow sectors to the west and south, as well as the IceSat-2 protected area to the north-east \cite{Kelly2017}. 

\subsection{Station plan}
The layout plan of the RNO-G station is shown on the right in \autoref{fig:array_and_station_layout}, which emerged from design studies presented in the whitepaper~\cite{RNO-G:2020rmc}. Each station is composed of 24 low-noise receiving antennas installed in a hybrid configuration with both near-surface and deep receivers.

The radio signals are received at antennas installed at different depths as shown in the right panel of \autoref{fig:array_and_station_layout}. In all polar ice, the density of the ice increases with depth within the firn ($\sim$80~m near Summit). It follows that radio waves from a deep source, such as the Askaryan-generated emission from an UHE neutrino, propagate along curved ray paths as they approach the surface so that antennas deployed deeper in the ice are sensitive to a larger volume of ice.  Antennas that are installed closer to the surface, while contributing less to the overall neutrino sensitivity, can be used to tag potential natural or human-made backgrounds originating above the ice. For the RNO-G station, the linear polarization angle of the Askaryan signal is used, in part, to reconstruct the neutrino arrival direction, so both the vertical and horizontal components of the electric field are measured with the borehole antennas.

The deep antennas are deployed in three boreholes down to 100 meters depth. Due to the signal propagation effects described above,  the main neutrino effective volume arises from the closely packed triggering subarray of four antennas at the bottom of what is called the \emph{power string}. The \emph{trigger threshold} on this subarray largely determines the neutrino detection sensitivity of an RNO-G station. This trigger threshold is defined as a transient signal-to-noise ratio, which relates the received signal voltage magnitude to the combined system and environmental noise, as discussed in~\autoref{sec:performance}.
The distributed antennas on the \emph{power string} have long vertical baselines, which drive the energy resolution obtainable with RNO-G \cite{Aguilar:2021uzt}. The two \emph{helper strings} provide horizontal baselines for both deep-receiver polarizations, which are particularly needed for reconstructing the neutrino arrival direction. Combined, the three borehole strings form an equilateral triangle with horizontal baselines of $\sim$\SI{35.5}{m} equipped with deep sub-arrays of antennas. These provide the baselines for time-domain pulse interferometry that are sufficient for sub-degree azimuth and zenith reconstruction of the impinging radio signal.~\cite{Plaisier:2023cxz}. In general, the \emph{neutrino} reconstruction will depend on the details of the event and, in most cases, will be limited by the measurement of the polarization angle~\cite{Bouma:2023koi}. 

The near-surface antennas are connected to a dedicated receiver and trigger block, which provides both additional neutrino effective volume and a tag for air shower backgrounds \cite{DeKockere:2022bto,Pyras:2023crm}. All of the receiver channels are connected to one instrument box that contains the data acquisition (DAQ) system, and provides power and the wireless communication links (\autoref{sec:instrument}).   

As the initial stations comprising RNO-G-7 are run solely through solar power, we required a station power budget of about 25~W to allow for science operations at least 50$\%$ of the year. The station design also required a low-power hibernation mode (<50~mW), in order to preserve the batteries during the polar night.

\subsection{Drilling}

\begin{figure}
\begin{minipage}[t]{.4\textwidth}%
\includegraphics[width=\textwidth,valign=T]{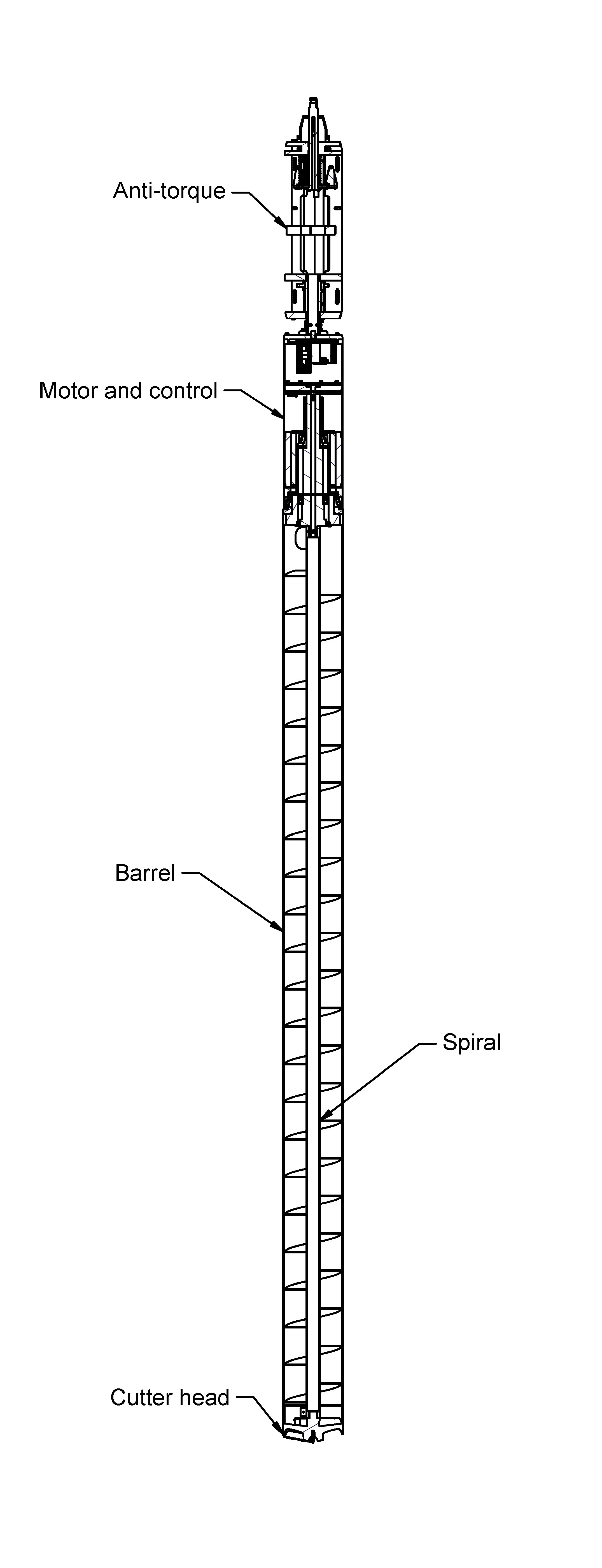}
\end{minipage}%
\hfill%
\begin{minipage}[t]{.6\textwidth}
\centering
\includegraphics[width=0.75\textwidth,valign=T]{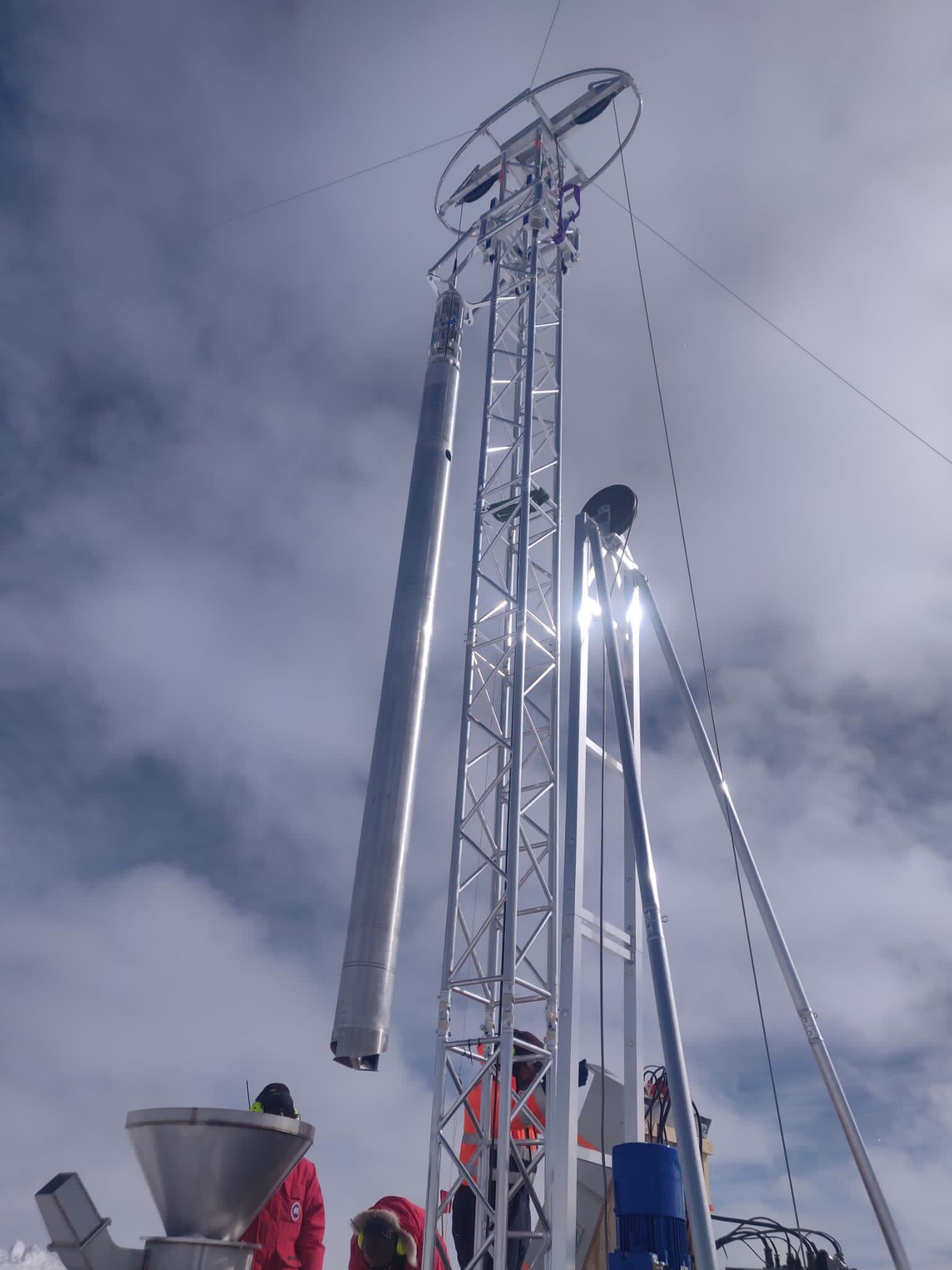}%
\medskip

\includegraphics[width=0.9\textwidth,valign=T]{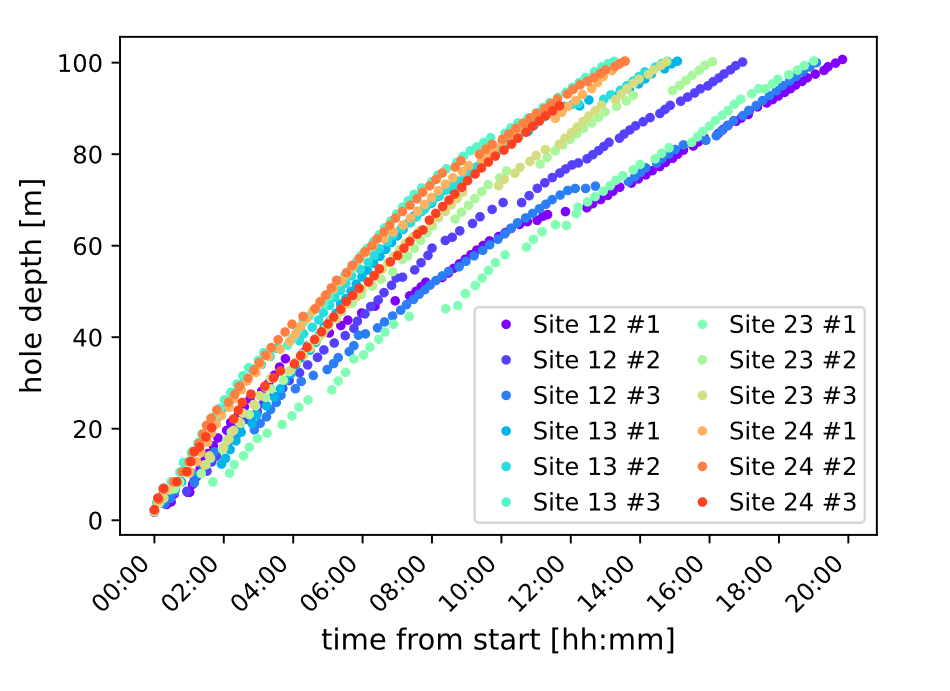}%
\end{minipage}%
\caption{The BigRAID of RNO-G. Left: Section-view drawing of the drill sonde, which is composed of the anti-torque, drill motor, barrel, and cutters/cutter-head sections. Right (top): Photo of BigRAID testing early in the 2021 summer season, showing the drill sonde supported with the tower structure without the protective tent. At the bottom left in the photo is the `snowblower', which catches and disperses the ice chips collected in the barrel after every drill run. Right (bottom): Performance of the BigRAID drill during the season of 2022 at Summit Station in Greenland, in which four complete RNO-G stations were drilled. The fastest hole was drilled within 13 hours. Further efficiency improvements are planned for future drilling seasons.}
\label{fig:drill_rates}
\end{figure}

The 100-m RNO-G boreholes were drilled using the BigRAID, an electromechanical drill that operates in a similar manner to a conventional ice-coring drill, but instead cuts the entire cylindrical ice-core section into small chips as opposed to preserving a pristine core. These chips are captured in the long drill-barrel section and returned to the surface where they are discarded (in our application), leaving a dry borehole ready for the installation of instrumentation. The BigRAID was developed from the original BAS Rapid Access Isotope Drill (RAID) design, which was optimized for efficient drilling speed and borehole access, instead of core retrieval~\cite{RIX_MULVANEY_HONG_ASHURST_2019}. With a diameter of \SI{28.5}{cm} the BigRAID delivers much larger boreholes than the \SI{8.2}{cm} diameter RAID.

The BigRAID sonde is detailed and pictured in \autoref{fig:drill_rates}.
The drill-sonde is suspended on a cable, traversing vertically in the hole as controlled by a custom-designed winch system from MacArtney. The winch is equipped with a 4-conductor \SI{260}{m} cable and has a safe working load of \SI{5}{kN}. In typical drilling operations, the depth of the hole increases by 1-2~m on any individual run (i.e.\ lowering, drilling, and raising), as the ice chips are collected within the \SI{4}{m} long barrel and then brought to the surface. The incremental drilling-depth per run depends on the structure of the ice and how quickly the chips fill the drill barrel. This typically decreases with depth, with drill runs of $\sim$2m common at the onset and runs of $\sim$1-1.5m at depth. After every drill run, the ice-chippings are discharged from the barrel through a snowblower, which dissipates the chips over the snow surface to avoid concentrated ice deposits at the RNO-G site.  The entire BigRAID drilling rig is housed on a 45 ft (\SI{13.7}{m}) long HMW-PE sled for transport, which is towable by a snowcat or similar tracked vehicle.

The drill-sonde is composed of four sections, from bottom to top: the cutters, the chip-collecting barrel, the motor, and the anti-torque sections. 
The brushless motor provides up to \SI{220}{Nm} continuous torque and its rotator is directly coupled to the barrel (without gearing). The barrel is equipped with two cutters at the lower end, which are similar to augers used for ice fishing. These cutters are manufactured from hardened-A2 tool steel that allows for field sharpening, while also remaining sharp for a useful duration (>\SI{100}{m} of drilling). During drilling, the chips, inside the rotating barrel, are pushed up the flights of a static auger. 
The motor rotation, direction, and speed can be changed during operations and an encoder allows the cutter position to be accurately set when stopped. 
At the top of the drill-sonde assembly, the (in-ice) anti-torque section provides engagement with the borehole wall using a set of retractable blades, which counteracts the torque from the driving motor and enables the sonde to drill ice. A separate above-surface anti-torque device is used when starting each hole, during which time the drill-sonde anti-torque section is either above the surface or in soft firn that does not reliably engage. 
 
The boreholes are extremely vertical due to the long length of the drill-sonde, with small deviations of $<$1$^\circ$ measured with initial inclinometer data. More systematic studies of the boreholes will be performed in upcoming seasons, as inclinometers were directly added to the drill sonde in 2024.  

During the 2022 season, the BigRAID and its operators delivered twelve 100~m deep instrumentation holes during an 8-week drilling season at Summit Camp, which includes constructing and packing up the drill setup, as shown with the per-hole completion rates in \autoref{fig:drill_rates}. 
The 2022 drilling effort also encountered two cases of a stuck drill sonde, one  at shallow depth ($\sim$4m) and the second deep ($\sim$86m). The sonde was successfully retrieved in both situations and modest redesign efforts are ongoing to mitigate stuck-drill issues going forward.
Efforts to automate the steering and drilling process are also ongoing, to streamline the number of holes that can be drilled in the upcoming field seasons. 

\begin{figure}
\centering
\includegraphics[width=0.99\textwidth, trim={8cm 0.1cm 0.05cm 5.2cm},clip]{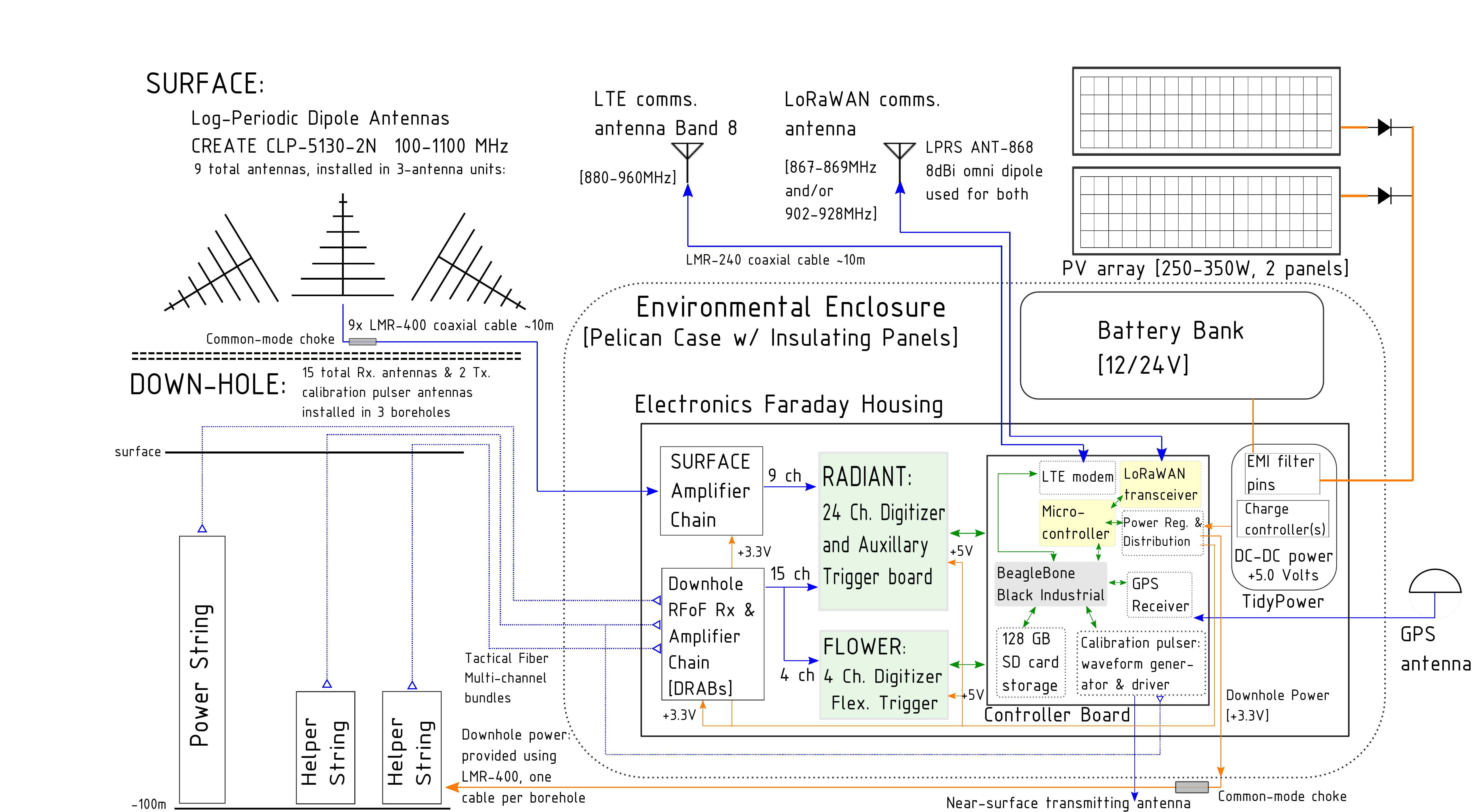}
\caption{System diagram of an RNO-G station}
\label{fig:sys_diagram} 
\end{figure}

\section{Instrument design}
\label{sec:instrument}
In this section, we will describe the RNO-G instrumental design in detail.
\autoref{fig:sys_diagram} provides a technical overview of the RNO-G station.

\subsection{Antennas}
\label{sec:antennas}
The antennas are the primary sensors for the RNO-G experiment, receiving electric fields generated by neutrino interactions, the background thermal noise from the ice, anthropogenic transmitters on the ice surface, and other radio sources.
Identifiable sources are reconstructed by using time-domain interferometry across the receiving antennas within the station, through the measurement of the radio-wave polarization, and through other properties such as the impulsivity and frequency content of the signal.

The antennas are characterized by a complex impedance and gain response. We distinguish between `in-air' and `in-ice' as these attributes are dependent on the dielectric environment in which the antennas are installed.  The in-air simulations allow a more straightforward verification of antenna performance in the lab, while the in-ice modeling describes the actual performance within the experiment.

The voltage standing wave ratio (VSWR) describes the impedance match between the output of the antenna and input of the front-end amplifier via the expression 
\begin{equation}
    \mathrm{VSWR} =  \frac{1+ \left| \rho \right|}{1 - \left| \rho \right|} 
\end{equation}
where $\rho$ is the frequency-dependent complex voltage reflection coefficient. The front-end amplifier units are designed to present a constant 50~$\Omega$ real-valued impedance across the RNO-G signal band (see \autoref{sec:rf_chain}), such that the frequency dependence of the VSWR is largely determined by the antenna impedance.  To maximize the transfer of signal power to the RF signal chain, the RNO-G antennas are designed to match this 50$\Omega$ impedance as best as possible.

The antenna gain describes the directive amplitude response, which corresponds to the effective power-collecting area of the antenna. 
As RNO-G directly measures the voltage waveforms, the antenna vector effective length (VEL) best characterizes the antenna gain response. The VEL, defined along the antenna E-plane axis, relates the incident electric field vector to the open-circuit output voltage $V_{OC}$ of the antenna and is given by the expression
\begin{equation}
    V_{OC} = \overrightarrow{E} \cdot \overrightarrow{\mathrm{VEL}}
\end{equation}
For a perfectly matched antenna and receiver system, the voltage actually measured is $V = V_{OC}/2$. More generally, we characterize a realized VEL (RVEL) that includes the effects of impedance mismatches between the antenna and receiver as described by the VSWR.
The RVEL corresponds to the realized gain pattern of the antenna.

The RNO-G station employs three different antenna types. A commercial log-periodic dipole array (LPDA) antenna is used near the surface, which provides a relatively high-gain directional response. For the deep receivers, a set of custom cylindrically-symmetric antennas were designed with the constraints imposed by the borehole-installation requirement. Of the borehole antennas, one is a vertically-polarized antenna (Vpol) that is sensitive to the vertical component of the electric field, and the other is a horizontally-polarized antenna (Hpol) that inductively couples to a magnetic field along the antenna axis, providing a measure of horizontal electric fields. The antennas were designed and characterized with a finite-difference time-domain antenna simulation software (XFdtd~\cite{XF}) and their as-built responses have been measured in the lab. In-ice measurements have also been performed, but will be presented in a detailed follow-on calibration paper.

The borehole antennas include water-jet cut nylon endcaps that have a diameter of 8 and 9.5 inches (20.3 and 24.1 cm) for the Vpols and Hpols, respectively. The endcaps provide a housing for the downhole front-end amplifiers and fiber transceivers, protection for fiber and cable routing, and serves as a minimal `spacer' between the antenna and the ice. The spacer functionality aids in centralizing the antennas within the borehole, which has been shown in simulations to reduce systematic uncertainties in the antenna response due to potential offsets in the borehole~\cite{RNO-G:2021upe}. Crucially, the endcaps also prevent the metal antenna elements from contacting the borehole wall, which was found to induce a severe azimuthal asymmetry in the antenna response in detailed simulations. The borehole antenna units span \SI{90}{cm} in vertical length when the frame, endcaps, and front-end amplifier enclosures are fully installed.

\begin{figure}
\centering
\includegraphics[width=1\textwidth, trim={0 2.5cm 0 0cm},clip]{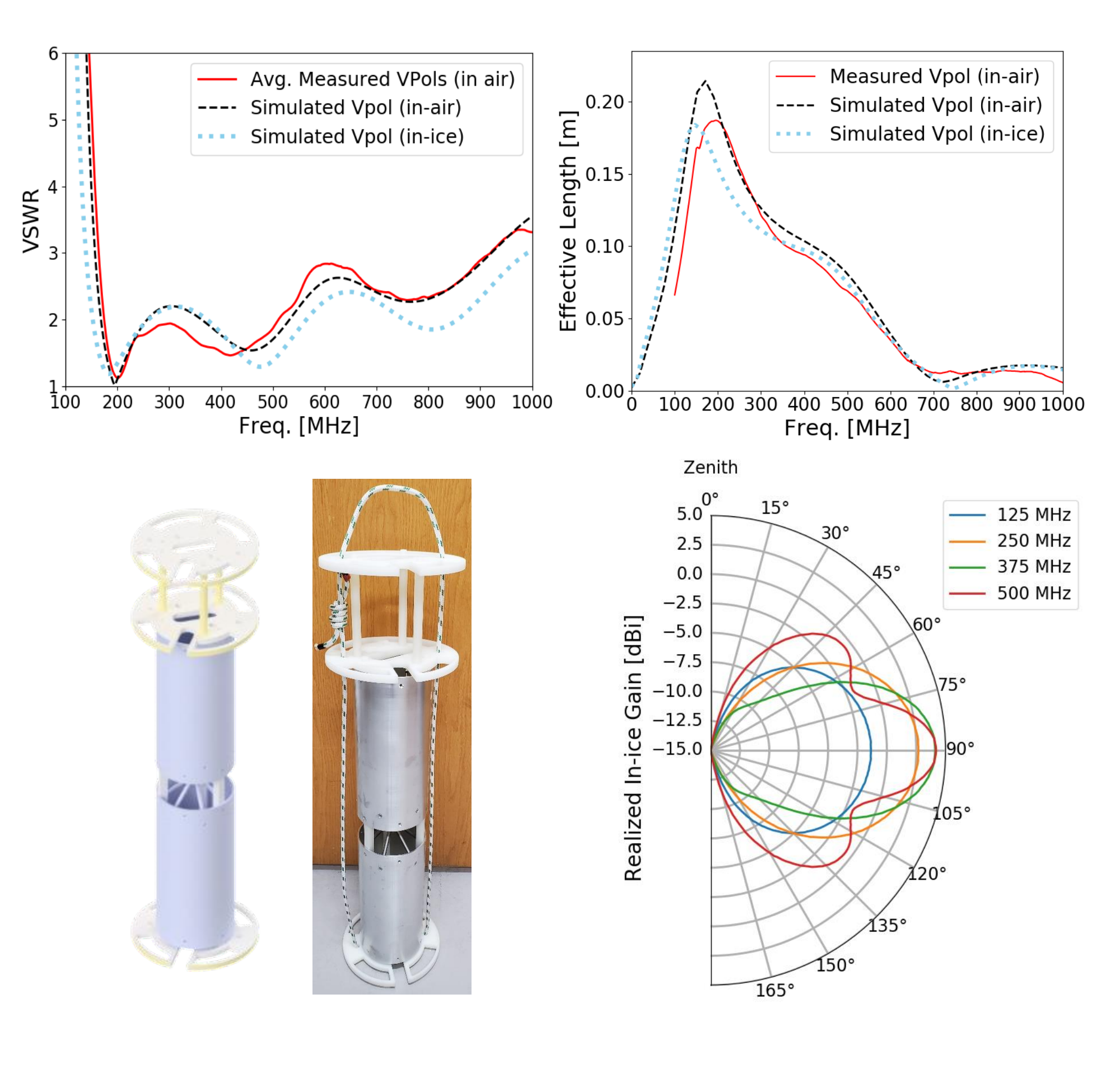}
\caption{The RNO-G vertically-polarized (Vpol) borehole antenna. Bottom left: CAD model of the Vpol and photograph of the assembled Vpol antenna in the lab, including the deployment frame and rigging rope. Top left: The average of the measured in-air VSWR of all assembled and shipped RNO-G Vpols is compared to XFdtd simulations in-air and deep ice.  Top right: Theta (co-polarized) component of the realized vector effective length measured in-air as compared to XFdtd simulations in-air and deep ice, shown at boresight. Bottom right: Realized gain pattern of the Vpol at representative frequencies, simulated in the deep ice borehole.}
\label{fig:vpol_display} 
\end{figure}

\subsubsection{Vpols}

The Vpol antennas, shown in \autoref{fig:vpol_display}, are a fat-dipole antenna design with a symmetric feedpoint between two 12.7~cm diameter cylindrical elements for a total length of 60~cm~\cite{RNO-G:2021upe}.
The wide diameter of the elements provides the broad bandwidth.
The feed transition design, following from the RICE~\cite{Kravchenko:2011im} and GNO dipole~\cite{Avva:2016ggs} antennas, uses a conical section screwed into the sides of the aluminum walls; the opening-angle of which was tuned to optimize the VSWR. The conical transition section for the RNO-G dipole is made from a set of 8 machined spokes to reduce the overall weight and cost. A 50~$\Omega$ SMA connector sits at the feedpoint and effectively provides the balanced-to-unbalanced conversion (balun) to the coaxial cable. A 12" RG316 double-shielded coaxial cable connects the antenna feed to the low-noise amplifier unit housed just above the antenna.
 
The Vpols have a 2:1 VSWR in-ice bandwidth of $\sim$\SI{140}-\SI{600}{MHz} as shown in \autoref{fig:vpol_display}. Before shipping to the field, each fabricated antenna is measured in the lab and shows excellent agreement to the in-air simulated VSWR. The zenith-angle response pattern has a clear primary boresight beam until $\sim$\SI{550}{MHz}, above which the response starts to go multimodal.

The simulated and measured RVELs are also shown in \autoref{fig:vpol_display}, with the measurement done in an anechoic chamber. There is good agreement between the measured and simulated in-air RVEL at the 10$\%$ level. 
Once embedded in the deep ice (n=1.8) borehole, the RVEL response curve shifts down by 30-40~MHz, which roughly tracks the same shift given by the simulated in-ice VSWR. 

Identical Vpols are also embedded in the station as transmitting antennas used for the calibration pulsers, which are installed on the two helper strings as well as near the surface. These are equipped with different front-end analog amplifier units that are designed for pulsed-radio transmission instead of reception.
 
\begin{figure}
\centering
\includegraphics[width=1\textwidth, trim={0 2.0cm 0 1.2cm},clip]{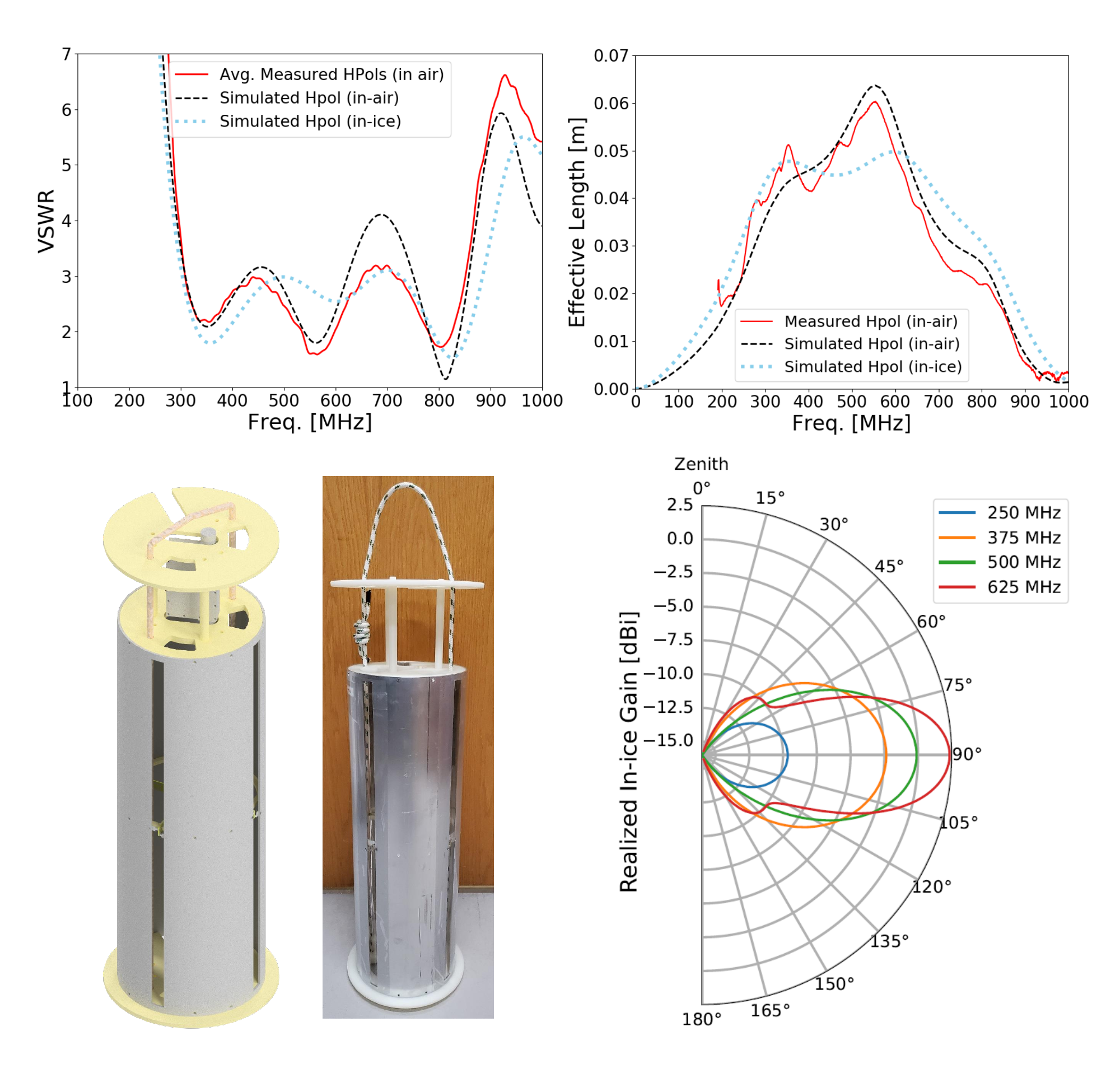}
\caption{The RNO-G horizontally-polarized (Hpol) borehole antenna. Bottom left: CAD model of the Hpol and the assembled Hpol antenna in the lab, including the deployment frame and rigging rope. Top left: The average of the measured in-air VSWR of all assembled and shipped RNO-G Hpols is compared to XFdtd simulations in-air and deep ice.  Top right: Phi (co-polarized) component of the realized vector effective length from XFdtd simulations in-air and deep ice, shown at boresight (90$^\circ$ zenith), compared to a lab measurement. Bottom right: Realized gain pattern of the Hpol at representative frequencies, simulated in the deep ice borehole.}
\label{fig:hpol_XFDTD}
\end{figure}

\subsubsection{Hpols}
The Hpol antennas, shown in \autoref{fig:hpol_XFDTD}, are a quad-slotted cylinder antenna design with a passive matching and feed network.
Slotted cylinder antennas offer a reasonably high gain-bandwidth solution for measuring the toroidal electric field component, given practical limitations of the borehole installations. The slotted antennas operate via Booker's extension of Babinet's principle to radiate/receive perpendicular to the vertical slots in the cylinder~\cite{RNO-G:2023xyg}.
The wider BigRAID \SI{28.5}{cm} boreholes allow for optimizing the quad-slot design: for a single-slotted cylinder, the antenna diameter increases the low-end bandwidth.
However, to maintain cylindrical symmetry in the beam response, four slots were implemented that get combined into a single feed-point. The slot multiplicity again reduces the low-end bandwidth due to decreasing the 
shunt inductance of the cylinder walls. 

The Hpol quad-slot design was informed by ARA, 
which solved this low-frequency
response issue by loading the inside of the cylindrical antenna
with a large amount of ferrite material~\cite{Allison:2011wk}. This improved the low-frequency response (down to $\sim$150~MHz), but proved difficult to accurately model in simulations and made for a rather heavy antenna unit.
The RNO-G quad-slot design did not consider adding ferrites, but rather
used a passive (reactive) matching network, feeding each slot, to 
get the best performance from the antenna. The matching network is formed using a hub-and-spoke printed circuit board design, with four spoke boards feeding each balanced slot with the matching network, and connected at the hub in parallel.
The hub incorporates an SMA connector in the center that serves as the balun. 
Similar to the Vpols, a 12" RG316 double-shielded coaxial cable connects the SMA balun to the low-noise amplifier unit housed just above the antenna.

The RNO-G quad-slotted cylinder is 60 cm long, matching the length of the Vpol antennas to get a similar gain-response pattern, and has an outer diameter of 20.3~cm (8"). 
 As with the Vpol antennas, each Hpol is measured in the lab before shipping to the field, and shows excellent agreement to the in-air simulated VSWR as shown in \autoref{fig:hpol_XFDTD}. 
The Hpols have a 3:1 VSWR in-ice  bandwidth of \SI{300}-\SI{850}{MHz}.
Overall the peak response for the Hpol is $\sim$2-3~dB lower than the Vpol, and sits predominately in a higher frequency band though there is considerable overlap in the two antenna bandwidths by design. 
The RNO-G experiment does not currently implement a trigger using the Hpols, but they are important for event reconstruction in the deep sub-array of the station.

\begin{figure}
\centering
\includegraphics[width=1\textwidth, trim={0 1.5cm 0 1.5cm},clip]{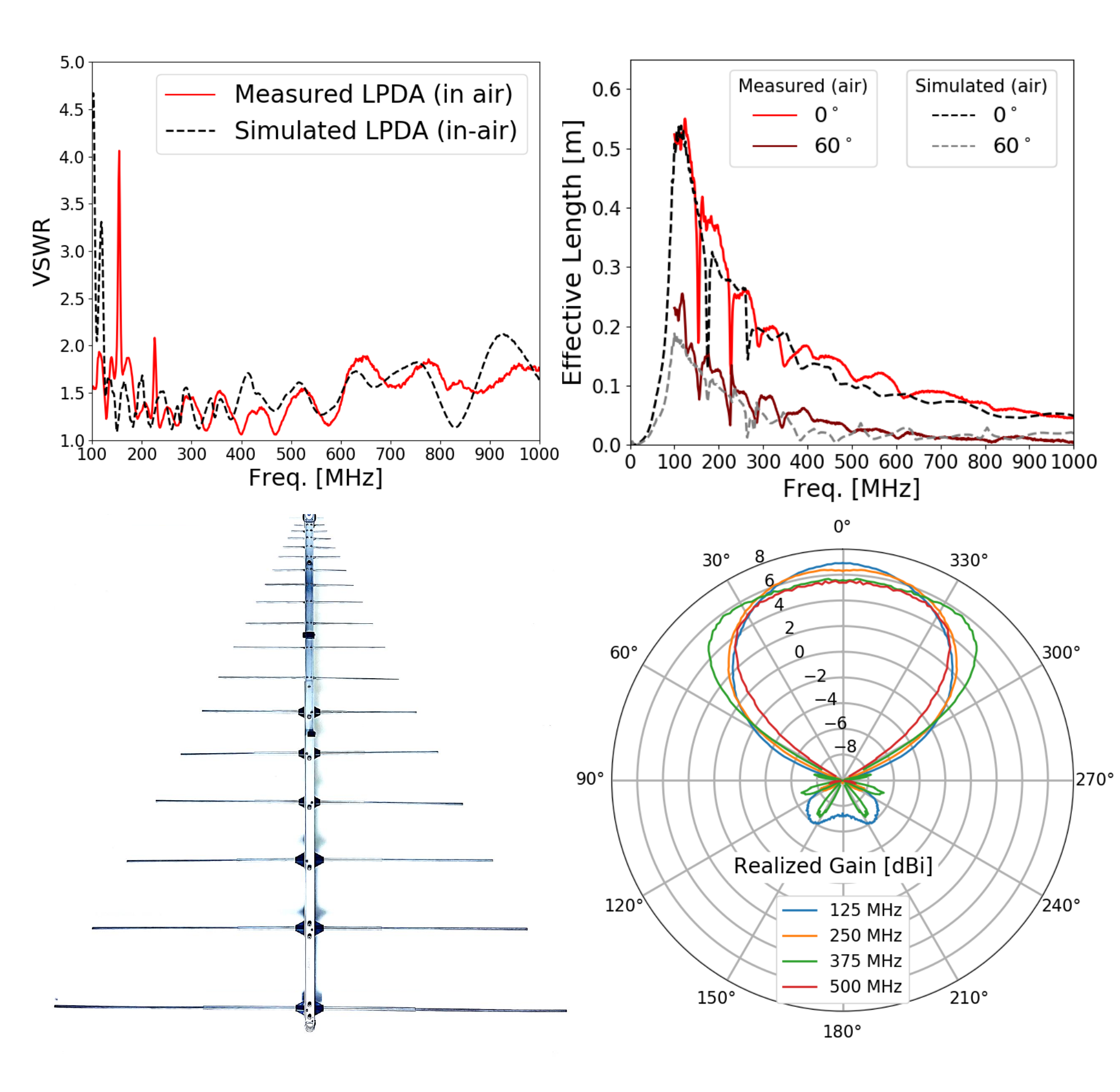}
\caption{
The Create CLP-5130-2N LPDA used for the  RNO-G near-surface antennas. Bottom left: Photo of the assembled LPDA in the lab. Top left: Measured in-air VSWR for a single LPDA compared to an in-air WIPL-D simulation.  Top right: Co-polarized component of the realized vector effective length from WIPL-D simulations in-air (dashed) compared to lab measurements (solid), shown at two different zenith angles including boresight (0$^\circ$ zenith). Bottom right: Realized H-plane gain pattern of the LPDA at representative frequencies, as measured in-air.}
\label{fig:LPDA_insitu_s11} 
\end{figure}

\subsubsection{LPDAs}

The LPDAs are commercially-procured antennas, manufactured by Create with model number CLP-5130-2N and are the same antennas as used in the ARIANNA experiment where they were characterized in detail~\cite{Barwick:2014boa,Barwick:2016mxm}. 
These antennas are shipped disassembled, for compact packing, and
assembled in the field.
The LPDAs are installed in trenches just below the snow surface in three separate rows that are centered on each borehole radial axis. Each row of LPDAs consists of three LPDAs, two that are canted downward at approximately a 60$^{\circ}$ angle and one that is directed upward (see top left of \autoref{fig:sys_diagram}).

The upward pointing LPDAs are intended to be used as a cosmic ray and surface background veto, while the two downward pointing antennas add a secondary contribution to the overall neutrino effective volume of the detector. As they are buried just below the surface, they are not limited by the geometric constraints of the boreholes and thus can be much larger than the borehole antennas. This affords a higher gain and lower turn-on frequency than the borehole antennas, as can be seen in the REVL shown in~\autoref{fig:LPDA_insitu_s11}. These LPDAs, an established broadband dipole-array design, also have a wide bandwidth from \SI{100} - \SI{1000}{MHz} with VSWR better than 2:1 over that range.

\subsection{RF system}
\label{sec:rf_chain}
The RF system was designed to balance the small power budget (<6~W for 24~channels) with the low-noise (<120~K noise temperature) requirements, and to fit within small and high-density form factors. 
With these instrumental constraints, the signal chains are fully custom as there were no commercially available options meeting these specifications. 

\subsubsection{RF signal chain}

\begin{figure}
\centering
\includegraphics[width=0.97\textwidth]{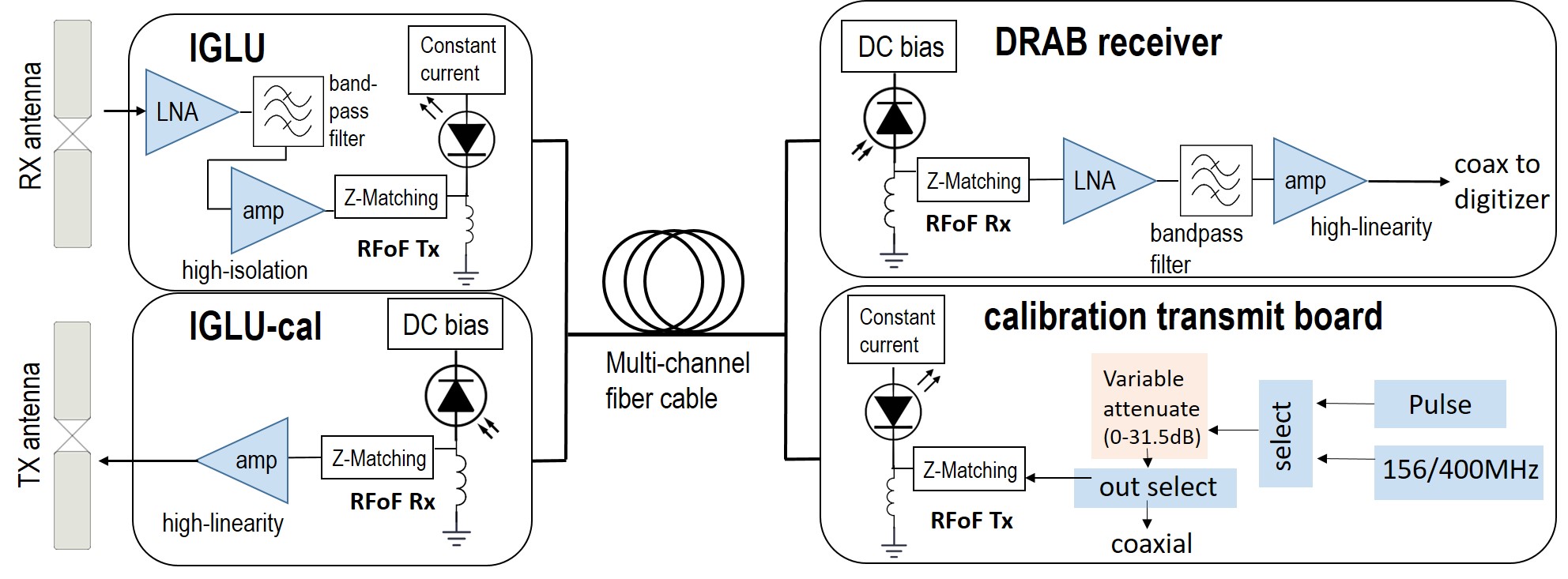}
\caption{Block diagram of the downhole receiver and calibration-transmit RFoF systems. On the left are the
in-ice \texttt{IGLU} and \texttt{IGLU-cal} units, which interface to the \texttt{DRAB} receivers and Calibration Transmitter board in the surface instrument box over 4- or 6-channel fiber bundles.}
\label{fig:dh_rf_diagram} 
\end{figure}

The deep borehole receiving antennas are equipped with a custom RF-over-fiber (RFoF) implementation for long-distance signal transport, which was the favored solution over long coaxial cable runs in order maintain a $\sim$flat ($\pm$2~dB) gain profile without a significant frequency-dependent slope  and to have a manageable borehole-string design. The In-ice Gain with Low-noise Unit (\texttt{IGLU}) is the front-end RF module that is installed directly above each borehole antenna, and includes a 50~$\Omega$-matched low-noise amplifier chip and custom amplitude-modulated RFoF transmitter circuit. The \texttt{IGLU} draws a total of 43~mA on a 3.3V supply voltage, meeting the ultra-low power requirement of a fraction of a Watt per channel. The Downhole Receiver and Amplifier Board (\texttt{DRAB}) is a quad-channel board that receives the optical signal from each \texttt{IGLU}, re-converts to an electrical signal, and further filters and amplifies the signal before sending it to the data acquisition boards. The \texttt{DRAB} is installed within the main instrument enclosure at the surface. 
A block diagram of the RNO-G RFoF systems, including the downhole calibration pulser system, is shown in \autoref{fig:dh_rf_diagram}.

The surface antenna receivers, called the \texttt{SURFACE} boards, are custom highly-integrated circuit boards that connect directly to the LPDA antennas via 10~m coaxial cables (LMR400). The five-channel \texttt{SURFACE} boards have three amplification and bandpass filtering stages per channel, using the same first-stage LNA chip as the \texttt{IGLU}, and are housed in a tight-fitting custom enclosure lined with RF-absorbing material to eliminate cross-talk. Unlike the \texttt{IGLUs}, the \texttt{SURFACE} receivers include an input RF power-limiter on each channel to prevent damage to the signal chain from large radio signals that could be generated in close proximity to the near-surface LPDA antennas. Both the surface and downhole signal chains are described in further detail in \cite{RNO-G:2023fkv}.  The total 24-channel RF system power draw for the RNO-G instrument is just under 6 W. 

\begin{figure}
\centering
\includegraphics[trim={2.5cm 1.5cm 2.0cm 2.4cm},clip,width=0.8\textwidth]{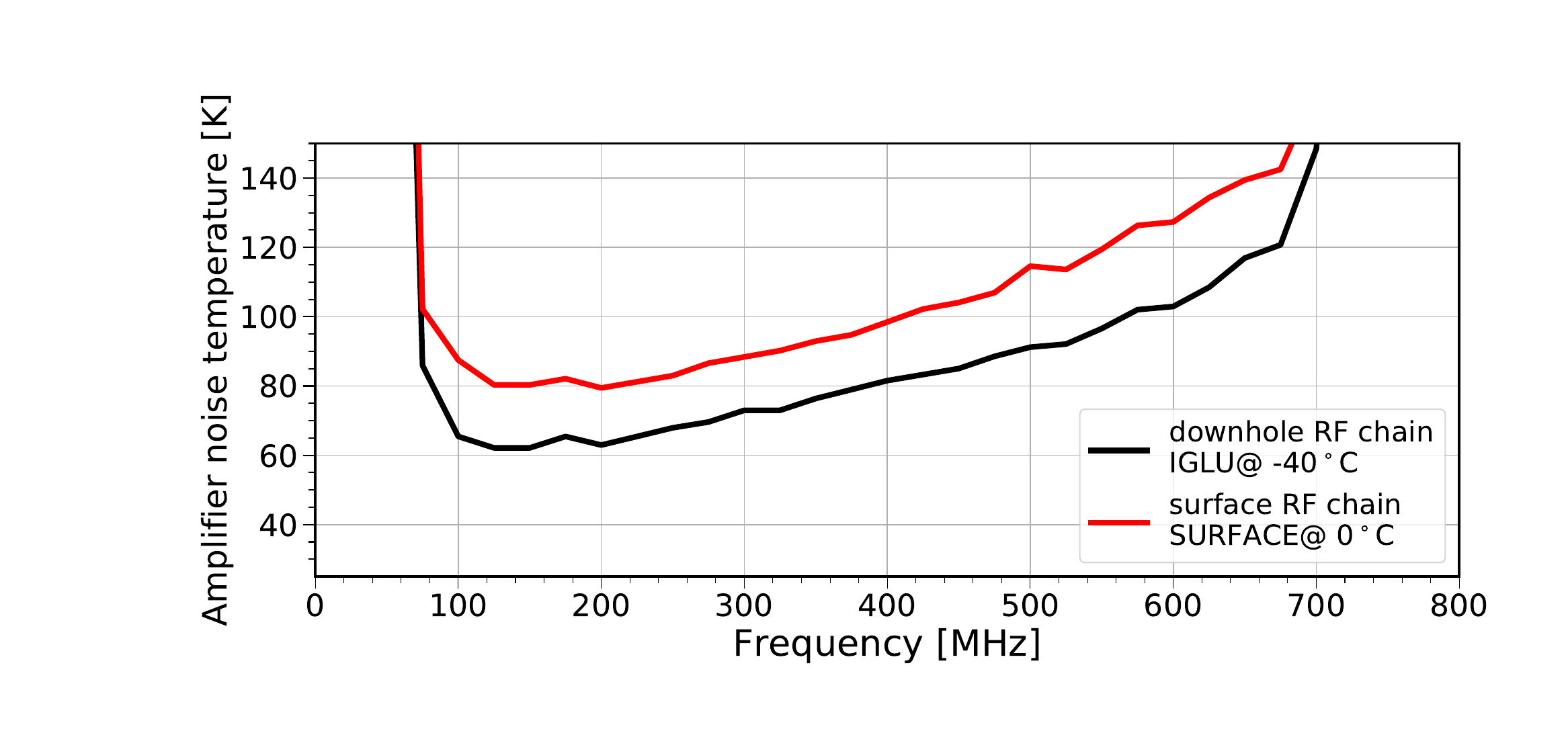}
\caption{Noise temperature of the borehole (\texttt{IGLU-DRAB}) and surface antenna signal-amplifier chains, at the relevant physical front-end temperatures, as measured in the lab using a climate-controlled chamber. }
\label{fig:chain_noise} 
\end{figure}

The additive noise temperature from the RF system was required to be sub-dominant to the ice temperature of $\sim$240~K, which is the primary contributor to the antenna temperature of both the borehole-installed Vpol/Hpols and downward-pointed surface antennas. 
This performance metric has been confirmed in the fielded stations (see \autoref{sec:ext_radio_sig}) and has also been measured in the lab to be below 140~K for both chains across the full receiving bandwidth as shown in \autoref{fig:chain_noise}. 
For the deep antennas, the noise temperature of the \texttt{IGLU-DRAB} chain is below 80~K in the 100-400~MHz bandwidth range that primarily contributes to the neutrino trigger.

The gain profile of the RF system has three major requirements: 1) boost the baseline antenna voltage-output signal to a measurable level of at least 5-10~mV$_{\mathrm{rms}}$, 2) preserve a linear response for signals with amplitudes at or greater than $30\times$ above the thermal noise level, and 3) sufficiently reject out-of-band noise, such as the wireless communications signals that sit just above 850~MHz. The thermal noise  power received by a perfectly impedance-matched antenna viewing a constant ice temperature is given by 
\begin{math} k_{B}  T  \Delta f \end{math}.
Over a 100-700~MHz bandwidth with an ice temperature of 240K, this is equivalent to a baseline thermal noise power of $\sim$2~pW presented to the input of the RF signal chain. 
This corresponds to an input-referred voltage level of 
$\sim$10~$\mu$$V_{\mathrm{rms}}$ in a 50~$\Omega$ system and sets a minimum gain
requirement of approximately 1000$\times$ in voltage amplitude, or 60~dB.

\begin{figure}
\centering
\includegraphics[width=0.9\textwidth]{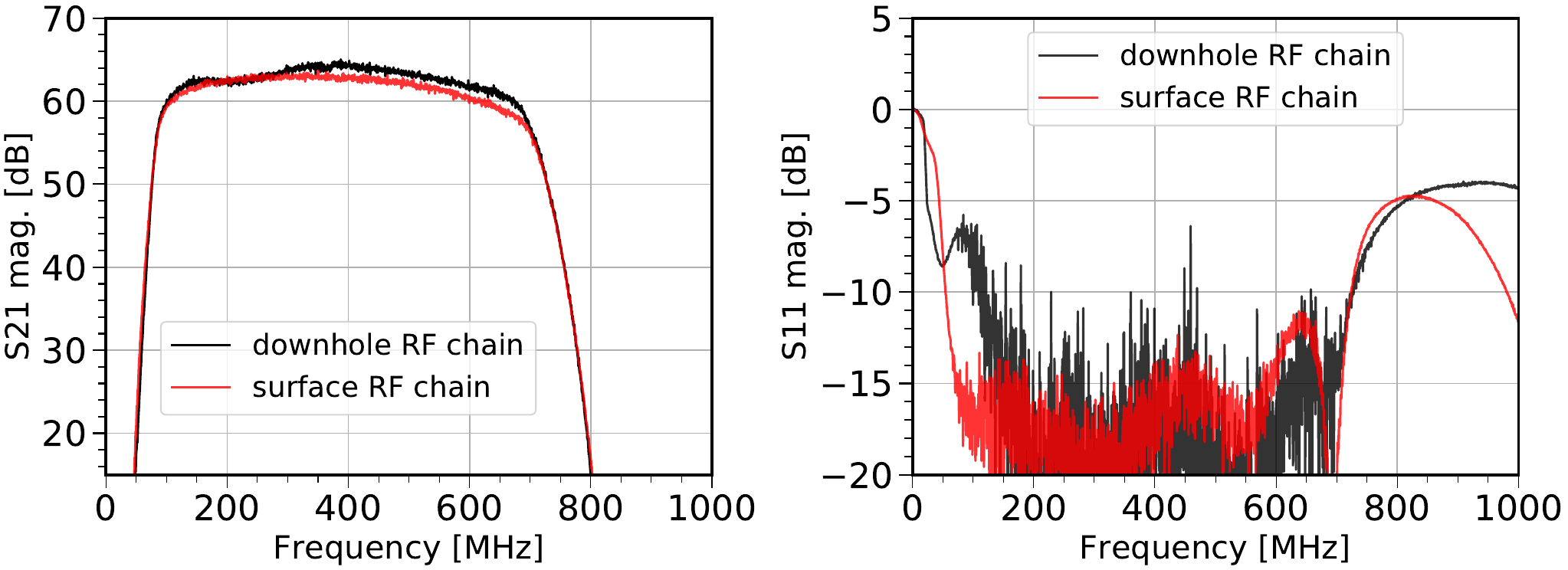}
\caption{Representative gain and input return loss performance of the RNO-G RF signal chains. Left: The forward gain (|S21|) of the two signal chains, meeting the >60~dB specification. The bandwidth profile of the two chains are fairly closely matched; the surface  profile is smoother without the fiber transceiver components.
Right: The S11 magnitude response, which is a measure of the return loss. Both chains are well matched to 50S$\Omega$ given by the -10~dB target specification: the downhole and surface chains meet 10~dB return loss specifications from 110-720~MHz and 70-720~MHz, respectively.}
\label{fig:chain_forward_gain}
\end{figure}

The gain and input return loss responses of both the downhole and surface signal chains are shown in \autoref{fig:chain_forward_gain}. Both chains have a gain in the range of 61-63~dB in-band and maintain a relatively flat frequency profile. 
Also, both chains are well matched to 50~$\Omega$ input over the RNO-G signal bandwidth.
The downhole chain has a return loss that exceeds 10~dB (2:1 VSWR) from 110-720~MHz, and the surface chain meets the same performance over a 70-720~MHz bandwidth. A lower frequency cut-off was designed for the surface chain to accommodate the wider bandwidth of LPDA antennas.
The full magnitude and phase responses for all RF signal-chain components are saved into a database as complex S-parameters for use in later calibration and data analysis efforts.

\subsubsection{RF calibration system}

\begin{figure}
\centering
\includegraphics[width=0.99\textwidth]{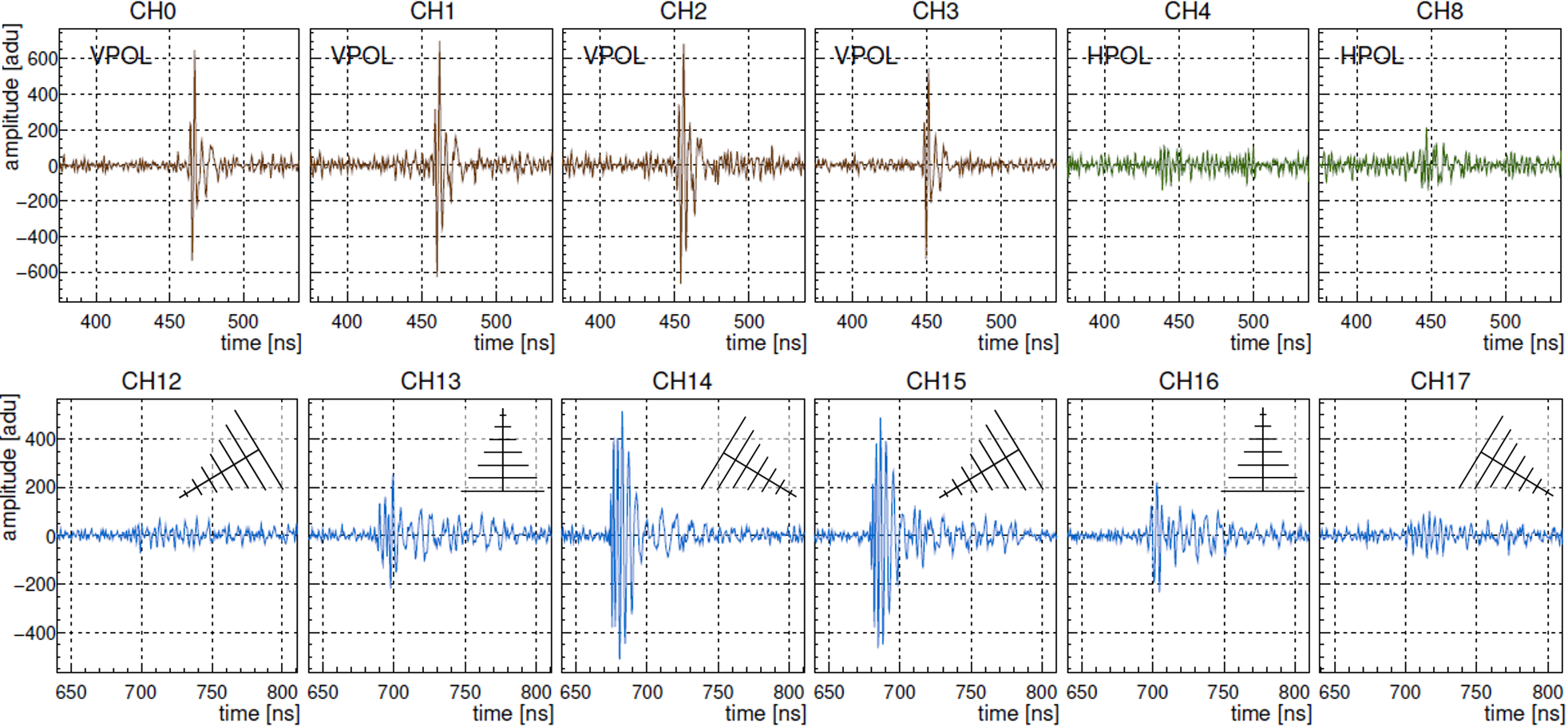}
\caption{In-situ calibration pulser waveforms from station 23. Top: Pulse transmitted from an in-situ deep Vpol pulser on a helper string and received by the lowest six antennas of the power string. The waveforms of the lowest four Vpols (the phased array) and two Hpols are shown. A small amount of cross-polarization response is visible in the Hpol channels.
Bottom: Pulse transmitted from a near-surface Vpol, installed in a hand-augered borehole at a depth of 19~m, and received by six surface-installed LPDA antennas.  This transmitting antenna is installed equidistant between two of the three LPDA trenches, and the corresponding received pulse amplitude response tracks the antenna gain pattern as expected from their orientation: CH14/CH15 are installed in separate trenches and both pointed towards the transmitter, CH13/CH16 are the upward-pointing LPDAs in each trench, and CH12/17 are pointed away from the pulser. The installed LPDA orientations are shown in the inset for clarity. }
\label{figure:calpulser}
\end{figure}

The RNO-G instrument incorporates a pair of deep calibration pulsers on the helper strings to test the in-situ performance of the receiver channels. These are essentially implemented as the optical reverse path to the \texttt{IGLU-DRAB} receivers, as shown in \autoref{fig:dh_rf_diagram}. Each borehole transmitting antenna is installed with an \texttt{IGLU-Cal} module that receives the optical pulser signal over fiber, re-converts to an electrical signal, and applies a last-stage signal amplification before sending to the antenna feed. The \texttt{IGLU}, used for the receiving antennas, and the \texttt{IGLU-Cal} are housed in identical RF enclosures to maintain a uniform module design for the borehole antennas.
All of the installed station transmitters use Vpol antennas. 

The Calibration Transmit Board generates the RF signals for the calibration pulsers, and includes a fast impulse generator and a pair of continuous sine-wave (CW) frequencies. The selected signal type can be attenuated over a 31.25~dB range via remote digital control.
Additionally, there are four selectable RF outputs from the Calibration Transmit Board: two RFoF transmitters, which are routed to the \texttt{IGLU-Cal} modules on the helper strings, and two coaxial outputs.  One of the coaxial outputs is sent to a near-surface transmitting Vpol antenna over standard LMR240 or LMR400 cable, while the other is typically only used in lab testing. Only one output from the Calibration Transmit Board can be active at a given time.
The Calibration Transmit Board is managed by the RNO-G station controller board (\autoref{sec:control}) and is powered down when not in use to prevent the possibility of spurious pulses being transmitted during science-data runs. This also minimizes the total power consumption of the instrument. 

The near-surface calibration transmit antenna is typically dug-in by hand, at most 2~m deep. At two RNO-G-7 stations (21 and 23), this pulser antenna was installed in a deeper augered borehole, between 10 and 20~m depth. Such depths allow more favorable
transmit-receive angles between the LPDAs and the upper Vpol antennas on the power string, and enable monitoring of the near-surface firn properties~\cite{Anker:2019zcx}. It is envisioned to install near-surface pulsers at such depths for future RNO-G stations.

The in-situ calibration system is used to validate the station operation immediately after installation and over the lifetime of the station. The resulting calibration dataset can be used to determine the relative positions of the antennas and, to some level, the index of refraction model that describes the ice as a function of depth. The pulsers are also used to validate the trigger design (\autoref{sec:trigger}), and to measure the in-situ time-domain system response (\autoref{sec:tdomain}).
Example calibration pulser event waveforms are shown in \autoref{figure:calpulser}, showing the deep antennas on the power string as received from the helper string pulser and a selection of the shallow LPDA antennas as received from the near-surface pulser.

\subsection{Borehole String Design}
\label{sec:string_install}

The design of the borehole antenna strings was carefully considered in order to reduce component count, ease the deployment process, and to maintain uniform antenna responses. 
This is largely accomplished through the design of the constituent borehole antenna `units' (\autoref{sec:antennas}), which are compact pre-assembled and tested modules that integrate the antenna, mechanical frame, and front-end \texttt{IGLU} (or \texttt{IGLU-Cal}) RF enclosure. 
Each antenna unit has a simplex fiber connection (FC/APC) for the RFoF signal and a coaxial connection (N-type) for power. A LMR240 through-power cable is included in the pre-assembled antenna unit, which passes power from  the \texttt{IGLU}, installed above the antenna, down through the internal unit exposing a bottom-side connector that is available to attach to either the next antenna or a run of LMR400 cable. Therefore, each antenna unit has the same conductive through-cabling power element and the same length of feeding coaxial cable, which maintains a common impulse-response for the Vpols and Hpols, separately. 
The fiber-optic signal cables, which are routed up the array on the outside of the antennas, are of negligible influence on the antenna response. 
This borehole string design concept is inherited from the ARA phased array~\cite{Allison:2018ynt}, which demonstrated well-matched impulse responses among many antennas in a compact string.

\begin{figure}
\centering
\includegraphics[width=0.92\textwidth]{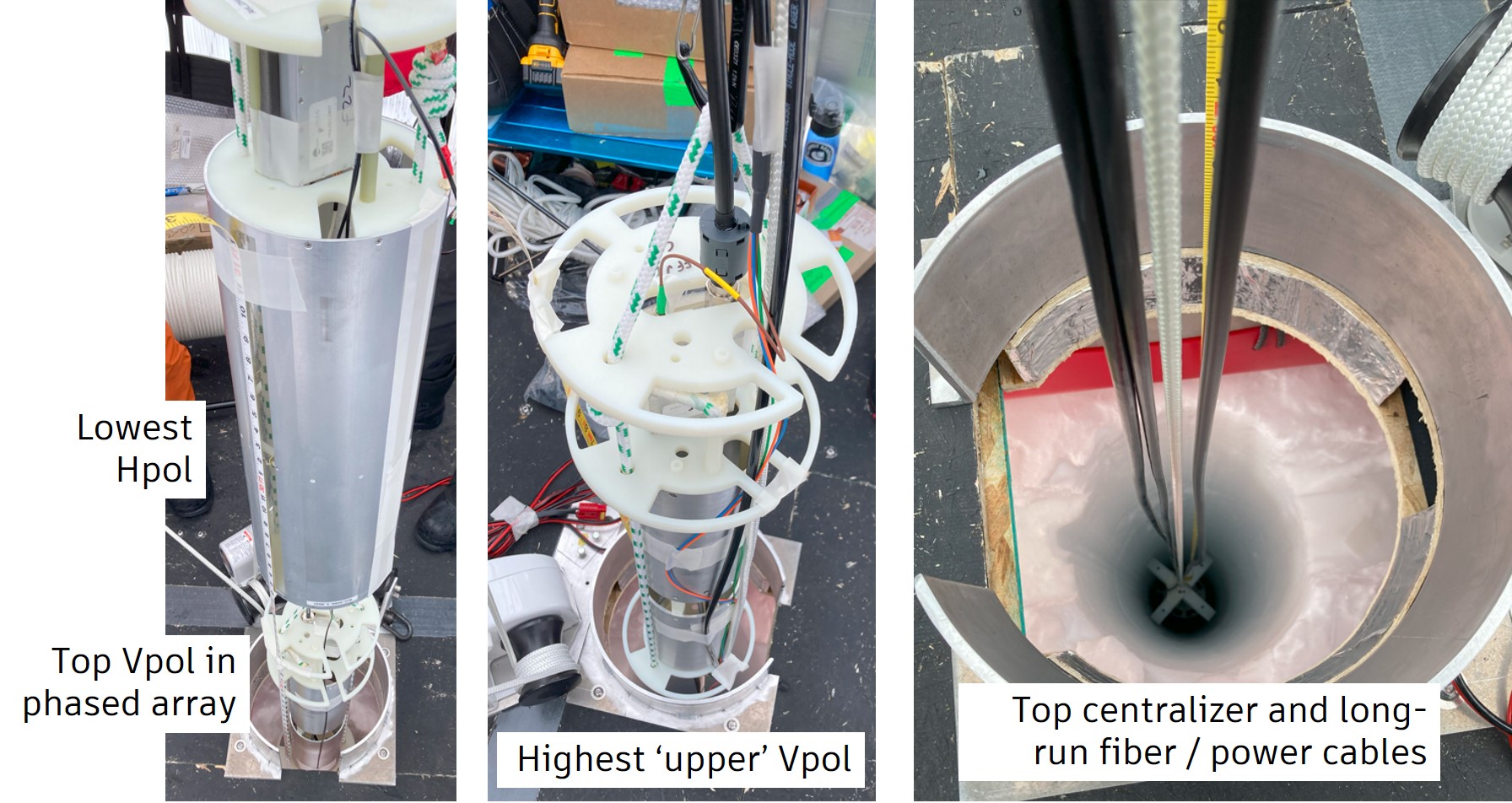}
\caption{Aspects of the power string design and borehole installation. Left: The compact lower antennas in the string with a 1~m vertical spacing, center-to-center. The highest Vpol in the phased array, just entering the hole, and the lowest Hpol antenna are pictured.
Middle: The uppermost Vpol antenna on the string, which is ultimately installed at a depth of 40~m as shown in \autoref{fig:array_and_station_layout}. The staggered-length fibers are visible as is the final power connection to the string via a LMR400 cable with a clamp-on ferrite.
Right: The string as it is lowered into place, with the last string centralizer visible that sits above the highest upper Vpol antenna. The power string uses three long-run cables, two ruggedized multi-channel fiber cables and a single power cable.}
\label{fig:powerstring} 
\end{figure}

The RNO-G borehole string installation process is shown at a few intermediate stages in \autoref{fig:powerstring}. These photos show the power string installation,
which includes both the compact deep set of 6-antenna (4 Vpols + 2 Hpols) 
deployed with 1~m center-to-center spacing
and the upper Vpol antennas with much longer baselines of 20~m. Also visible is the separate
string-centralizer device, which is an extremely simple part made from 
flexible HDPE that further minimizes antenna offsets within the borehole. 
These are installed
directly above the deep set of antennas on each string, and also above
each upper Vpol in the power string. The depths of all components are 
recorded using an open-reel tape measure that is spooled out and
installed with the string.

Rugged single-mode fiber cables are used for the RFoF signals. 
The simplex fiber from each antenna unit is connected to a bulk 
multi-channel fiber bundle that is routed to the instrument box on the surface.
These cables are included in the RF signal chain calibrations
to reflect the added group delay from the cable length.
A thermal coefficient of delay for these fibers has been measured at
$\mathcal{O}$(\qty{100}{ps \: K^{-1} \: km^{-1}}), which is a significant, but constant delay adjustment
from the room temperature lab measurements to the fielded installation
and will be discussed in a follow-on detailed calibration paper.
All fiber cables are rated to -55$^{\circ}$C.

When operating in the low-power winter-mode (\autoref{sec:power_stuff}), the downhole \texttt{IGLUs} are powered off until sufficient power becomes available to return to science operations. Each borehole string has a dedicated power switch on the station controller board and the current draw can be independently monitored. For RNO-G-7, the strings have been powered down for 2+ winters, all returning to nominal condition 
without a failure. 

\subsection{Data Acquisition and Trigger}
The RNO-G instrument records time snap-shots of full-bandwidth voltage waveforms on all 24 receiver channels at a sampling rate between 2.4 and 3.2~Gigasamples-per-second~(GSa/s) with a sample resolution of 12-bits. Events are recorded based on a set of hardware-level triggers, taken typically at an overall per-station rate of $\sim$1~Hz. There is a back-end station controller board, based on a low-power single-board computer (SBC), that controls the data acquisition (DAQ) and trigger systems, and saves the waveform data to a local micro-SD card. These DAQ hardware systems and associated embedded software are described below.

\subsubsection{\textsc{Radiant} and \textsc{Flower} Boards}
\label{sec:radiant}

The full-band waveforms for all 24 antennas within a station are digitized using the RAdio DIgitizer and Auxiliary Neutrino Trigger (\textsc{Radiant}) board. Each channel is equipped with a LAB4D chip~\cite{Roberts:2018xyf}, a custom switched-capacitor array application-specific integrated circuit (ASIC) with a 
sampling rate of up to $\qty{3.2}{GSa/s}$ and an analog memory of 4096 samples, 
addressable in 128-sample blocks, which allows for the capability of on-chip multi-event buffering to reduce readout-induced dead-time.  

For the purposes of RNO-G, the LAB4D operates with two buffers of 2048~samples each, providing a 
$\qty{640}{ns}$ recording window per event when operating at 3.2~GSa/s. These two analog buffers are utilized in a back-and-forth manner for each subsequent recorded event.
The LAB4D has an on-chip ramp-compare ADC that provides digital data with 12-bit resolution and this conversion takes $\sim$10~$\mu$s per event. The digital conversion and readout of the LAB4D data is only performed when the system is triggered.

The \textsc{Radiant} Artix-7 field-programmable gate array (FPGA) controls most of the functionality on the board, including managing the control and readout of the 24~LAB4D chips. Due to pin-count limitations on the FPGA, some of the board control lines are handled by a pair of lower-level complex programmable logic devices. The waveform data are sent to the station controller board over a dedicated serial link, while slow control is handled by a separate on-board microcontroller for configuring the board.

The \textsc{Radiant} has a set of on-board signal sources that enable in situ calibration of the LAB4D ASICs without a need for external signal inputs. There is an on-board sine-wave generator circuit that uses a direct-digital synthesizer chip with a configurable output filter bank, which suppresses the higher harmonics.
These sine waves are distributed to the LAB4Ds in groups of 8 channels via an RF switch on each input channel and are used to calibrate the LAB4D time-base.
Each channel also has a programmable DC-offset voltage, which is common for all channels, to set the bias voltage at the input of the LAB4D, which can record signals between 0.1 and $\sim$2.0~V. The bias level can be scanned (`bias scans') over this range to calibrate the ADC transfer function for each sample within  the LAB4D chips. For normal data acquisition, this bias voltage is set at a static level that optimizes the linear range of the ADC.
These board-level calibrations are run at regular intervals within the instrument to keep the \textsc{Radiant} response up-to-date in the field.

The \textsc{Radiant} version 2 board (v2) was developed for the initial 2021 season and is used in the RNO-G-7 array described in this paper. During 2023, a new revision 3 board (v3) was designed, fabricated, and deployed in the 2024 field season to partially upgrade the RNO-G-7 array. 
The \textsc{Radiant} v3 enhances the stability and cold operation of the board, added LAB4D on-board calibration options, and improved sensitivity of the on-board \textsc{Radiant} trigger. This paper will focus on the RNO-G station performance with the \textsc{Radiant} v2, and the upgraded stations will be discussed in a future publication.

The FLexible Octal WavEform Recorder (\textsc{Flower}) board is the other component of the DAQ and trigger system. For RNO-G, this board is hardware-wired with four input channels such that RF signals are captured using two 8-bit streaming digitizers (HMCAD1511) at a sampling rate of \SI{472}{MSa/s}, slightly below the original design target of \SI{500}{MSa/s} due to an issue with the ADC-clock generating chip. 
The digitized samples are streamed to a Cyclone~V~FPGA where they are  deserialized at a parallel clock rate of 118~MHz, which drives the core logic. The \textsc{Flower} can be used to save the waveforms on these four channels, with a record length of up to 20~$\mu$s/channel, but
its primary function is to provide the deep neutrino trigger for the RNO-G instrument. The signals for the four lower Vpols  on the power string (`phased-array' as denoted in \autoref{fig:array_and_station_layout})
are sent through an RF power splitter, low-pass filtered at $\sim$\SI{240}{MHz} to prevent aliasing from the lower sampling rate, and sent to the \textsc{Flower} where they are digitized and processed through custom trigger firmware.

Both the \textsc{Radiant} and \textsc{Flower} have a common link to the RNO-G station controller (\autoref{sec:control}). In addition to sending waveform data, these links also allow for the application FPGA firmware on each board to be remotely updated, although a lab-verified and programmed base firwmare image is also maintained on the boards. The RNO-G station has an internal low-drift 10~MHz reference clock that is distributed to each board, allowing each station to be individually synchronized. During science operations, the \textsc{Radiant} draws between 13-14~W and the \textsc{Flower} draws 3~W, with baseline firmware, up to 5~W with more complicated trigger algorithms.

\paragraph{Voltage and Timing Calibration of the \textsc{Radiant}} 

The LAB4D requires amplitude and time-base calibrations, which are common for such custom waveform sampling ASICs. 
With the internal on-chip ADC, each of the 4096 analog samples has a voltage-to-adu (adu = ADC count) transfer function.
The LAB4D time-base also is calibrated, primarily through tuning configurable delays in the on-chip timebase generator that drives the 128-cell primary sampling array.  Individual sample delays along the 128-sample switched-capacitor array, subject to process variations during chip fabrication that induce sample-to-sample timing errors, each have a delay element that can be tuned to reduce this error. Separately, the  `wraparound' time interval, which is the seam that closes the primary sampling delay-locked-loop from 
\verb|sample-128| back to \verb|sample-1|, has an independent delay control block. The LAB4D calibration process is described in detail in~\cite{Roberts:2018xyf}.

\begin{figure}
\centering
\includegraphics[width=0.99\textwidth]{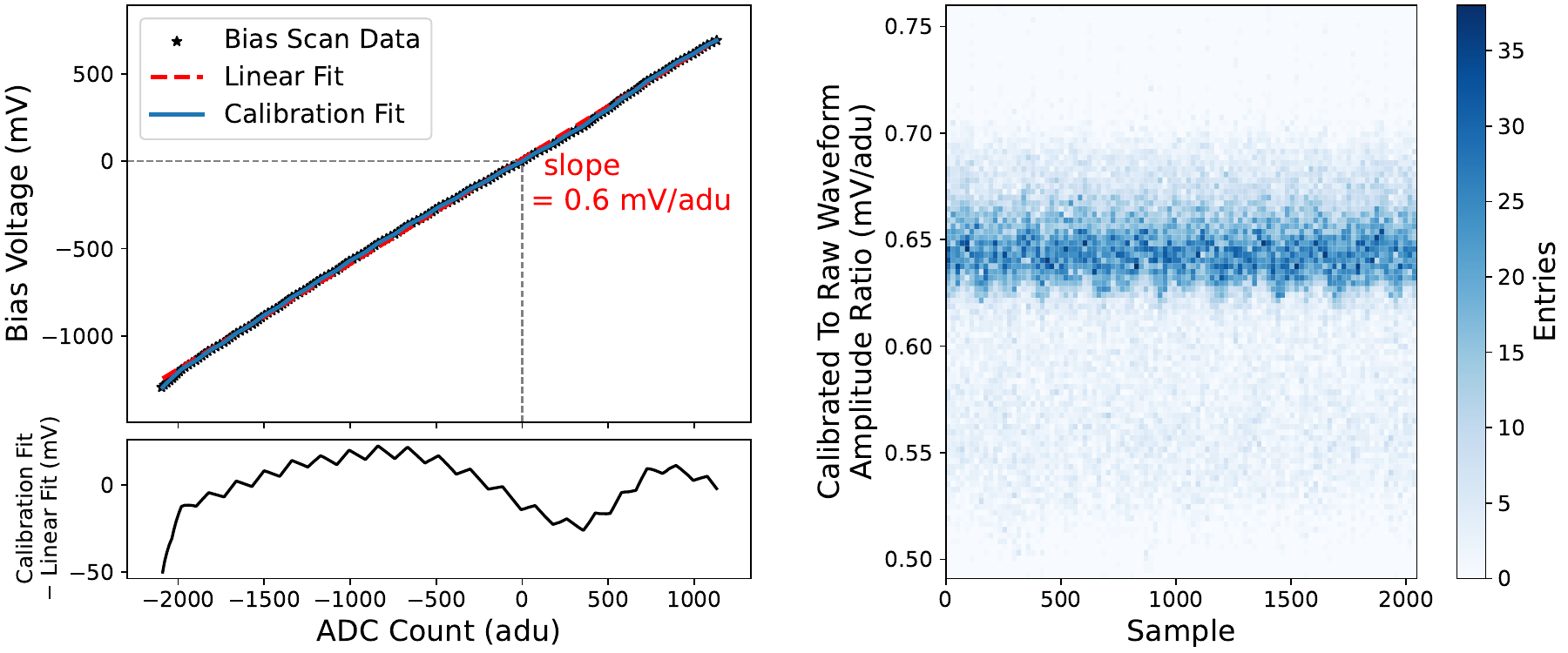}
\caption{Left: Bias voltage scan as function of ADC count for a sample cell and the corresponding calibration fit (9th-order polynomial). The bottom panel shows the residuals from a simple linear fit. The step-like pattern in the linear-fit residuals is due to quantization from the 12-bit digital-to-analog converter used to generate the bias voltage.  Right: Uniformity of the voltage calibration. The ratio of \textsc{Radiant} calibrated-to-raw waveform amplitudes are histogrammed for all 24~channels in an event with the nominal 2048-sample record length.}
\label{fig:ADC_to_voltage} 
\end{figure}
\begin{figure}
\centering
\includegraphics[width=0.5\textwidth]{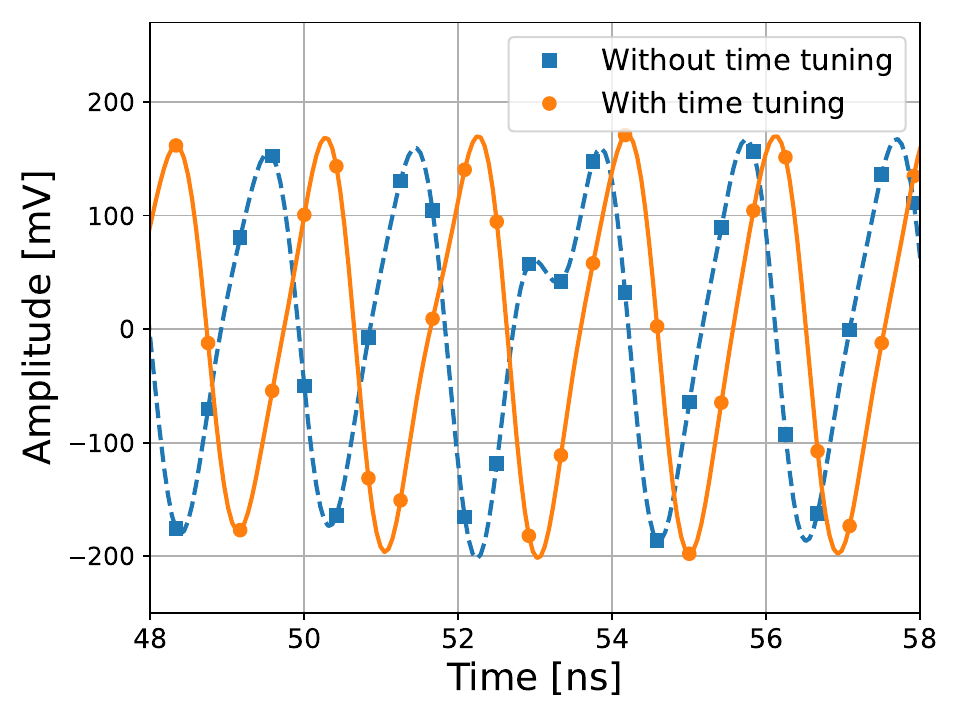}
\caption{\textsc{Radiant} waveforms, 
before (blue dashed) and after (orange solid) time-tuning the LAB4D. 
On \textsc{Radiant} start-up, the wrap-around time-step, which is the timing interval from the last primary sample cell back to the first sample, needs to be set to the correct value to establish a uniform timebase.
These calibration data were obtained with the on-board signal generator providing a \SI{510}{MHz} sine wave. }
\label{fig:RADIANT_wave}
\end{figure}

The built-in functionality of the \textsc{Radiant} allows for the LAB4D calibration to be done in situ. In normal operations, bias scans are run every 26~hours to prevent against any diurnal systematic effects, and the extracted ADC transfer functions are used to calibrate the voltage amplitude for science runs during the same period. We find that the resultant ADC transfer function varies at most by 1\% between successive bias scans, indicating that this calibration interval is sufficient.  Science data are saved in raw format as adu for efficient use of disk space and the voltage calibration is applied offline. An example bias scan is shown in the left panel of \autoref{fig:ADC_to_voltage} as taken in situ in Greenland, for a single analog sample within a LAB4D chip. The bias scan data (discrete data) maps between voltage and adu for one sample of one channel. A calibration fit function (9th-order polynomial, the blue curve in the figure) is used to convert an input adu to an output voltage.  
Raw waveforms, given by adu vs.\ time,  are presented in many figures in this paper.
The right panel of \autoref{fig:ADC_to_voltage} shows the ratio of calibrated waveform amplitude to raw waveform amplitude vs.\ time, histogrammed samples from all 24 channels in a recorded event, which has been found to be generally stable.

The \textsc{Radiant} sine-wave generators are used to calibrate the LAB4D timebases. An example waveform, before and after time calibration, is shown in~\autoref{fig:RADIANT_wave} in which the corrected alignment of the wrap-around sample delay is clearly visible.
After hardware calibration, the LAB4Ds have been demonstrated to have sample-to-sample timing of $\mathcal{O}$(\qty{10}{ps}), and $\mathcal{O}$(\qty{20}{ps}) sample-to-sample time-step variations have so far been achieved with the \textsc{Radiant} as shown in \cite{Roberts:2018xyf, thesis_dansmith}.

\subsubsection{Trigger options}
\label{sec:trigger}

\begin{figure}
\centering
\includegraphics[width=0.7\textwidth]{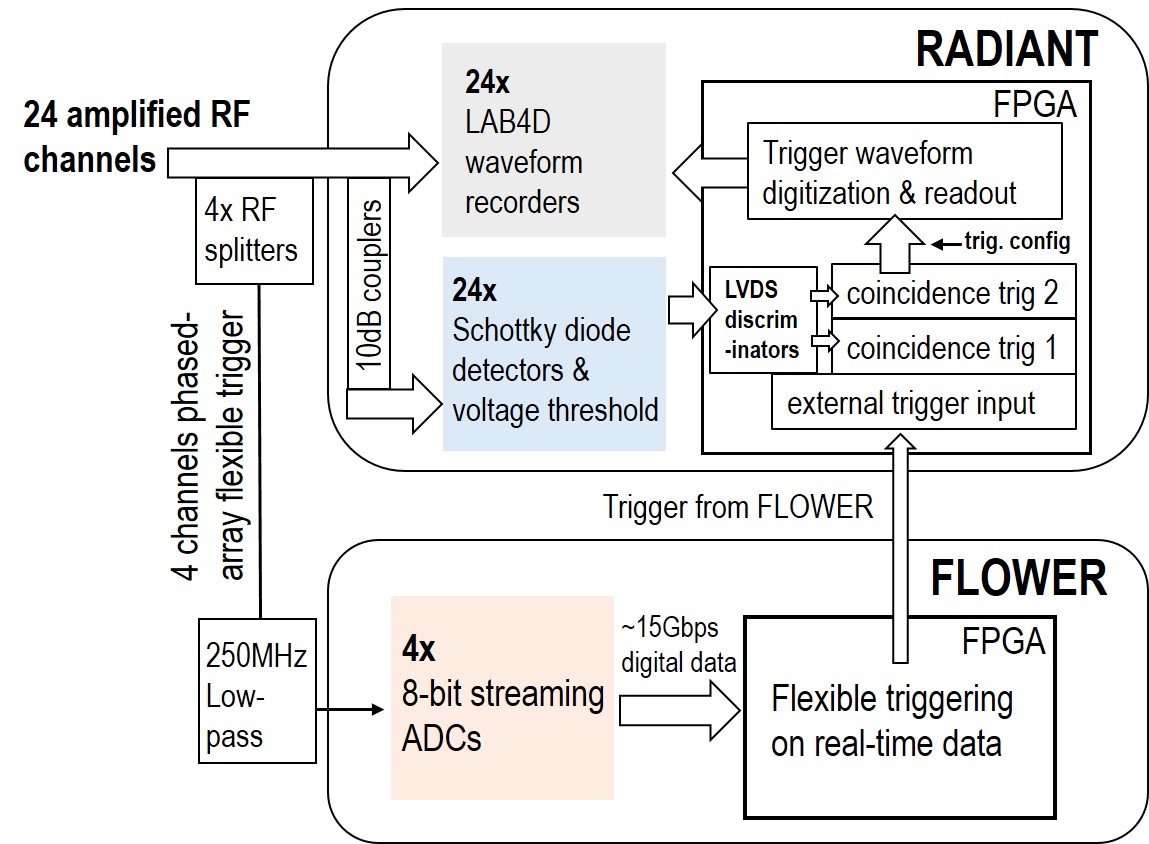}
\caption{RF trigger diagram. The three built-in self-trigger paths are shown, which form the flexible station trigger using streaming digital data on the 4-channel \textsc{Flower} and the two configurable \textsc{Radiant} coincidence triggers. This represents the \textsc{Radiant} v2 design that was used until the 2024 season; the 10~dB couplers in the \textsc{Radiant} trigger path were swapped out for 3~dB splitters in the \textsc{Radiant} v3.} 
\label{fig:trigger_block} 
\end{figure}

A number of trigger options are built into an RNO-G station, including the \emph{\textsc{Flower} trigger} used for neutrino detection by deep antennas, the \emph{\textsc{Radiant} trigger} used for surface antennas, the \emph{GPS pulse-per-second trigger} to sync to external calibration sources, and the \emph{software trigger} to monitor background noise levels. All of these triggers can be enabled simultaneously while taking data,
and the target rates for each type can be set in a configuration file.
An overview of the RF-specific trigger architecture is shown in \autoref{fig:trigger_block}, 
which describes the \emph{\textsc{Flower} trigger} and \emph{\textsc{Radiant} trigger} interfaces, 
 comprising the \textsc{Flower} and \textsc{Radiant} boards, which will be described in the following.


\paragraph{\textsc{Radiant} trigger} The \emph{\textsc{Radiant} trigger} is based on a low-cost analog diode-detector circuit using a surface-mount Schottky diode (Skyworks SMS7621) and has been characterized in detail in \cite{Pyras2024}. A small DC bias ($\sim$100~$\mu$A) is applied to the diode circuit to stabilize the response over temperature~\cite{AN1090}.  The diode output is restored and conditioned using $\sim$40~dB of amplification and low-pass filters, respectively, such that the bandwidth of the signal-envelope sent to the discriminator (e.g. the video bandwidth) is $\sim$\SI{0}-\SI{30}{MHz}, ensuring that the primary RF signal band is sufficiently rejected.

Each \textsc{Radiant} channel is equipped with this diode detector circuit, which
is coupled off the mainline LAB4D signal path with a -10~dB coupler. 
This was adjusted to a -3~dB splitter in the \textsc{Radiant} v3 to increase the signal power in the trigger path.
The nine surface channels have an added 200~MHz low-pass filter (weak roll-off, 2nd order) in the trigger path, after the coupler and before the diode, which improves the trigger-level signal-to-noise ratio (SNR) for cosmic-ray air showers as received by the LPDAs, which have stronger low-frequency signals~\cite{masterthesis-pyras}.

The output of each diode detector is fed to one of the differential inputs of a
LVDS (Low-Voltage Differential Signaling) input of the \textsc{Radiant} FPGA.
The opposite input is connected to a voltage threshold that is generated from
an digital-to-analog converter (DAC) made from an integrated pulse-width modulation circuit.
This architecture effectively employs the LVDS input block of the FPGA as a threshold-discriminator circuit and is easily scaled to high-channel count density with low power, as demonstrated
in the ANITA experiment~\cite{Varner:2006ec}.
 The design goal was for each diode trigger to be thermal-noise riding with adjusted thresholds to provide a stable coincident trigger rate $\ll$\qty{1}{Hz}. However, the observed reduced sensitivity of the \textsc{Radiant} v2 diode trigger required these to be operated typically with a noise-avoiding threshold well above the thermal noise, where the threshold was dynamically adjusted to keep the single-channel rate well under approximately 10 Hz.  Within the FPGA,
 the digital output from each discriminator is sent to two coincidence trigger blocks, where channels can be masked and arbitrary trigger delays can be applied to form the \emph{\textsc{Radiant} trigger}, as shown in~\autoref{fig:trigger_block}.

In science operations, the RNO-G instrument employs the first \textsc{Radiant} coincidence trigger on the three upward-facing LPDA antennas as an air shower veto. This trigger is typically configured with a 2-of-3 coincidence that must be fulfilled within \qty{60}{ns}. The second \textsc{Radiant} coincidence trigger utilizes the six downward-canted LPDA antennas and must typically meet a 2-of-6 criterion. This reduces triggers due to random fluctuations, since the assumed plane wave generated by a particle cascade is expected to be measured by at least two antennas. These coincidence triggers are
sometimes applied, via remote configuration,
to the deep Hpol antennas, for example, when doing calibration pulsing.

Lab measurements show that the efficiency of the trigger circuit depends on the pulse shape, the temperature and the diode bias.
A trigger efficiency of 6.2~SNR was achieved in the field (v2) using the surface pulser, though a large threshold variance is observed across channels. 
Lab testing on the \textsc{Radiant} v3 demonstrates improved performance of the diode trigger circuit, which is expected to be able to operate stably at much higher single-channel rates. The updated design was tested on air-shower-like signals in the lab, which resulted in a trigger efficiency of 4.1~SNR and a much more stable variance between channels. 

\paragraph{\textsc{Flower} trigger} 
The primary neutrino trigger is provided by the \textsc{Flower} board,
using streaming digitized data into an FPGA 
as received from the four deep phased-array Vpols on the power string. 
These four signals are split off from the main \textsc{Radiant} path using broadband 3dB splitters, 
which are uniquely applied to these channels. 
The \emph{\textsc{Flower} trigger} operates on a \SI{100}-\SI{240}{MHz} bandwidth, which
was found to improve the trigger-level SNR for a wide range of Askaryan off-cone view angles, as this corresponds to the frequency range (<\SI{300}{MHz}) where in-ice Vpol RVEL provides maximum voltage response~(\autoref{fig:vpol_display}). 
As outlined in the RNO-G whitepaper,
this `low-band' trigger is the primary driver for the array's 
trigger-level sensitivity to deep neutrinos~\cite{RNO-G:2020rmc}. 

The \emph{\textsc{Flower} trigger} is a flexible design that allows for updating the 
FPGA algorithms, as applied to the streaming data, remotely. 
The design goal is to build a phased-array trigger on these compact
4~Vpol antennas within the FPGA, as demonstrated previously in ARA~\cite{Allison:2018ynt},
and was first deployed on the \textsc{Flower} during the 2024 season. 
For the first few deployment seasons, the as-implemented \emph{\textsc{Flower} trigger} in the RNO-G-7 array comprised a simpler trigger algorithm.
This initial trigger imposes a high-low requirement on each signal channel within a tight time window, which identifies for bipolar impulsive signals similar to the trigger described here~\cite{Kleinfelder57371}, but in the digital domain. 
Once a trigger decision is formed, an external trigger signal is sent to the 
\textsc{Radiant} to initiate a full-station event readout 
of the `full-band' waveform as shown in~\autoref{fig:trigger_block}.
The performance of this trigger is discussed in~\autoref{sec:trig_performance}.

\subsubsection{Software and Station Control}
\label{sec:control}

Each station contains a modular Station Controller board which is responsible for interfacing with all the subsystems and eventually transmitting data off of the local station. One of the primary functions of the Station Controller is to act as a carrier for the BeagleBone Black Industrial~\cite{BBB} single board computer (SBC), which is responsible for science data collection from the \textsc{Radiant} board.  Additionally, the Station Controller incorporates a Global Navigation Satellite Systems (GNSS) receiver, the wireless communication systems (\autoref{sec:comms}), and a microcontroller (Microchip SAMD21J18A) responsible for managing the controller board, including collecting housekeeping sensor data and turning subsystem power rails on and off. 

The SBC runs a Debian Linux operating system, which is built using a custom RNO-G image. It has Serial Peripheral Interface (SPI) and Universal Asynchronous Receiver Transmitter links to the \textsc{Radiant}, for data transfer and slow control, respectively, and an SPI link to the \textsc{Flower} board (see \autoref{sec:trigger}).  Software on the SBC is used to configure the digitizer boards for data taking, and then perform data acquisition. After initial configuration (including  tuning of LAB4D parameters), the data acquisition daemon (\texttt{rno-g-acq}) is started, which is fully steered by configuration files. In addition to recording the data and writing it compressed to disk, \texttt{rno-g-acq} is also responsible for dynamically adjusting trigger thresholds to attempt to maintain an approximately constant rate. Standard runs are 2 hours in length, although shorter runs may be taken during calibration campaigns or while tuning configuration parameters. The SBC has a 4 GB embedded Multimedia Card used for the operating system and other programs and a \SI{128}{GB} industrial SD card to temporarily store data.  
Data is written out in a simple custom binary format and compressed using \texttt{zlib}. Due to a lack of random access, this format is not convenient for analysis so data is typically converted to ROOT format \cite{ROOT} after leaving the detector.

The GNSS receiver is a single-band u-blox LEA-M8T module (upgraded to a dual-band u-blox ZED-F9T in 2024) and is enabled via a general purpose input-output signal from the SBC, and also has a serial connection to the SBC for configuration and readout. The pulse-per-second signal from the GNSS receiver is distributed to the \textsc{Radiant} and \textsc{Flower} boards, the calibration driver board, and (in newer revisions) the SBC itself as common timing reference signal across the RNO-G array. At each RNO-G station, the pulse-per-second signal is registered on the \textsc{Flower} board using the 118~MHz local FPGA clock (derived from the 10~MHz station reference clock), which captures the time-offset of the local trigger relative to each pulse-per-second rising-edge. Software on the SBC is used to configure the GNSS receiver and can also record GNSS data which can be used, for example, for GNSS reflectometry to measure the ice height \cite{Komjathy2000}.

The SBC controls and communicates to a number of other peripherals.  
The single USB link on the SBC is used to transmit and receive data from the LTE modem (\autoref{sec:comms}), and a USB2.0 4-port hub chip is employed on a newer controller board revision that allows for additional USB interfaces, such as industrial-rated flash drives for redundant storage.
The SBC also controls the calibration transmit board, allowing remote selection of the output channel, the pulser waveform source, and the pulser attenuation.
The SBC furthermore has a serial connection to the controller board microcontroller, which can be used to request sensor data, turn power rails on and off, or in-situ program the microcontroller. 

The microcontroller firmware is designed to robustly operate the station even with the SBC off (as well as enabling power cycles of the SBC). It incorporates a LoRaWAN communications stack for remote control independent of the SBC and can also be configured  to automatically turn the station into low power mode when battery voltage falls below a configurable threshold. External SPI flash memory holds configuration parameters, as well as four slots for microcontroller firmware. The custom bootloader for the microcontroller can be instructed to load any of those 4 slots on a reboot, allowing for safe in-situ reprogramming of microcontroller functionality. A special low-power mode designed for winter operation turns off most devices (\autoref{sec:power_stuff}) and causes the microcontroller to sleep most of the time. 

\subsection{Communication systems}
\label{sec:comms}

RNO-G uses a private LTE cellular network for science data transfer and remote access to the SBC. LTE was chosen to reduce the power requirements at each station and avoid use of highly directional antennas that may become misaligned during storms. The LTE network operates on LTE band 8 (uplink \SI{880}-\SI{915}{MHz}, downlink \SI{925}-\SI{960}{MHz}), currently using one eNodeB on the lowest \SI{10}{MHz}  channels, though additional eNodeB's may be added in the future to increase capacity. The LTE backend is run on a virtual machine on the RNO-G Summit Station Central Server, a Dell PowerEdge 7515 2U rack-mounted server.  The LTE system was procured from Star Solutions. Two large 13 dBi base station antennas were installed on top of the Big House, the main building at Summit Station, that send and receive the LTE communications to/from the RNO-G sites. 

\begin{figure}
\centering
\includegraphics[width=0.60\textwidth]{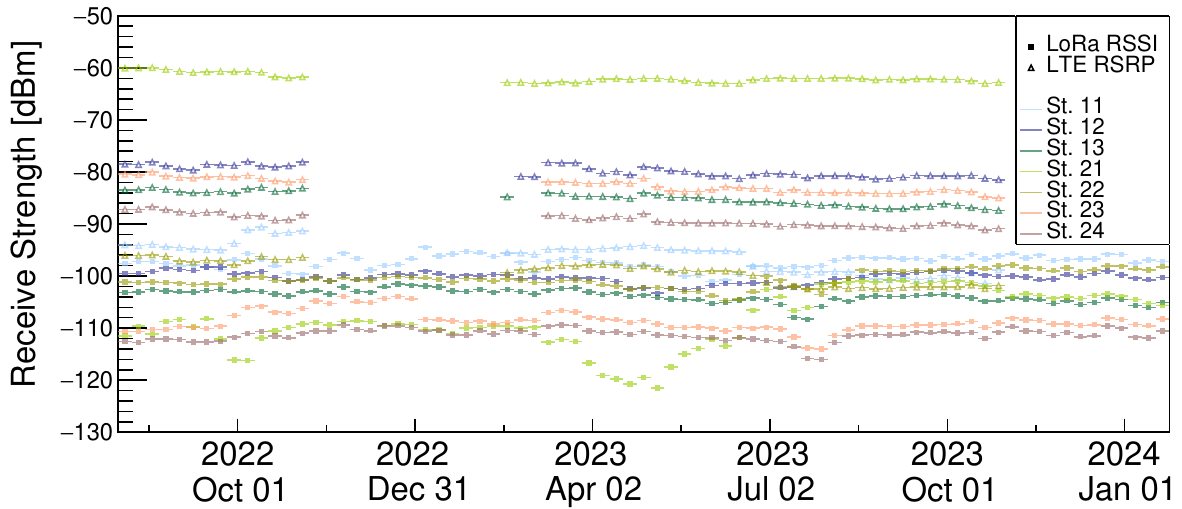}
\includegraphics[width=0.38\textwidth]{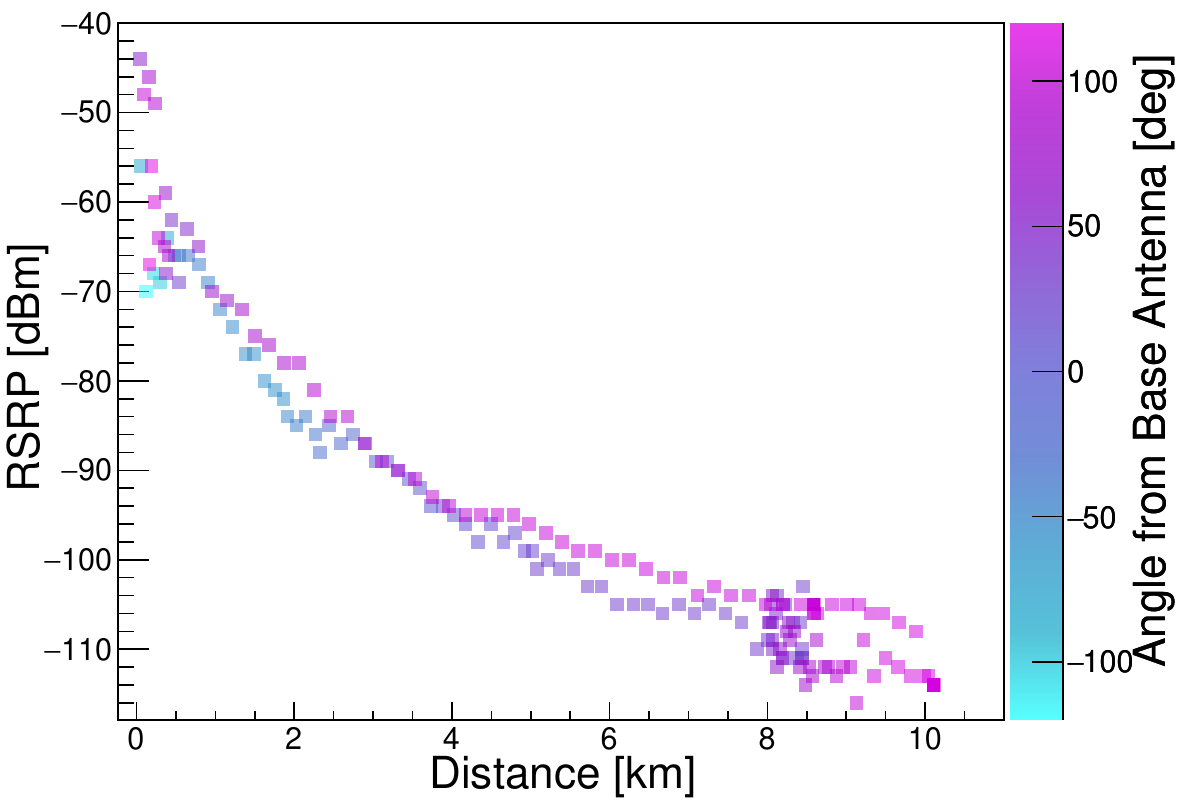}
\caption{Performance of the communications systems of RNO-G. Left: Received signal strength, defined by the Reference Signal Received Power (RSRP), as function of time. The LTE network is generally turned off during the winter, as long as wind power is not available. Right: Received signal strength as function of distance to the station. Data are taken from a survey conducted during the first installation season. The color scale depicts the effects of moving out of the field-of-view of the base antenna.}
\label{fig:rssi} 
\end{figure}

Each station has an 8 dBi omnidirectional \SI{900}{MHz} antenna (LPRS-ANT-868) mounted on a solar panel. This is connected via a 15~m coaxial cable (LMR240) through a \SI{900}{MHz} ceramic resonator bandpass filter (MiniCircuits CSBP-A940+) to a Nimbelink cellular modem (NL-SW-LTE-TC4EU), which communicates with the SBC using USB2.0. The modem also has a serial connection to the microcontroller, used for monitoring the LTE connection quality. A software daemon running on the SBC verifies connectivity and automatically power cycles the modem if connectivity is lost. 
The LTE network performance is summarized in \autoref{fig:rssi}. Generally, performance is stable, though susceptible to strong snow storms. A post-deployment survey indicates that the LTE network is indeed capable to cover the entire 35-station array. The design requirement was to provide sufficient bandwidth for a sustained data rate of 1 Hz.

A Long Range Wide Area Network (LoRaWAN~\cite{lorawanspec}) network on the \SI{915}{MHz} ISM radio band was also deployed at Summit for monitoring of sensor data collected by the controller board as well as remote control of the microcontroller via predetermined commands. This second network is required because the LTE modem uses too much power to be enabled in low-power mode. LoRaWAN uses much less power but is also only capable of very limited bandwidth. Sensor data  (temperatures, currents, voltages) are sent once a minute during normal operation and once an hour while in low-power mode. Additionally, LTE and LoRaWAN connectivity data are also sent. Using LoRaWAN commands, we can put the stations into and out of low-power mode, power cycle the LTE modem and SBC, reset the microcontroller, or turn a small heater on or off.

The Summit Central Server collects all data from both LTE and LoRaWAN. Science data from LTE is converted to ROOT format and stored on redundant disks. A random subset is sent via satellite to RNO-G insitutions for monitoring purposes. The full dataset is transferred by hand-carrying the hard drives. In the future, online analysis will run on the server in order to generate multimessenger alerts. Sensor data from LoRaWAN as well as other types of monitoring data (local weather station, server, network and UPS housekeeping) are inserted into a \texttt{postgresql} database, which is then replicated off-Greenland.  

\subsection{Power system}
\label{sec:power_stuff}

\begin{figure}
\centering
\includegraphics[width=0.8\textwidth]{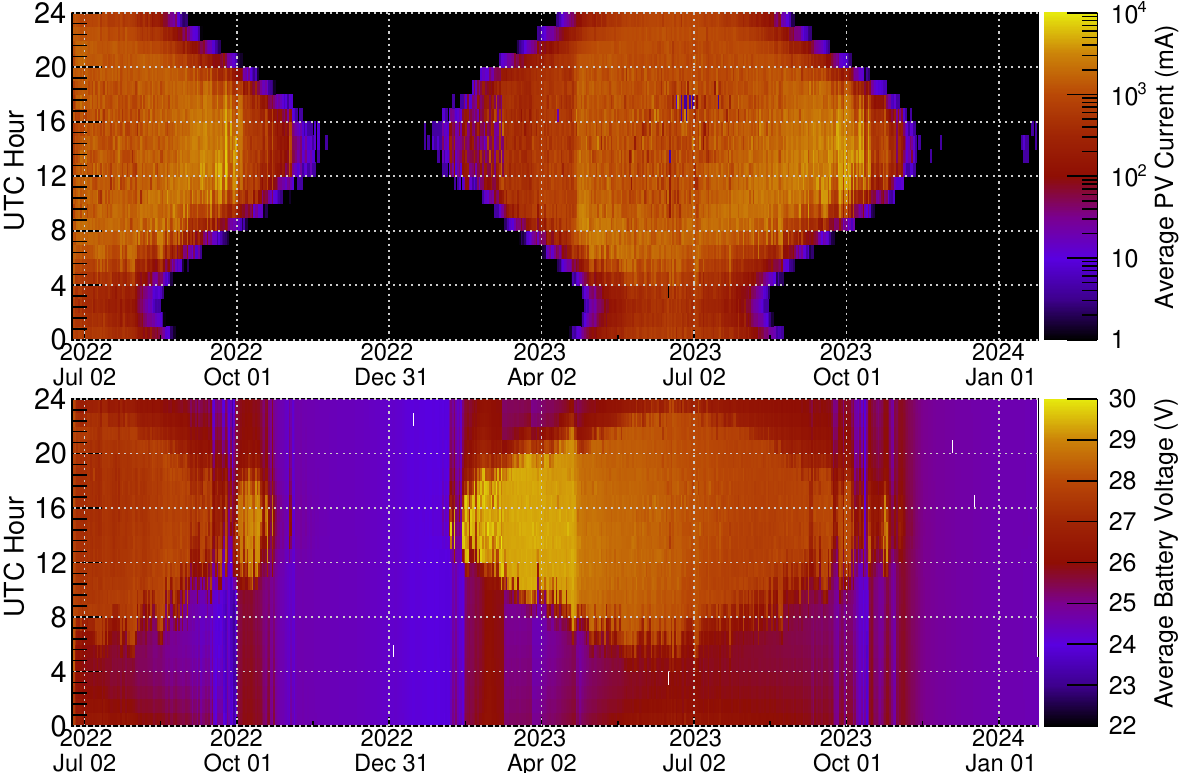}
\caption{Performance of the solar powered system of RNO-G. Top: Average photovoltaic current as function of date and UTC hour at one RNO-G station. Clearly visible are the effects of the position of the Sun throughout the year. Bottom: The same figure for the average battery voltage. The periods in Oct.2022 and Mar. 2023 that display an elevated battery voltage show the batteries receiving brief charge from the PV panels, while the instrument is in low-power mode. The batteries have minimal load in this mode. } 
\label{fig:solar_power} 
\end{figure}

RNO-G stations are not grid-connected and must generate their own power from renewable systems. The RNO-G-7 stations are primarily solar-powered, with wind-generation options undergoing development and testing in order to prolong operations into the polar night present at RNO-G's latitude. \autoref{fig:solar_power} displays the solar power generation over a period of 1.5 years at a single RNO-G station equipped with a 24~V battery bank. 

The solar power system consists of two solar panels delivering up to 350 W. The solar panels are connected through Electromagnetic Interference (EMI) filter pins to a Flexcharge NC30L solar regulator~\cite{flexcharge}. A battery bank comprising four sealed-lead-acid batteries (DEKA 8G31, 12V 108Ah) provides approximately 5 kWh of battery capacity, which practically de-rates by a factor of 2-3 at typical operating temperatures and to avoid deep-discharge cycles.  The first 2021-deployed stations were built with 12~V battery banks, while the later 2022 stations were wired at 24~V to decrease the charging currents and for improved compatibility with potential future wind turbines. There is a custom power filtering and routing board within the instrument box called the `TidyPower', which interfaces to the solar panels and batteries, simplifying the internal cabling. The solar panels are connected to the instrument using a pair of shielded four-conductor 14AWG cables, with lengths up to 10~m, to accommodate maximal charge currents of 13~A (25~A) for the 24~V (12~V) systems. The RNO-G controller board includes sensing circuits to record the solar panel voltage and charge current, as well as to monitor the battery voltage and temperature.

The RNO-G instrument uses a pair of wide-voltage input (9-36~V) and high-efficiency (>90$\%$) step-down DC-DC converters to generate the internal voltage rails used by the instrument subsystems from the primary battery voltage. An input-side EMI filter circuit is implemented before these DC-DC switching converters that sufficiently rejects undesired radio emissions that might be picked up in our receivers. 
A dedicated +3.3 V micropower DC-DC converter is used to power the controller board microcontroller, LoRaWAN wireless transceiver, and low-level controller board sensors. 
This +3.3V rail is always enabled, providing power to the microcontroller and its peripherals used to continuously monitor the system, provide resets if necessary, and run in the over-winter ultra-low power mode.
A separate high-current +5V DC-DC converter is used to power everything else in the instrument, including the digitizer boards, amplifiers, SBC, and LTE modem. This converter can provide up to 8~A, and can be remotely enabled and disabled by the microcontroller via LoRaWAN. All downstream RF amplifier boards have a local low-dropout regulator and additional input filtering, which suppress any residual power-supply noise on these sensitive components from this primary +5V rail.

In low-power \emph{winter} mode, the RNO-G instrument draws less than 100mW, allowing the station microcontroller to remain on and transmitting low-duty cycle station monitoring information throughout the polar winter period. The primary power draw in winter-mode is the self-consumption of the Flexcharge solar controller ($\sim$3~mA). We note that all stations in the RNO-G-7 array have successfully operated in this mode over at least two winters without a failure, re-engaging to full-power science operations in the spring.

Significant effort is now being made to develop a wind power system that has minimal in-band EMI emissions and is robust to the extreme weather conditions at Summit Station. With a successful development of these new wind turbine systems, we expect to be able operate during 70-90\% of the year given the recorded wind conditions at Summit.

Power for the LTE base station and RNO-G computer server, both installed in powered facilities at Summit Camp, is drawn from the primary station generator with a total of 1~kW allocated for RNO-G equipment.

\subsection{Instrument Faraday Housing and Construction}
\begin{figure}
\centering
\includegraphics[width=0.6\textwidth]{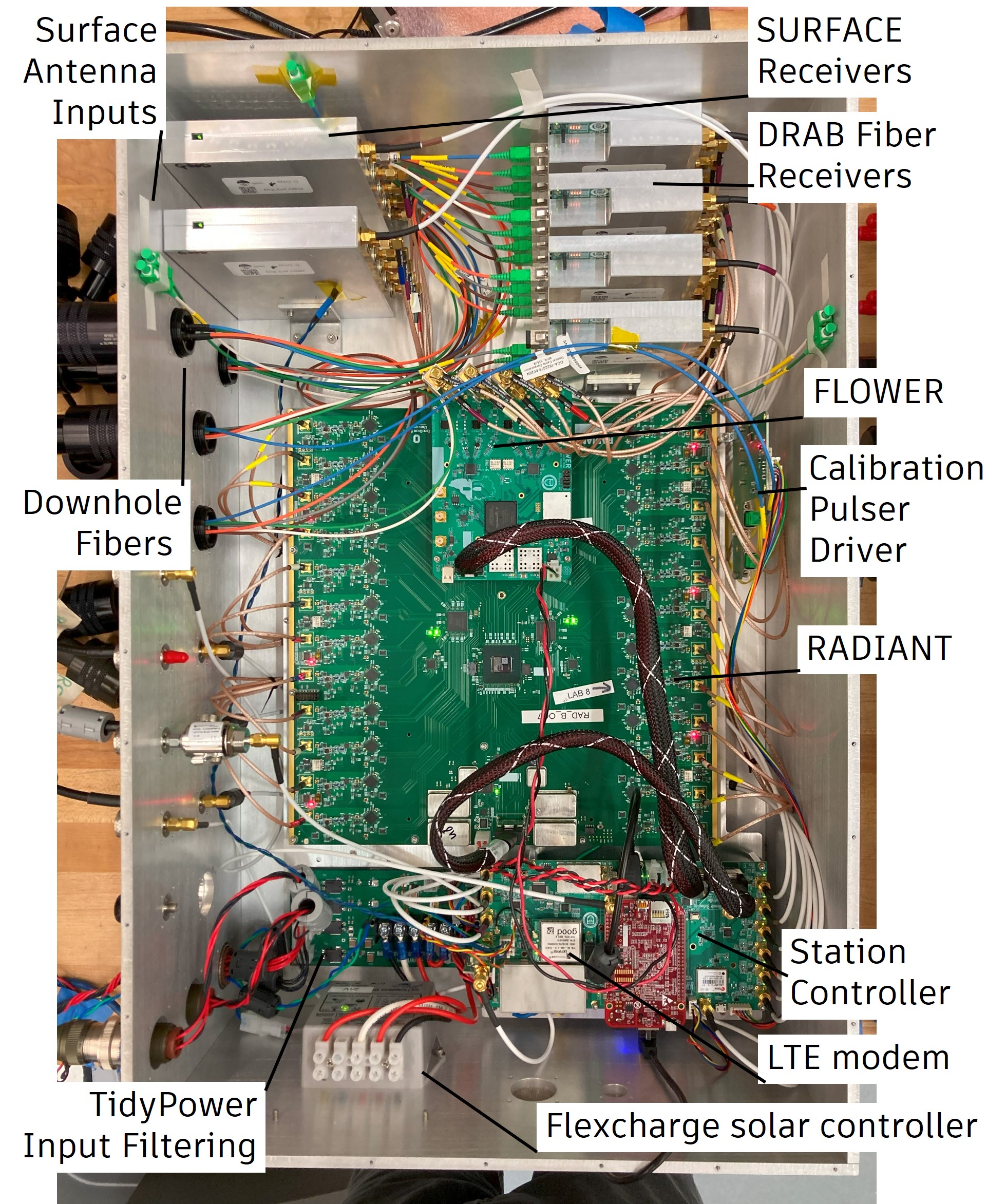}
\caption{The RNO-G instrument box with the lid removed, exposing the internal electronics. A single Faraday enclosure houses all of the surface-side electronic subsystems described in this section. }
\label{fig:rnog-box}
\end{figure}

The RNO-G-7 instrument box, shown in \autoref{fig:rnog-box},
holds all of the digital, power, and surface-side RF electronics
described in this section within a single RF-shielded enclosure. 
The custom 24"x16"x8" enclosure is machined from aluminum and subsequently applied with a chemical-conversion coating that enhances the conductivity for RF shielding purposes. All external I/O connectors are all placed on a single side of the box for ease of cable management and installation within the environmental enclosure in the field.
This tightly integrated design enables the scalable production of the
RNO-G instrument.

\subsection{Instrument calibration database and monitoring}
\begin{figure}
\centering
\includegraphics[width=0.55\textwidth]{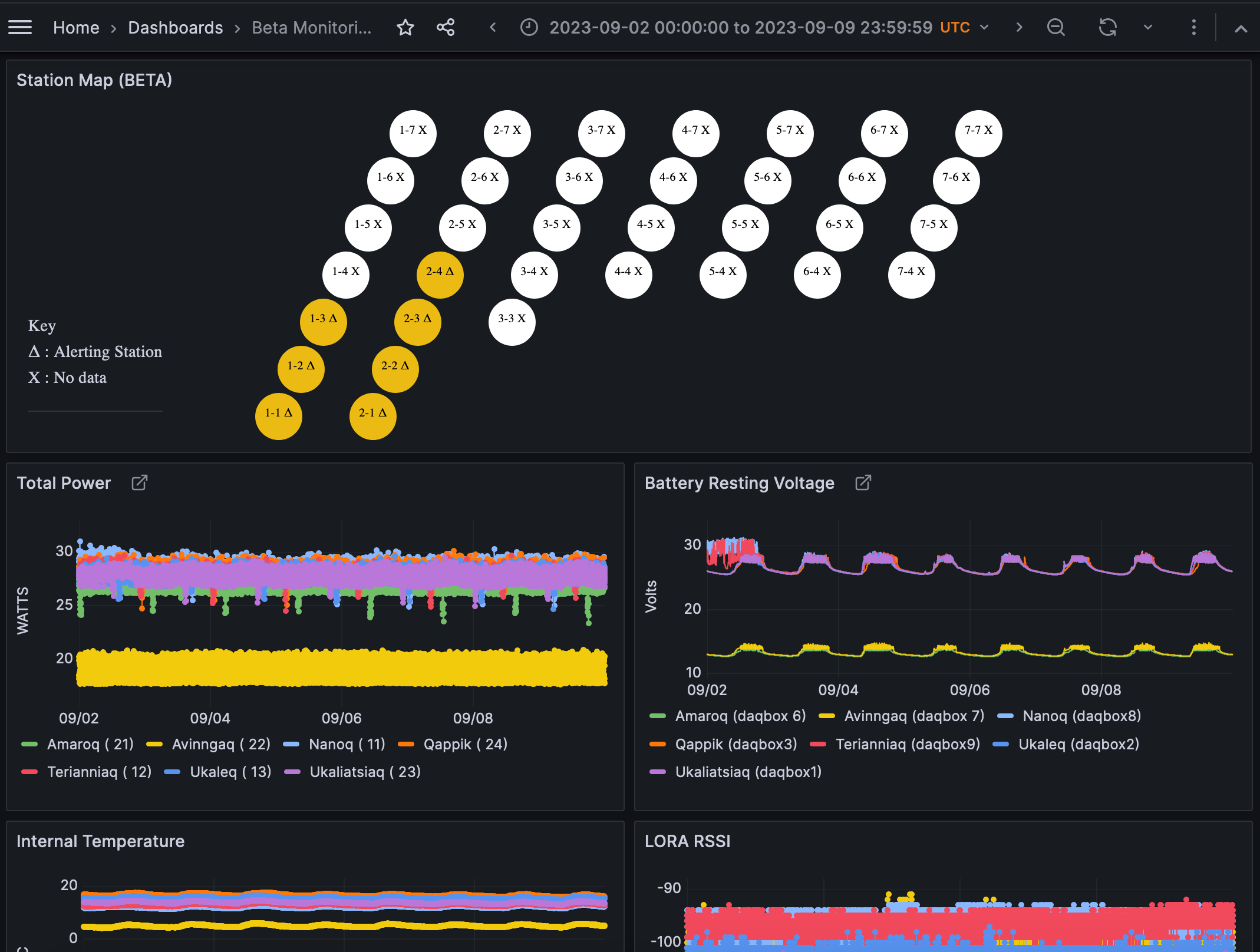}
\includegraphics[width=0.44\textwidth,trim={13cm 1cm  1cm 1cm },clip]{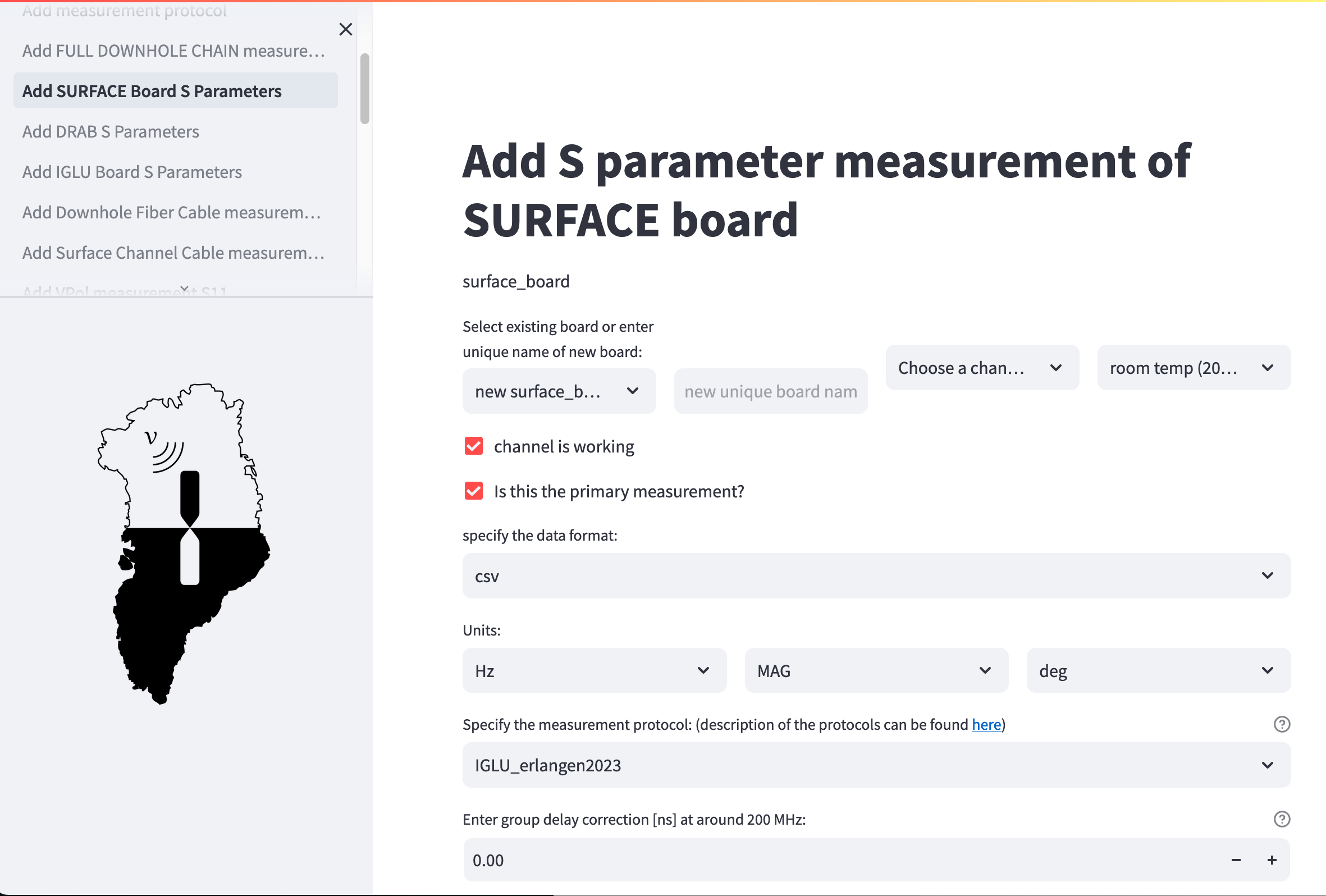}
\caption{Database interfaces for the operation, monitoring and calibration of RNO-G. Left: The \texttt{grafana} webinterface depicting (a fraction of) the monitoring data transferred from every station. Automated alerts are in place. Right: The webinterface to the calibration database, that is being used during production and testing. The underlying database is directly usable for reconstruction. }
\label{fig:dbs} 
\end{figure}

All electronic hardware components are cold-soaked and temperature tested in the lab before installation, involving several rapid stress-testing  temperature cycles (-50$^{\circ}$C to 50$^{\circ}$C) and functional tests at -40$^{\circ}$C. Furthermore, all relevant performance characteristics are measured (e.g.\ S-parameters, pulser waveforms) before shipping to Greenland. The calibration measurements are stored in a database (MongoDB, see \autoref{fig:dbs}, right), which is directly accessible for reconstruction and detector simulations. The open source code \texttt{NuRadioMC}~\cite{Glaser:2019cws} is used for reconstruction in RNO-G, with only RNO-G-specific modules being private to the collaboration. 

The \texttt{grafana} package is used to visualize the  \texttt{postgresql} database containing sensor data from stations, with instances operating both in Greenland (for the deployment team as well as to generate near-realtime alerts) and down South (for general monitoring and historical inspection.)  A set of rich \texttt{grafana} dashboards have been created to help visualize the state of the detector (\autoref{fig:dbs}, left). Alerts of housekeeping parameters deviating from expected values can be relayed by e-mail or \texttt{Slack}.

\subsection{Installation}
\label{sec:installation}

\begin{figure}
\centering
\includegraphics[width=0.49\textwidth]{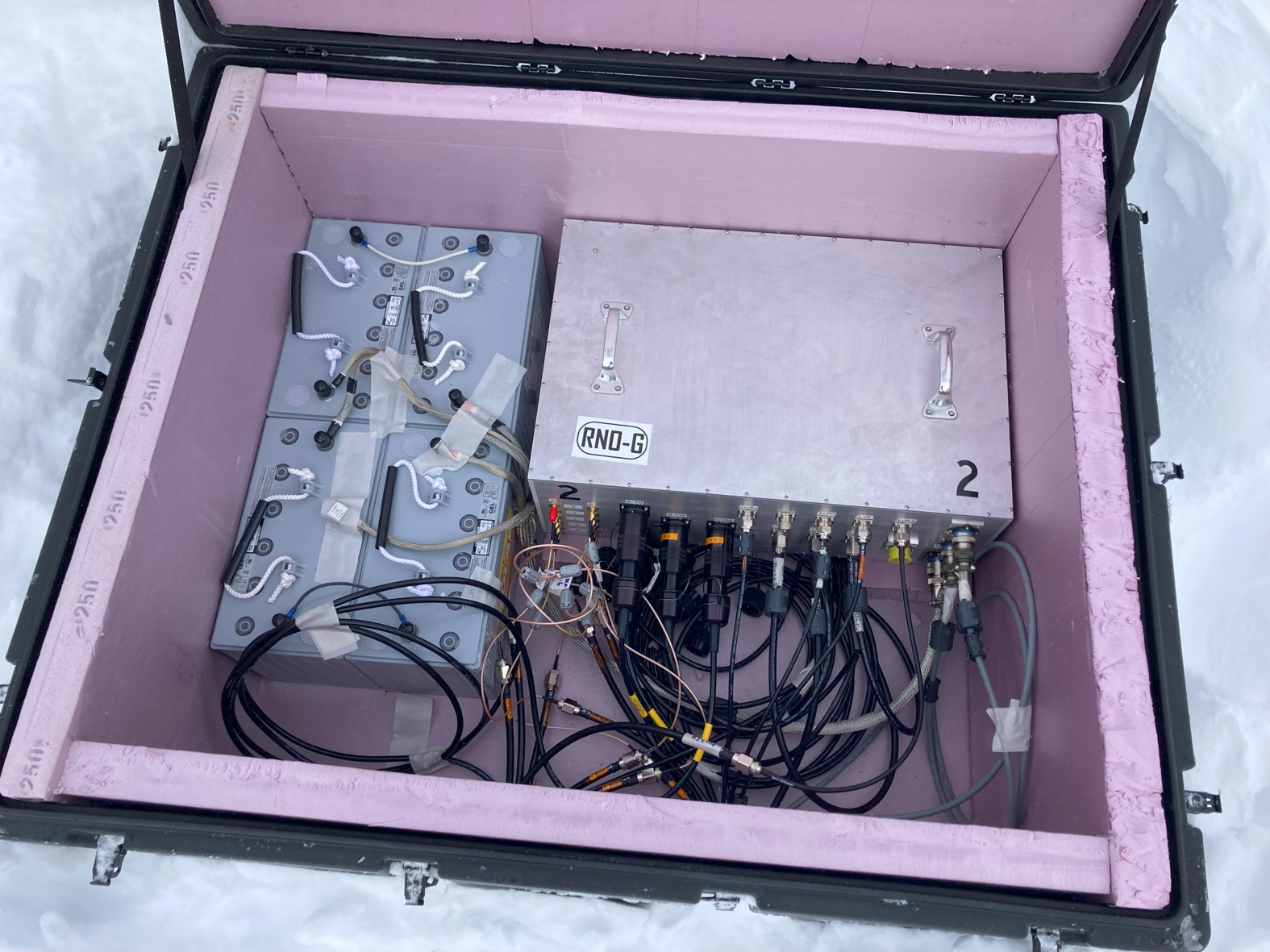}
\includegraphics[width=0.49\textwidth]{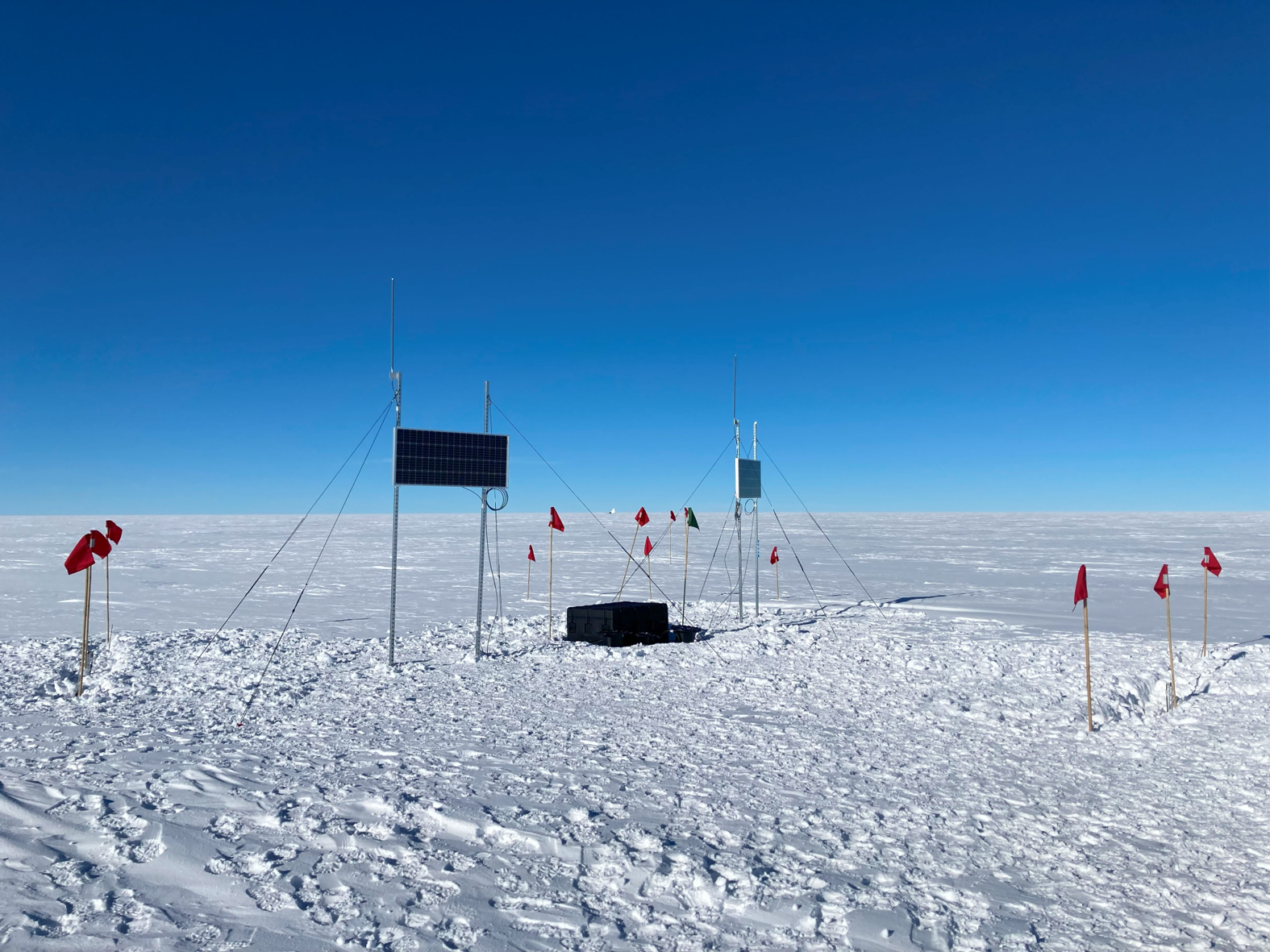}
\caption{Fielded RNO-G station. The instrument vault is pictured on the left, with the battery bank and the RF-shielded instrument box. On the right, the station-site is shown shortly after completing the installation with the central vault flanked by two south-facing vertically-mounted solar panels. The communications and GPS antennas are mounted on the top of these structures.}
\label{fig:installation} 
\end{figure}

The installation of the science instrumentation at a new RNO-G station is a 3-4 day process, not including drilling. It involves string deployment, near-surface antenna installation, construction of the solar-power and communications structures, and the early validation of performance.
First, the instrument box is installed in an insulation-lined environmental enclosure with the battery bank, and wired to the solar panels and communications antennas.
The LTE and LoRaWAN links are then verified, bringing the instrument online. 
The environmental enclosure and solar panel installations are shown in~\autoref{fig:installation}.

The antenna installation then follows, concurrently deploying a borehole string and the corresponding 3-LPDA trench along the same spoke in the station layout. Each string and set of surface antennas is connected to the instrument immediately after deployment to verify channel-level operation and observation of thermal-noise backgrounds.
In the case a bad channel is found, the deployment team can consider raising the string to repair or replace, and a procedure for this action has been defined and implemented in the field with good success.
The helper string takes $\sim$3~hours to deploy, while the power string takes 5-6~hours due to the added complexity of the upper Vpol antennas, which require more complicated load-transfer operations and cable-management in the string deployment.
A 3/8" (\SI{0.95}{cm}) diameter poly rope is used for rigging the borehole strings, which is loaded from the borehole instrumentation at significantly less (<10$\times$) than its 2,400~lbs (\SI{1100}{kg})  breaking strength rating.

\section{Station performance}
\label{sec:performance}
This section describes the initial performance of the first 7 stations of RNO-G. We will highlight measurements taken during installation and those taken with early science data. These data are used by RNO-G to develop methods and tools to reliably operate and calibrate the detector. 

\subsection{Instrument uptime}
\begin{figure}
\centering
\includegraphics[width=0.8\textwidth]{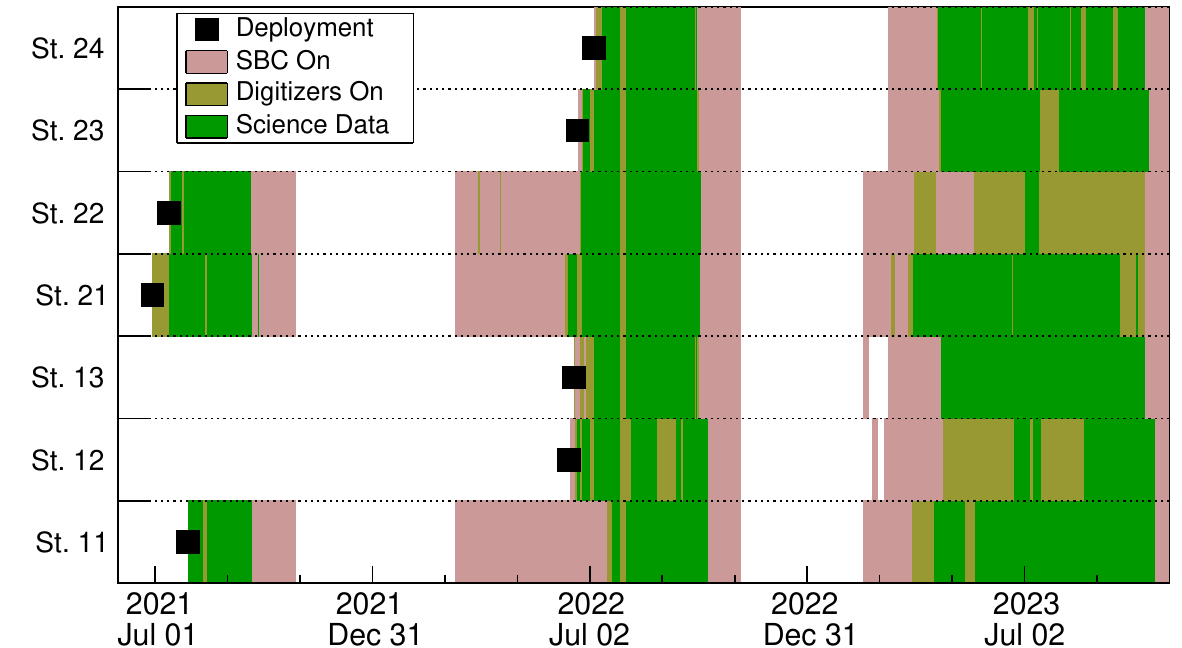}
\caption{Uptime of the stations as function of time, spanning the three summer seasons of 2021 to 2023. The figure indicates when stations were installed (black squares), when the stations were in a low-power mode with the SBC and LTE-link powered on (dark pink), when the digitizers were on (olive), and when they were taking science data (dark green).}
\label{fig:uptime} 
\end{figure}

The livetime of the RNO-G-7 array is shown in \autoref{fig:uptime}, which notes the station deployment date and distinguishes the operational modes including science data taking.  The best performing station (13) delivered a science data-taking uptime of 51\% and an overall uptime (SBC on) of 75\% since it was installed. The SBC on-time, a power mode that draws roughly 4~W, is relatively similar for all stations, meaning that the solar power system has been working reliably under conservative operation parameters. 
The livetime performance is condensed in~\autoref{tab:uptime}, which shows the time-fraction in which each station was fully powered on and when each station was engaged in science operations. The 2023 season was primarily dedicated to operations so there is a general improvement in station livetimes when compared to the 2021 and 2022 seasons, in which drilling and installation activities were ongoing.
In 2023, five of the RNO-G-7 stations delivered science data for more the 43\% of the year and all stations were fully powered on between 44\% and 56\% of the year. The exceptions to the science operation uptime were stations 12 and 22. Station 12 was compromised due to electromagnetic interference from a prototype wind turbine power-system and station 22 was not fully operable due to a faulty \textsc{Radiant} board that did not properly re-start after the winter. Both issues were resolved in early 2024. 

\renewcommand{\tabcolsep}{6pt}
\begin{center}
\begin{table}
\caption{Uptime of the RNO-G-7 stations. Two live-times are tabulated from~\autoref{fig:uptime}:  the fractional time when each station is fully powered on and when the station is recording science data. The uptimes are shown for the 2023 calendar year, which was a year dedicated to operations instead of field-installation work.}
\label{tab:uptime} 
\centering
\begin{tabular}{|p{1.3cm}|p{2.4cm}|p{2.4cm}|}
\hline
\textbf{Station} &  \textbf{full-power uptime} &  \textbf{science-ops. uptime} \\
\hline
11   & 56\%   & 48\%  \\
12    & 48\% & 22\%  \\
13   & 47\% & 47\%  \\
21   & 55\% & 48\% \\
22   & 44\% & 3\%  \\
23   & 48\% & 43\%  \\
24   & 48\% & 43\% \\
\hline
\end{tabular}
\end{table}
\end{center}

\begin{figure}
\centering
\includegraphics[width=0.9\textwidth]{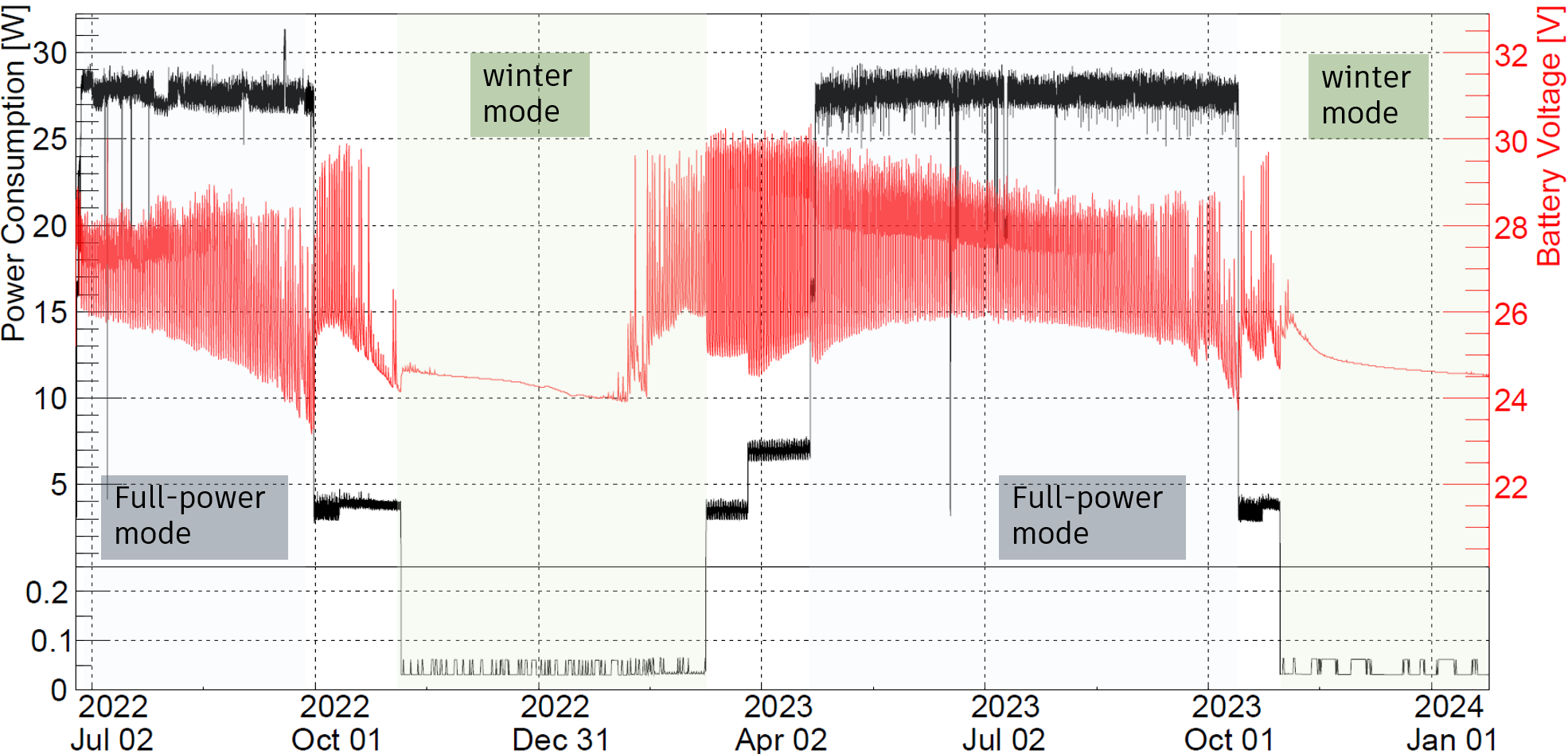}
\caption{All-year operation of the solar-power system at RNO-G station 23. Shown are both the power consumption (left axis, black) and battery voltage (right axis, red) as function of time. The power consumption shows the different modes of the experiment, including the full-power mode comprising science operations and the low-power winter mode. The transitions between these two modes are also visible, in which the primary DAQ boards are powered-off while keeping the SBC and LTE communication link fully on. The battery voltage is determined by the availability of solar power (see \autoref{fig:solar_power}) and the power consumption of the system. }
\label{fig:over-winter} 
\end{figure}

\autoref{fig:over-winter} shows the year-round operation of one station. At the end of fall, the digitizers are turned off first to ensure that the stations enter winter with full batteries. In spring, the stations are slowly turned on to not immediately drain the batteries again and to start warming up the stations. This is needed to avoid booting the DAQ hardware at or below $-40^{\circ}$C, where performance is less reliable. It is anticipated that with optimized processes and improved experience with the system, the science data-taking time can be increased.

\subsection{Time-domain system response}
\label{sec:tdomain}

\begin{figure}
\centering
\includegraphics[width=0.99\textwidth]{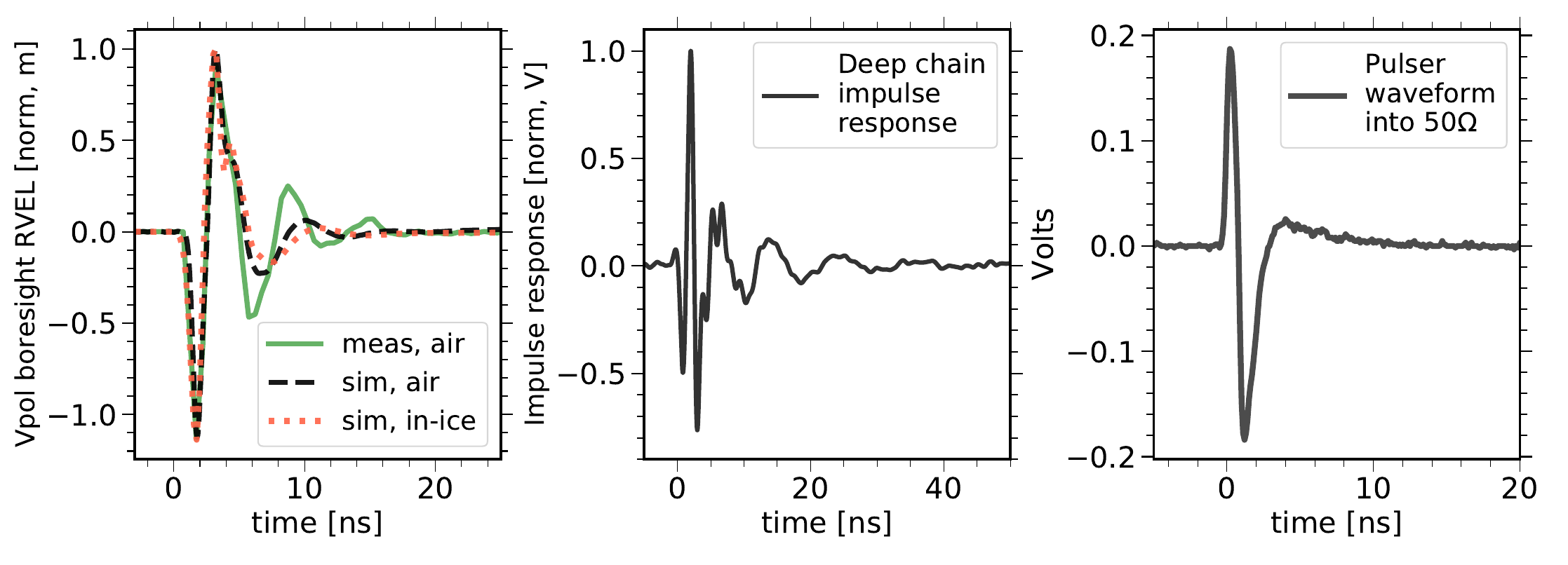}
\caption{Component-level time-domain responses. Left: RVEL for the Vpol antenna as provided by the in-air and in-ice Xfdtd simulations, compared to an anechoic chamber measurement. Middle: Impulse response of the full deep-receiver signal chain, as measured from the \texttt{IGLU} input and recorded through the \textsc{RADIANT}. Right: Deep calibration pulser waveform, at the output of the \texttt{IGLU-Cal}, as measured on an oscilloscope.}
\label{fig:sysresponse1}
\end{figure}

\begin{figure}
\centering
\includegraphics[width=0.99\textwidth]{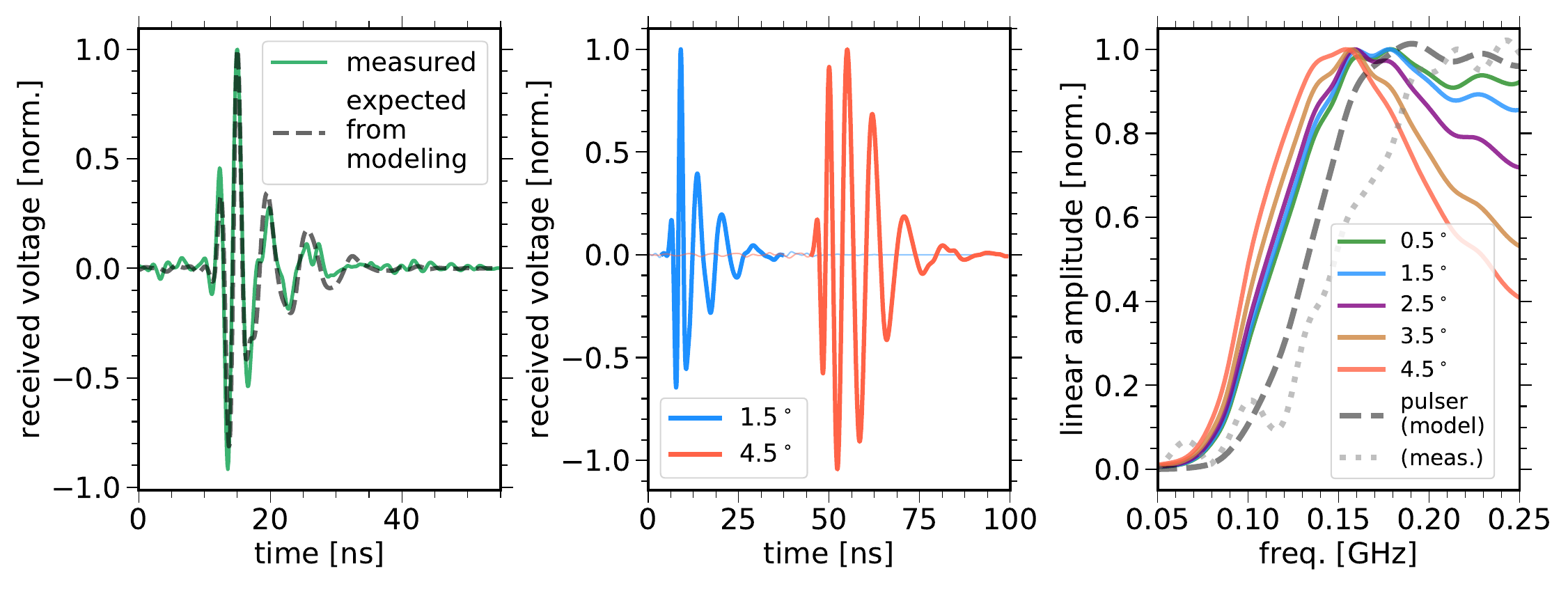}
\caption{Full-system received signals from calibration pulsers and simulated Askaryan electric-fields. Left: Received calibration pulser waveform at station 11 (green solid) compared to expected waveform from cascading component-level models (black dashed). Middle: Expected waveforms for 1.5$^{\circ}$ and 4.5$^{\circ}$ off-cone Askaryan signals, as received on-boresight at a Vpol antenna and propagated through the full instrument response.  
Right: The received frequency response of various off-cone angle Askaryan signals compared to the received signal response arising from calibration pulser events, shown within the \emph{\textsc{Flower} trigger} bandwidth.}
\label{fig:sysresponse2}
\end{figure}

The time-domain response of the RNO-G system predominantly  
determines the characteristics of the received Askaryan signal
at the instrument, and has implications for trigger design and reconstruction capabilities.  These responses have been measured
for each component in the lab, and at a full-system level using
both fielded data from in-situ calibration pulsing, using the transmitting antennas on the helper strings, and lab data taken
in anechoic chamber measurements. Additional in-situ system-timing performance metrics are presented in this section, including the system timing resolution and the reconstruction of vertical pulser scans of the power string receivers.

\autoref{fig:sysresponse1} shows the component-level 
time-domain responses in the RNO-G instrument. 
The on-boresight phase response of the Vpol receiving antenna RVEL (the amplitude response is described in~\autoref{fig:vpol_display}) is shown on the left, comparing a measurement in the lab to simulations in-air and within an in-ice borehole. There is mostly good agreement with some discrepancy between measurement and simulation after the main pulse cycle, which might be attributed to undamped reflections in the anechoic chamber measurement. The impulse response of the full electronics chain, including the interconnected RF and DAQ components, is shown in the middle graphic. By design, this response has minimal group-delay dispersion with most of the signal power present within the first 10~ns, although the ringing in the pulse tail is driven by  the dispersion in the high-pass filter edge.
On the right, the impulse waveform provided by the calibration pulser system is shown, as presented to the transmitting Vpol antenna, at a mid-range attenuation setting.

The time-domain response of the full system is examined in ~\autoref{fig:sysresponse2}, using the in situ calibration pulser and simulated Askaryan signals. In the left frame, an average pulser waveform on station 11 is compared to an expected waveform that is generated by cascading the component-level responses in~\autoref{fig:sysresponse1}. The expected (amplitude-normalized) waveform is generated using the formulation outlined in~\cite{Barwick:2014boa}. We find a peak cross-correlation coefficient of 0.94 between the measured and modeled received pulser waveforms, suggesting a good understanding of the time-domain system response.  Askaryan electric-field signals are simulated at two off-cone view angles and sent through the receiver system response in the middle plot. This shows the expected relative enhancement of the low-frequency tail of the received waveform for more off-cone events. In the right frame, the \emph{\textsc{Flower} trigger}-band pulse spectra are shown for simulated Askaryan signals received at a variety of off-cone angles, compared to the pulse spectrum from events provided by the in situ calibration system.   Although the calibration pulser provides an impulse that is fairly well-matched to an on-cone Askaryan signal, the marginally higher turn-on frequency for received waveforms from the calibration pulser has implications when evaluating the trigger using these events, as will be discussed in~\autoref{sec:trig_performance}.

\subsubsection{Two-channel timing}
\begin{figure}
    \centering
    \includegraphics[width=0.6\textwidth]{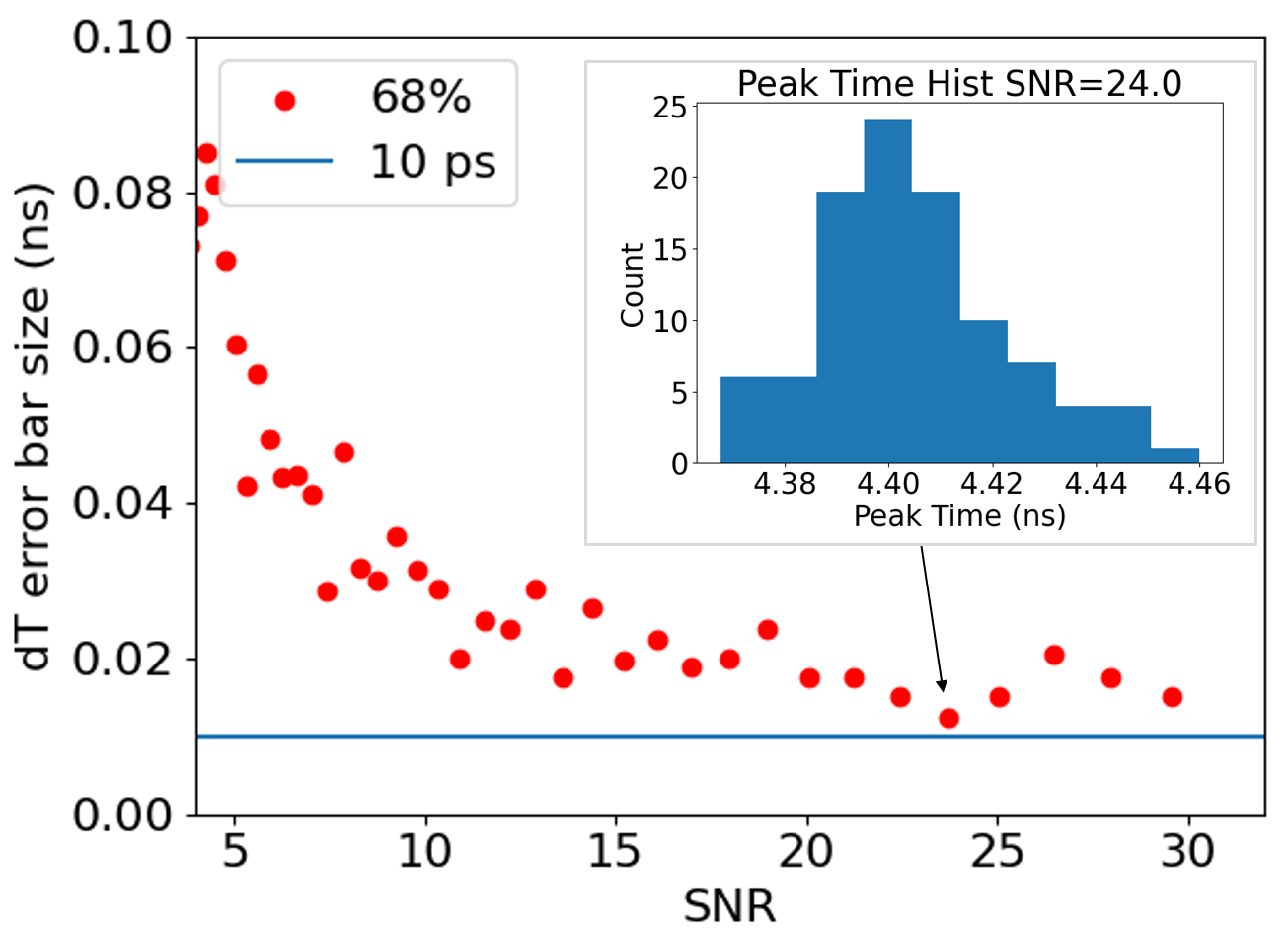}
    \caption{Two-channel relative timing resolution of an RNO-G station as function of signal-to-noise-ratio (SNR). The two-channel statistical timing resolution is measured using the maximum cross-correlation (dT) of in-situ calibration pulses between two channels on the helper string opposite to the pulser. The inset histogram shows a representative distribution of measured time-differences at each SNR, from which the  the 68\% containment interval defines the timing resolution.
    Pulses from the transmitting antenna were generated using different attenuation settings on the Calibration Transmit board.}
    \label{fig:timing-resolution}
\end{figure}

The time-difference resolution of an RNO-G station can be measured using the calibration system as shown in \autoref{fig:timing-resolution}.
The time-difference resolution of the system depends on 
the received SNR, which can be varied
at the transmitter as implemented for the trigger evaluations. 
To measure this timing resolution, we study the pulses received by two adjacent 
Vpol antennas on a helper string opposite the pulsing string. 
For each SNR step, the waveforms of each event are processed by upsampling to a~5~ps grid 
and the time difference is measured by finding the time of maximal cross-correlation between these upsampled pulses. 
The statistical time-difference resolution is defined as the 68\% containment interval on 
the dataset taken at each SNR. For high SNR ($>$10) signals, the time-difference resolution 
is found to be 10-20~ps, which characterizes the timing resolution inherent to the 
RNO-G instrument. At lower SNR values, this timing error increases to 
80-90~ps for signals near the trigger threshold level.

For clarity, we define the SNR metric used for evaluating the RNO-G timing and trigger performance. The SNR is defined as the maximum peak-to-peak amplitude divided by twice the standard deviation of pure noise $V_\text{noise rms}$, which is the square root of the average of the squared deviations from the mean, i.e.,
\begin{equation}
\label{eq:snr}
	\text{SNR} = \frac{V_\text{pp}}{2 \cdot V_\text{noise rms}} = \frac{V_\text{pp}}{2 \cdot \sqrt{\frac{1}{n_\text{noise}} \sum^{n_\text{noise}}_{i=1} (V_i - \bar{V})^2}}, \ \text{with} \ \bar{V} = \frac{1}{n_\text{noise}} \sum^{n_\text{noise}}_{i=1} V_i.
\end{equation}
The index $i$ denotes the sample and $n_\text{noise}$ the number of samples of the considered noise.

\subsubsection{Zenith pulser scans}
\begin{figure}
\centering
\includegraphics[width=1.00\textwidth]{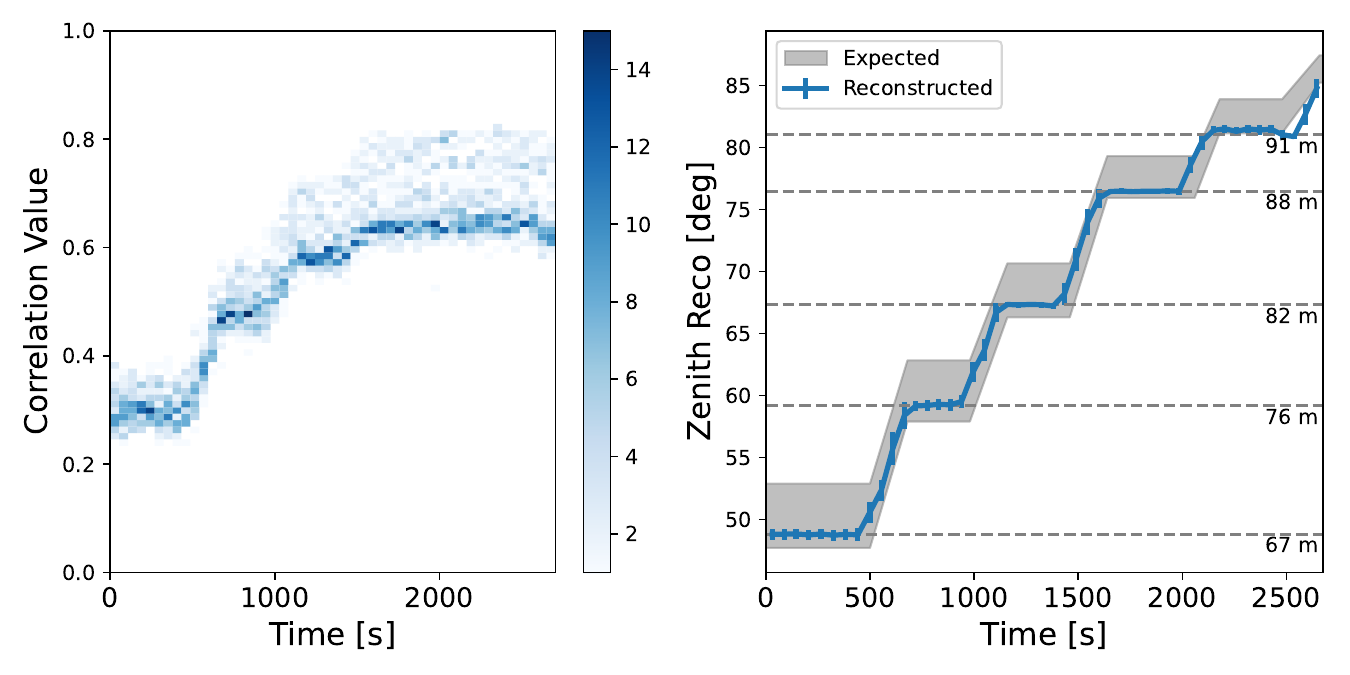}
\caption{Calibration data taken with Station 12 using the pulser on helper String B as it was lowered into its final position. Left: Maximum cross-correlation value between pulses recorded in the different Vpol channels on the deep phased-array Vpol antennas as function of time. The maximum correlation was used to reconstruct the incoming zenith angle and compared to the measured depths. Right: Reconstructed and expected incoming zenith angle as function of time. Also indicated are the measured depths at which the pulser was paused during the dropping process.}
\label{fig:pulser_drop} 
\end{figure}

During the deployment of two stations, the calibration pulser of the final helper string was enabled and transmitting while the string was lowered into place. The instrument was recording the signals on the power string and the other helper string throughout this procedure. The resulting signals show an increase in similarity (cross-correlation) with increasing depth as shown in \autoref{fig:pulser_drop}, which speaks to the improved quality of the ice and the cleaner trajectories, less affected by firn and surface effects. Furthermore, the signals reconstruct to zenith angles between $49-85^\circ$, which is in agreement with expectations from geometry. The expected zenith angle is presented as a band to account for small uncertainties in the hole depth, hole tilt, and lateral position. The expected and reconstructed zenith angles show good agreement over the full vertical extent of the transmitting antenna. 

We plan to incorporate further vertical pulser scans in upcoming station deployments. In addition to the zenith reconstruction measurement, these datasets allow for full in-situ mapping on the antenna complex gain response and constraining measurements on the index of refraction gradient in the local firn.

\subsection{Deep trigger}
\label{sec:trig_performance}

At the outset of RNO-G deployment, the \emph{\textsc{Flower} trigger} was initially installed with a simple high-low coincidence requirement, which established a baseline  trigger for science operations. To push towards higher sensitivity, an improved delay-and-sum beamforming trigger was deployed to the \textsc{Flower} during the summer of 2024 for first field testing.  For both triggers, the \textsc{Flower} ADCs are gain-controlled to set the thermal noise V$_{\mathrm{rms}}$ to 5~adu on each channel, to maintain equal trigger contributions from each Vpol receiver.
Both triggers have been evaluated in situ and are described here. The \textsc{Flower} board draws~2.7~W when operating the high-low coincidence trigger, and closer to 5~W when running the beamforming trigger.

\begin{figure}
\centering
\includegraphics[width=0.9\textwidth]{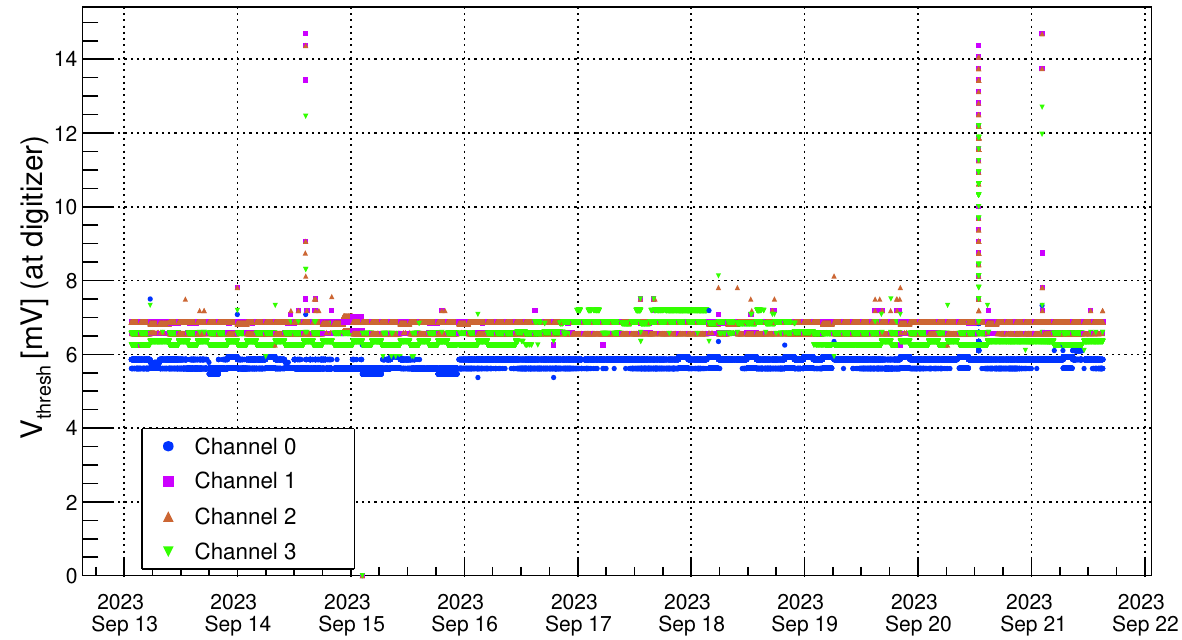}
\caption{Trigger thresholds, as used by the \textsc{Flower} board when operating the high-low coincidence trigger, on the channels forming the triggering `phased-array' subarray at the bottom of the power string. The thresholds, used for each channel in the coincidence trigger, show an excellent stability with only few periods of elevated noise, most likely from anthropogenic sources above the surface. The thresholds are set by the \textsc{Flower} board to obtain a constant trigger rate of \SI{1}{Hz}.}
\label{fig:threshold_stability}
\end{figure}

The simple coincidence trigger imparts a per-channel condition that requires the digital data to exceed a pair of symmetric high- (above noise baseline) and low- (below noise baseline) amplitude thresholds within a $\sim$14~ns window. A system-level trigger is subsequently defined if a 2-of-4 condition is met within a 42 ns coincidence window. These trigger conditions are configurable in the DAQ.
The symmetric positive and negative thresholds are controlled by software, as shown in \autoref{fig:threshold_stability} for station 13, and are set so that the resulting deep-antenna trigger rate is $\sim1$~Hz. The noise-riding thresholds are found to be stable for the deep phased-array Vpols. Each channel's trigger thresholds are dynamically adjusted based on the single-channel trigger rate at a lower ``servo" threshold, typically set to 80\% of the trigger threshold. The servo threshold rate is substantially higher than when noise-riding, allowing for more rapid feedback on the trigger threshold. The firmware records the rate at the servo threshold each second, and that is compared with a rate goal to determine if the threshold should be modified.

The firmware for the delay-and-sum beamforming trigger is based on the design of the trigger system first deployed on the ARA Phased Array \cite{Allison:2018ynt}, although with some distinct features for the $\textsc{Flower}$ architecture. Due to the lower \textsc{Flower} sampling rate, a first stage of 4$\times$ upsampling is used to generate sufficient delay-steps to have full beam coverage over the elevation angle range-of-interest. A finite-impulse response upsampling approach is implemented, using a 31-coefficient low-pass filter with a critical frequency at \SI{238}{MHz}. After upsampling, 12 delay-and-sum beams are formed, which are constructed on a plane-wave signal hypothesis and are evenly spaced to cover an elevation angle range of -60$^{\circ}$ to +60$^{\circ}$. 
Each beam operates as an independent trigger channel, with a dedicated threshold and servo control via software. The threshold criterion for each beam is computed using a power integration within a 16.9~ns window, which is re-evaluated at $\sim$4.2~ns intervals.

\begin{figure}
\centering
\includegraphics[width=0.95\textwidth]{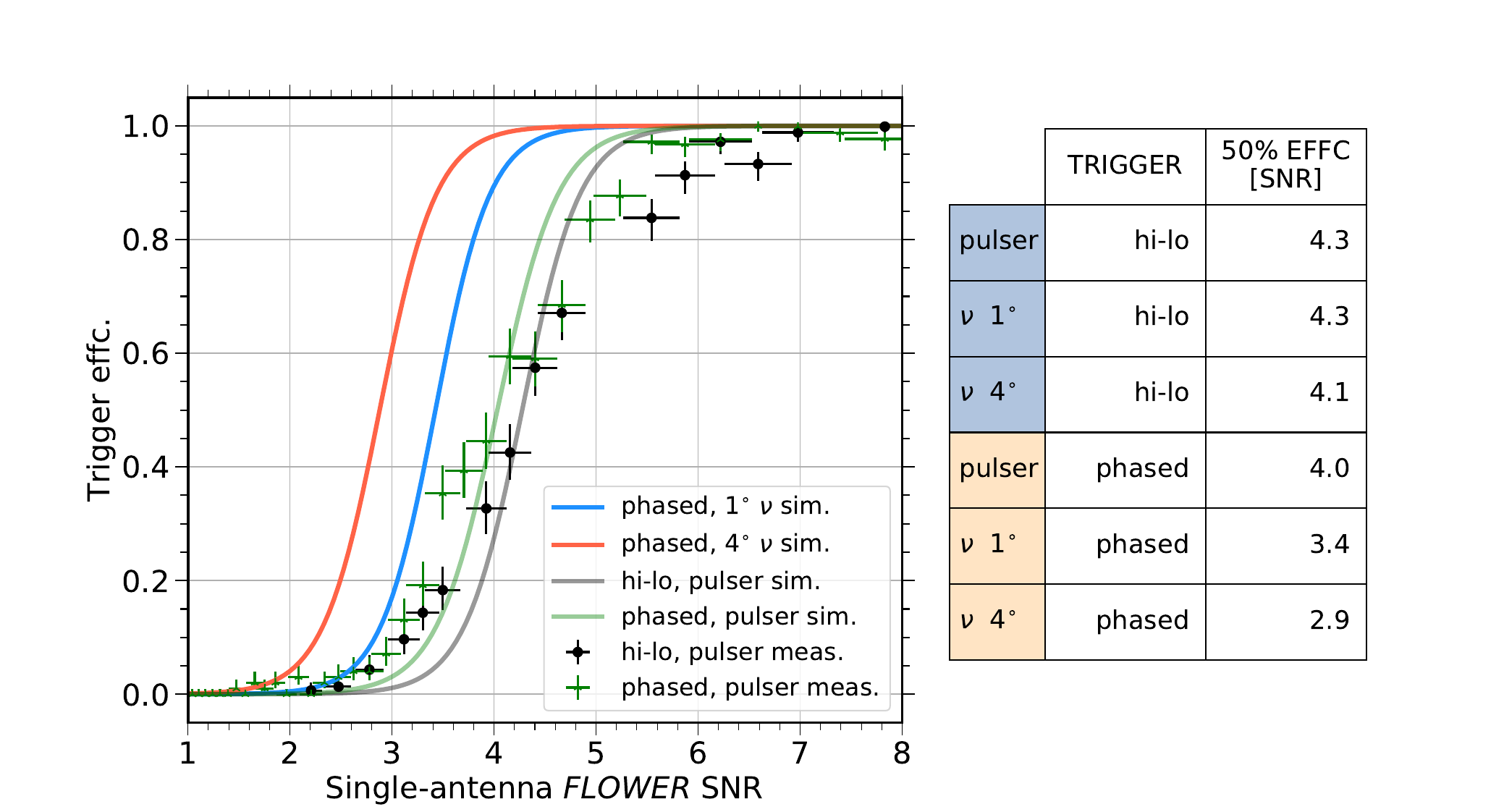}
\caption{Measured and simulated efficiency curves for the \emph{\textsc{Flower} trigger}. 
The performance of the initially-deployed high-low coincidence trigger (black data points)
and the 2024-deployed phased array trigger (green data points) are shown 
as measured at station 11. Each of these have
an associated trigger simulation that shows the expected performance using the calibration pulser, which closely matches the measured 50\% efficiency points. A significant improvement in the trigger performance is found when running the phased-array trigger simulation on neutrino events (1$^{\circ}$ and 4$^{\circ}$ off-cone Askaryan signals), due to the higher-frequency turn-on of the calibration pulser signal and the optimization of the power-integration window duration for off-cone Askaryan signals. A table of the 50\% efficiency points, given by the \textsc{Flower} SNR, is shown on the right for the two triggers as applied to the pulser and simulated neutrinos.}
\label{fig:trigger_effc}
\end{figure}

The performance of the deep trigger is shown in~\autoref{fig:trigger_effc}.
Measurements of the coincidence and beamforming triggers were taken
on station 11 using the calibration pulsers on the helper strings, which can be swept through a wide range of attenuation values. 
The 50\% trigger efficiencies are found at SNRs of 4.3 and 4.0 for the coincidence and beamforming triggers, respectively, when measured using the calibration pulser. Simulation modules for each trigger have been developed, and these find similar thresholds when run using the received pulser waveform (left panel \autoref{fig:sysresponse2}).
A marked improvement in the threshold sensitivity is found when running the beamforming (`phased') trigger simulation on Akaryan events generated with \texttt{NuRadioMC}~\cite{Glaser:2019cws}. 
For simulated neutrino events, we find a 50\% efficiency point at \textsc{Flower}-SNRs of 3.4 and 2.9 for 1$^{\circ}$ and 4$^{\circ}$ off-cone Askaryan signals, respectively. For further off-cone signals, the 50\% efficiency point plateaus around an SNR of 2.7.

The discrepancy between the trigger performance on the calibration pulser vs. simulated Askaryan signals is attributed to the higher turn-on frequency for the pulser signal as shown in the right panel in~ \autoref{fig:sysresponse2}. 
The modeled (measured) pulser signal has a turn-on frequency that is $\sim$\SI{35}{MHz} (\SI{45}{MHz}) higher than the $\sim$\SI{100}{MHz} turn-on for the Askaryan signals, which fully populate the $\sim$\SI{100}-\SI{240}{MHz} \emph{\textsc{Flower} trigger}-band for most off-cone view angles. As such, the calibration pulser has higher relative frequency content and is not an optimal signal source for evaluating the low-band trigger -- the calibration pulser events will integrate more (thermal) noise than signal in a material fraction of the trigger band.

The symmetric bipolar shape of the pulser waveform, shown in \autoref{fig:sysresponse1}, is the practical issue here. This pulse is derived from a broadband unipolar pulse, which is differentiated using a small-value inline capacitor (6.8$\pm$0.5~pF) on the Calibration Transmit Board that is housed in the instrument box. This pulse shape was chosen in order to maximize the linear amplitude that could be delivered to the \texttt{IGLU-cal} at the deep transmit antennas via the AC-coupled RFoF link. However, the spectrum of the pulse is not well-tuned for properly evaluating the \emph{\textsc{Flower} trigger} performance in situ, as the pulse turn-on frequency is higher than the turn-on frequency of the receiver.
We plan to modify this pulse shape in future instruments, in order to shift the pulse turn-on frequency well below the receiver band, and also possibly in retrofits of the existing instruments if such opportunities are presented.

\subsection{Observation of external radio signals}
\label{sec:ext_radio_sig}
RNO-G can observe a variety of external radio signals that can be used to validate, commission, and, in some cases, calibrate the array. 
Early results showing impulsive radio signals from cosmic ray air showers have already been presented \cite{RNO-G:2023kag} and will be followed up with a dedicated publication.

\begin{figure}
    \centering
    \includegraphics[width=0.8\linewidth]{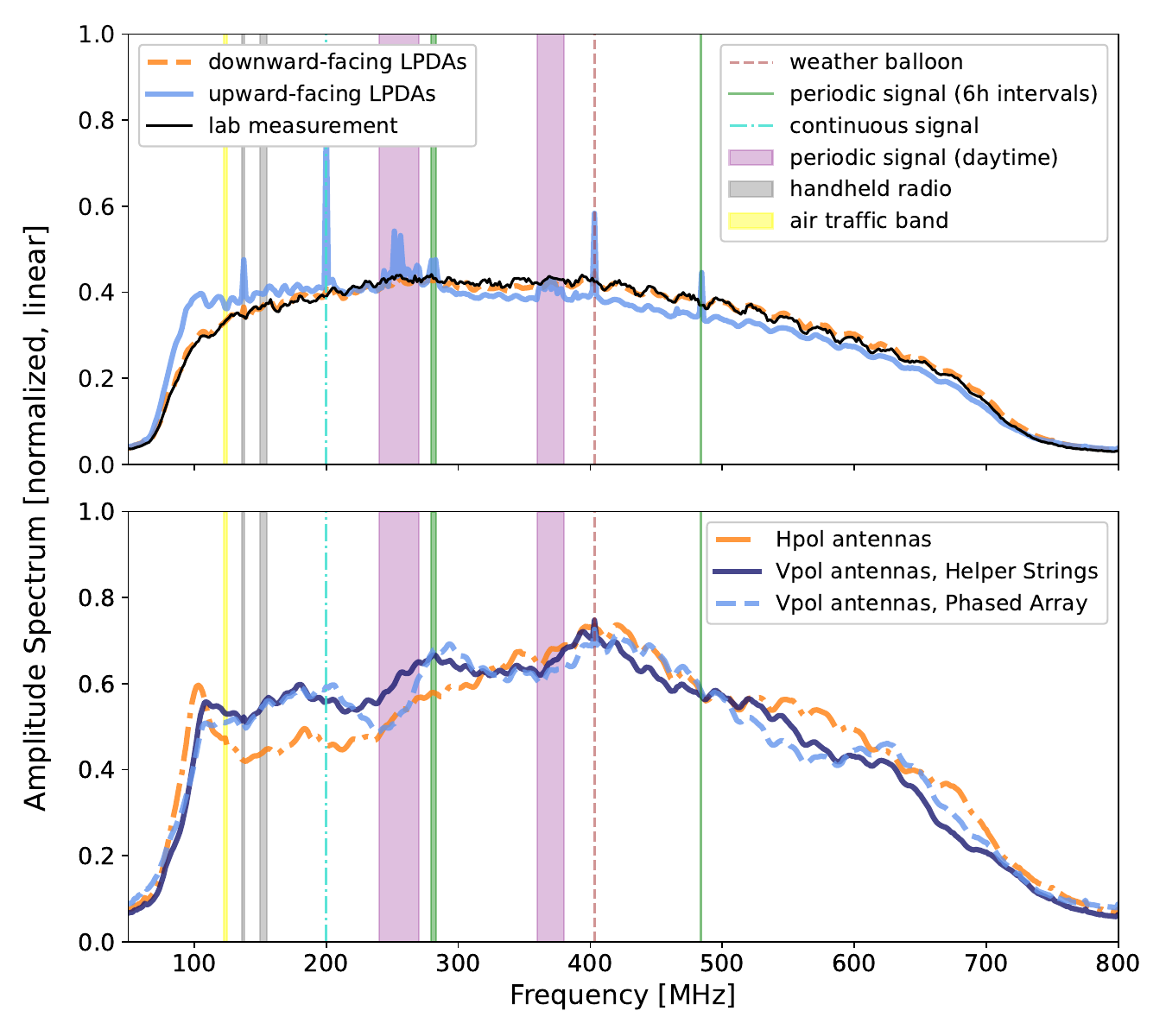}
    \caption{Average amplitude spectrum as recorded with the LPDAs (top panel) and different sets of antennas in the boreholes (bottom panel) of RNO-G over a period of 4 days in summer of 2022. 
    No absolute calibration has been applied here and the spectra have been normalized. Vertical lines and regions highlight regions with known/observed radio backgrounds and their sources. }
    \label{fig:antenna_noise}
\end{figure}

An average noise spectrum for an RNO-G station is shown in 
\autoref{fig:antenna_noise}, obtained by stacking
triggered events over the course of four days of science operations in 2022.
The data were taken during relatively low-wind days. It has been observed previously by all radio neutrino experiments, including RNO-G, that periods of high winds generate additional backgrounds~\cite{Aguilar:2023tba}. 
The overall shape is dominated by the bandpass of the RNO-G RF signal chain (see also \autoref{fig:chain_forward_gain}) that rejects signals below \SI{100}{MHz} and above \SI{700}{MHz}. 

The top panel shows the spectra measured by the upward-facing and the downward-facing LPDA channels, as well as a terminated \texttt{SURFACE} amplifier measurement in the lab that has been scaled to match the amplitude level of the downward-facing antennas.
Overall, there is good agreement between measurements in the lab and from the downward-facing antennas. 
The upward-facing LPDA channel shows a different spectral shape, measuring an excess below \SI{200}{MHz}, from the galactic radio signal (see \autoref{sec:galaxy}), and a deficit in amplitude above this frequency. Above $\sim$\SI{200}{MHz}, the galactic radio noise temperature becomes subdominant to the ice noise temperature, so that the up-pointing antenna temperature arises from some combined reception of the ice and the quiet sky. This is complicated to model precisely, due to near-field effects from the ice-air interface, but a simple model assuming 50\% contributions from each reproduces this fairly well, as shown in~\cite{RNO-G:2023fkv}. The ability to distinguish between
the up- and down-pointing antenna spectra highlights the low-noise performance of the RF signal chain design.

The upward-facing LPDAs also measure a number of narrow-band lines, stemming from human-activity on station (e.g.\ handheld radio frequencies), satellite constellations, and, to a small degree, are self-induced from the solar charging system, which will be rectified in an updated version of the power system. In general, the spectrum is mostly uncontaminated and free from anthropgenic EMI.

Spectra obtained with a selection of deep antennas are shown in the bottom panel of \autoref{fig:antenna_noise}. Similar to the spectrum obtained with the LPDAs, this spectrum still carries the instrument response characteristics. 
All spectra are arbitrarily scaled for this figure.
There is a clear deficit in power in the Hpol spectrum below 350~MHz, which is expected due to the low efficiency at these frequencies relative to the Vpol antennas. 
The deep antenna spectra are remarkably clean. Of the narrow-band lines visible with the surface antennas, only the weather radiosonde at \SI{403}{MHz} is marginally identifiable, illustrating the shielding effect of the ice and the reduced sensitivity of the deep antennas to radio sources at and above the surface.

\subsubsection{Galaxy}
\label{sec:galaxy}

The diffuse radio emission of the Milky Way has been observed by many radio detectors for air showers~\cite{Nelles:2015gca,PierreAuger:2023hun} and neutrinos~\cite{Barwick:2016mxm,TAROGE:2022soh}. Despite some existing systematic uncertainties \cite{Busken:2022mub}, the Galaxy provides an excellent absolute calibration tool \cite{Mulrey:2019vtz,deAlmeida:2023K5}, in particular between different experiments \cite{Mulrey:2020oqe}. 

\begin{figure}
    \centering
    \includegraphics[width=0.55\textwidth]{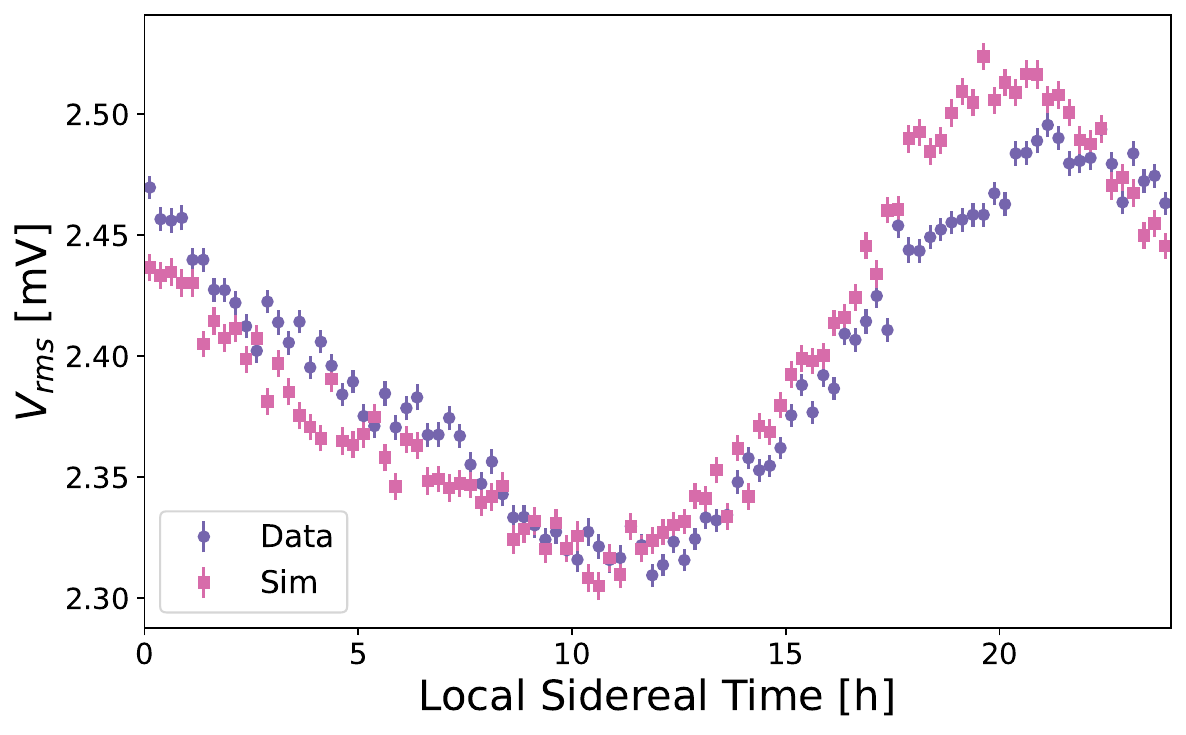}
    \includegraphics[width=0.44\textwidth]{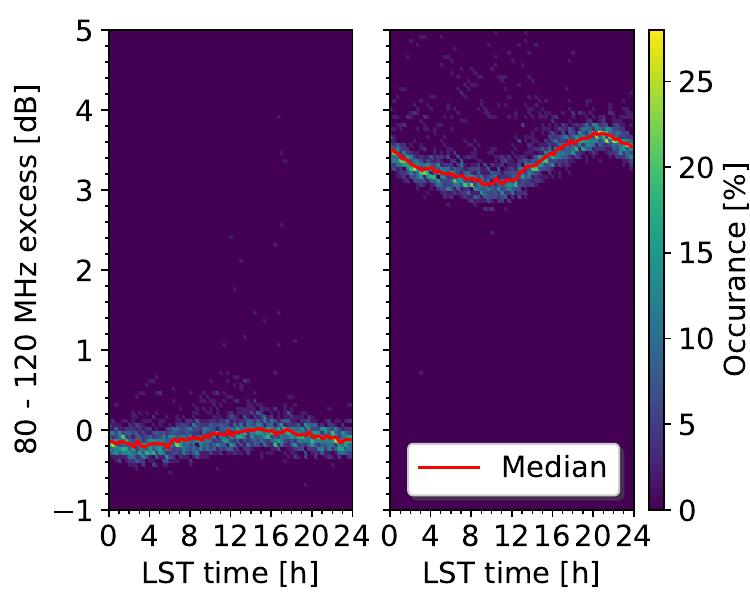}
    \caption{Time-variable signal caused by the Galactic diffuse emission in RNO-G. Left: Average root-mean-square of the voltage ($<120$ ~MHz) of the background noise in an upward facing LPDA as function of local sidereal time. Shown is both the recorded data (from June 2022 to August 2022) and the prediction of simulations that include the Galactic emission, the ambient temperature and the system response. Due to the currently incomplete absolute calibration, the simulations are scaled to match the data. Right: Comparison of the excess measured between 80 and \SI{110}{MHz} at Summit station compared to a terminated laboratory measurement. Shown are both a downward facing LPDA (middle) and an upward facing antenna (far right). The downward facing antenna shows a small residual oscillatory signal, which is compatible with no excess. }
    \label{fig:galaxy}
\end{figure}

As shown in \autoref{fig:galaxy}, the upward facing LPDAs measure a signal that is periodic in local sidereal time. The signal is seen both in the $V_{\mathrm{rms}}$ of the time-domain waveform (after a Butterworth filter with a low-pass of 110~MHz) as shown on the left, and as spectral excess (subtracting the terminated measurement) in the range of 80 - 120~MHz as shown on the right. As already shown in \autoref{fig:antenna_noise}, the spectral excess is only observed in the upward facing LPDAs. 
The signal matches the expected shape from simulations including the diffuse radio emission from the Milky Way. As done by previous experiments, the emission based on an all-sky map \cite{deOliveira-Costa:2008cxd} is calculated and fed-through a full system simulation. Work is on-going to understand the exact match between the prediction from simulations, since there is still a small systematic offset. 

In any case, these results indicate that an absolute calibration based on the Galaxy is feasible for RNO-G and that the system noise is sufficiently lower than the received galactic noise (in the low-frequency range of the RNO-G bandwidth), thereby meeting design specifications. 

\subsubsection{Radiosonde on weather balloon}
\begin{figure}
    \centering
    \includegraphics[width=0.49\textwidth]{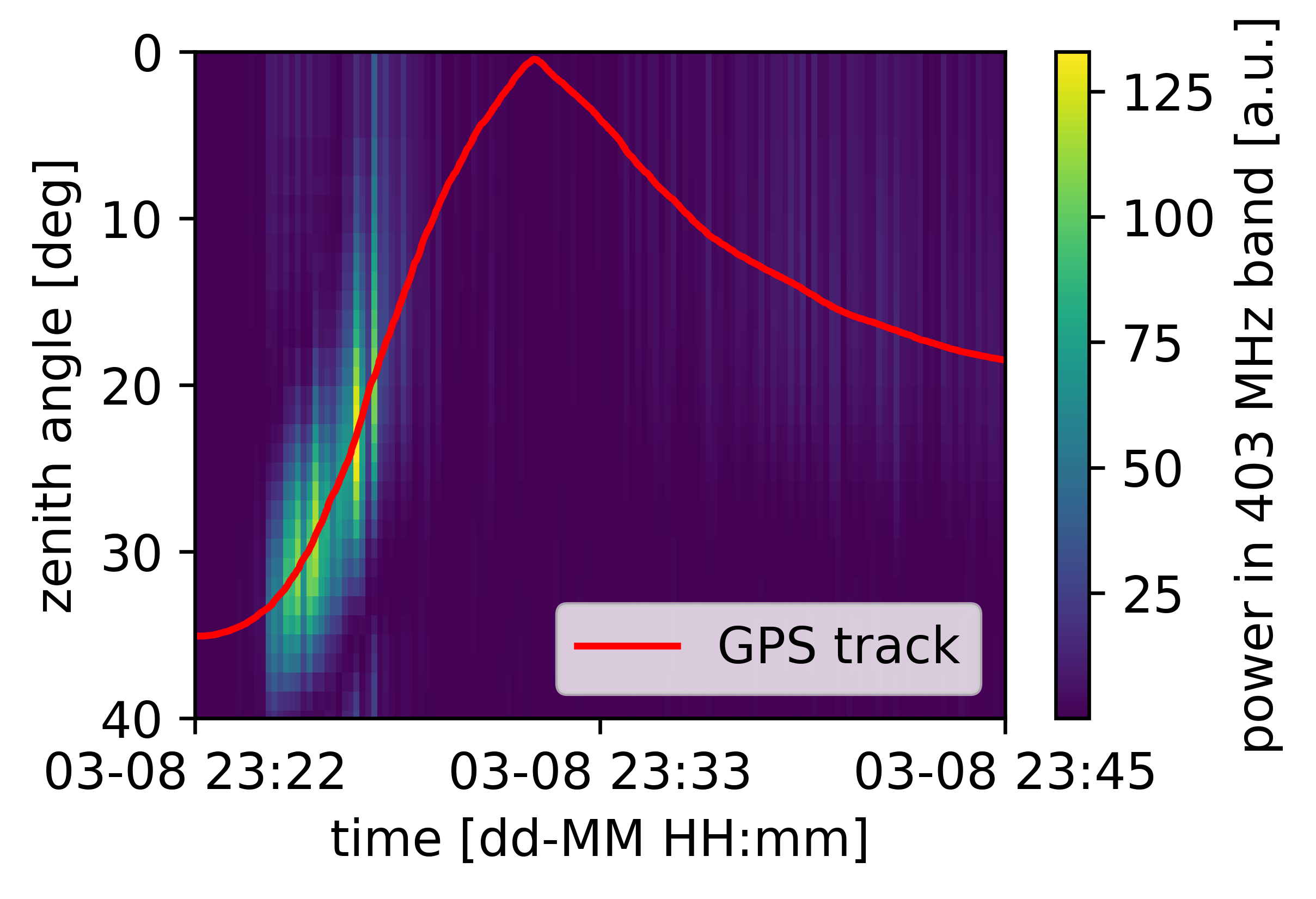}
    \includegraphics[width=0.49\textwidth]{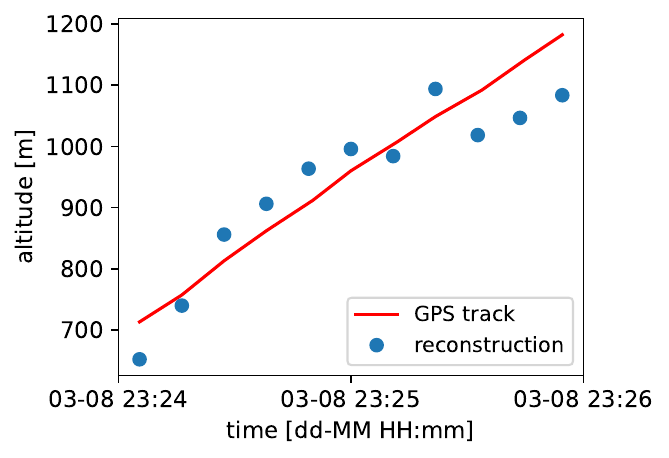}
    \caption{The radiosonde, a payload on the twice-daily weather balloon launched at Summit Station, as seen in RNO-G. Left: Interferometric map of the signal at \SI{403}{MHz} as function of time as detected in the phased array antennas of one RNO-G station, corrected for the refraction into the ice. Overlaid is the expectation from the GPS track of the emitter. Right: Altitude as function of time as reconstructed with all 7 RNO-G stations. Shown is only the best fit position, not taking into account uncertainties from reconstruction and ice model.}
    \label{fig:radiosonde}
\end{figure}

There is a regular radio background signal observed by the RNO-G array, originating from the radiosonde that is launched twice daily at Summit Station as part of the global weather monitoring (cf.~\cite{Radiosonde}). It transmits its data at a frequency of \SI{403}{MHz}, which is in the middle of the RNO-G receiving band as shown in \autoref{fig:antenna_noise}. The signal can be observed in all stations, in particular, if the wind drifts the radiosonde over the array. While the shallow antennas are most sensitive to this signal, the deep antennas also typically record the signal. The antennas of the phased array can reconstruct the zenith angle of this signal, as shown in \autoref{fig:radiosonde}. The altitude of the radiosonde can be reconstructed in certain balloon launches that result in favorable flight paths, such that the signal can be tracked in all of the RNO-G-7 stations. The altitude reconstruction results, shown in \autoref{fig:radiosonde}, use uncalibrated antenna positions, which explains the large variance, but illustrates the potential for using this signal as regular calibration source. 

\subsubsection{Observation of airplane signals}
\begin{figure}
    \centering
    \includegraphics[width=0.48\textwidth]{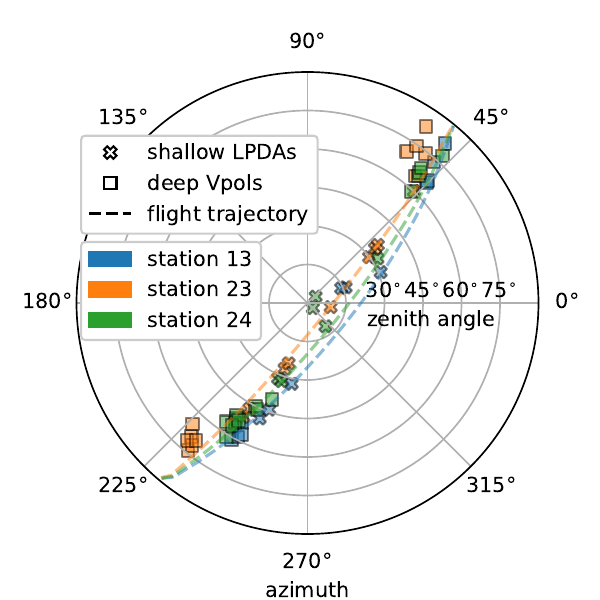}
    \hfill
    \includegraphics[trim={0 1.6cm 0 2cm},clip,width=0.51\textwidth]{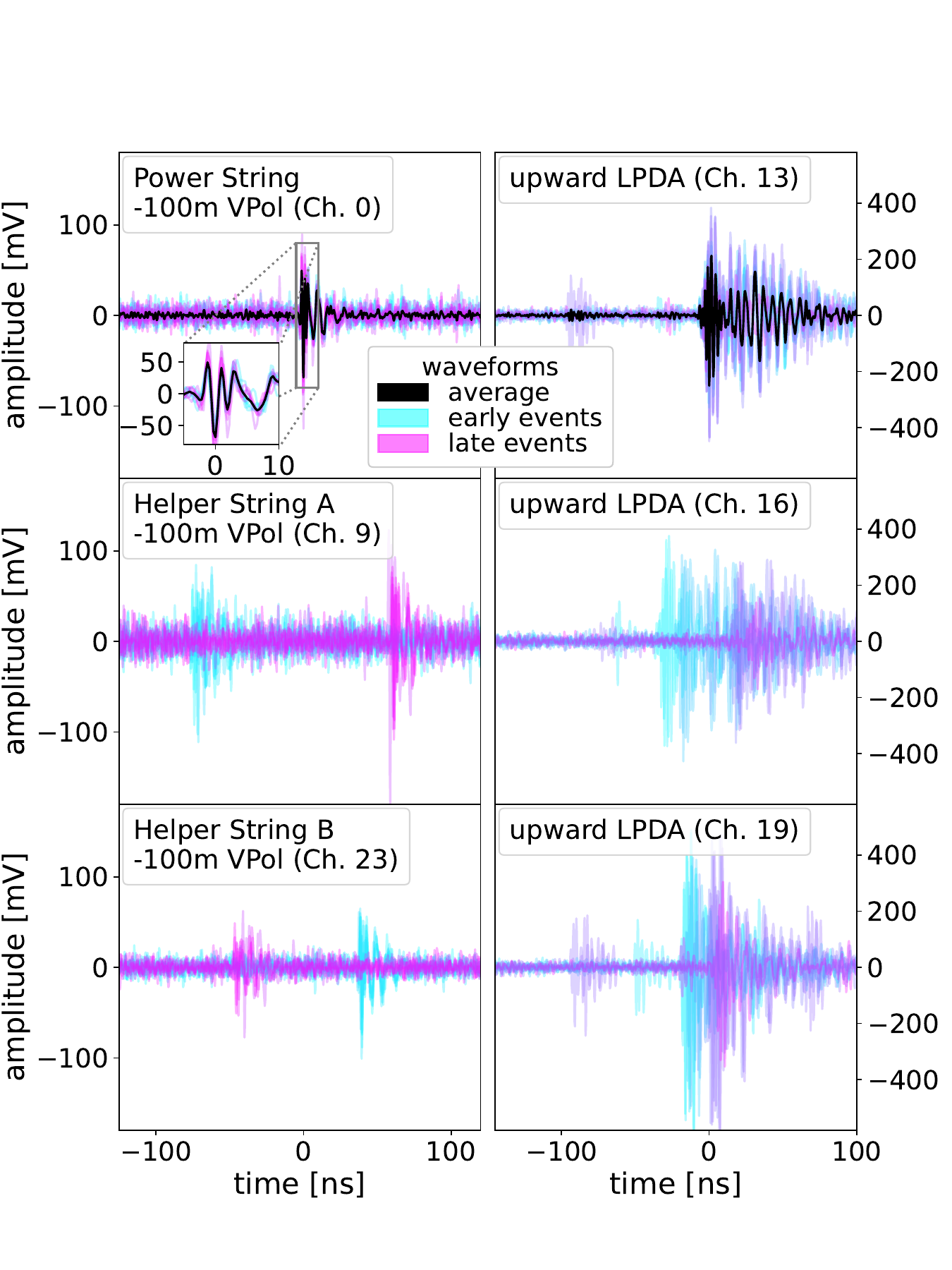}
    \caption{Observation of signals associated with an airplane overflying the RNO-G array. Left: Shown are reconstructed data from three stations (markers) that view the trajectory as transmitted by the airplane under sightly different angles (dashed lines). The pulses are captured in the deep antennas when the airplane is further away due to the lower trigger and better viewing angle for the Vpol antennas. It is then captured by the LPDAs and their trigger when the plane is overhead. The figure shows a very early reconstruction using a simplified ice model and preliminary calibration. Right: Waveforms which are obtained from the flight. These waveforms still include the antenna response, which means that they should be different between two antennas. For the same antenna, they are remarkably similar and average to a short impulse (see inset). The waveforms from the arriving flight (early) have been offset from the ones from the departing flight (late). }
    \label{fig:airplane}
\end{figure}

As observed by previous experiments~\cite{PierreAuger:2015aqe}, airplanes are a source of measurable radio emission in our band. Although the precise emission mechanism is unclear (e.g.\ unintended emission, altimeter pulsing or reflection of environmental RF signals), airplanes are obvious backgrounds~\cite{Southall:2022yil,Monroe:2019zkp}. In anticipation of these backgrounds, RNO-G installed a flight-tracker at Summit Station in 2021 that records timing, position (latitude, longitude, height), speed, and airplane details to tag potential signals in our array. Also, airplanes are excellent far field sources that may be used for calibration. In the occasional crossing of a trajectory directly above (or very close to) the array, the signals are clearly visible as pulsed signals and their direction can be reconstructed, as shown in \autoref{fig:airplane} for the overflying flight CPA085 on 7-May 2023. The signals are visible both as a brief increase in trigger rate, as well as in various anomalous-signal studies of RNO-G data~\cite{RNO-G:2023psm}. 

These aircraft-initiated signals also illustrate the capabilities of RNO-G in terms of trigger performance and angular acceptance. For distant planes, the low threshold trigger, based on the deep Vpol power-string antennas, dominates the trigger as it is sensitive to low-SNR signals and views the plane under a better angle, through the ice. For airplanes directly overflying the array, the \textsc{Radiant} trigger on the shallow LPDA signals initiates the trigger due to the signals reaching the LPDAs earlier than in the deep ice. For initial science data,  \textsc{Radiant} waveforms captured with the surface-antenna trigger
do not have sufficient record length to also capture the deep antenna signals, so that near- or direct-overflying aircraft can only be reconstructed using the surface antennas.
This issue was addressed with \textsc{Radiant} firmware upgrades in 2024, in which adjustable waveform time-delays for different trigger-types were introduced.

In future work, a comprehensive study will use airplane signals for calibration purposes, in particular to verify absolute timing across the array and across all channels. Preliminary studies indicate that the absolute timing between station is sufficient to reconstruct the direction of the airplanes from absolute trigger times in different stations alone, requiring no further correction of the GPS timing.

\begin{figure}
\centering
\includegraphics[width=1\textwidth]{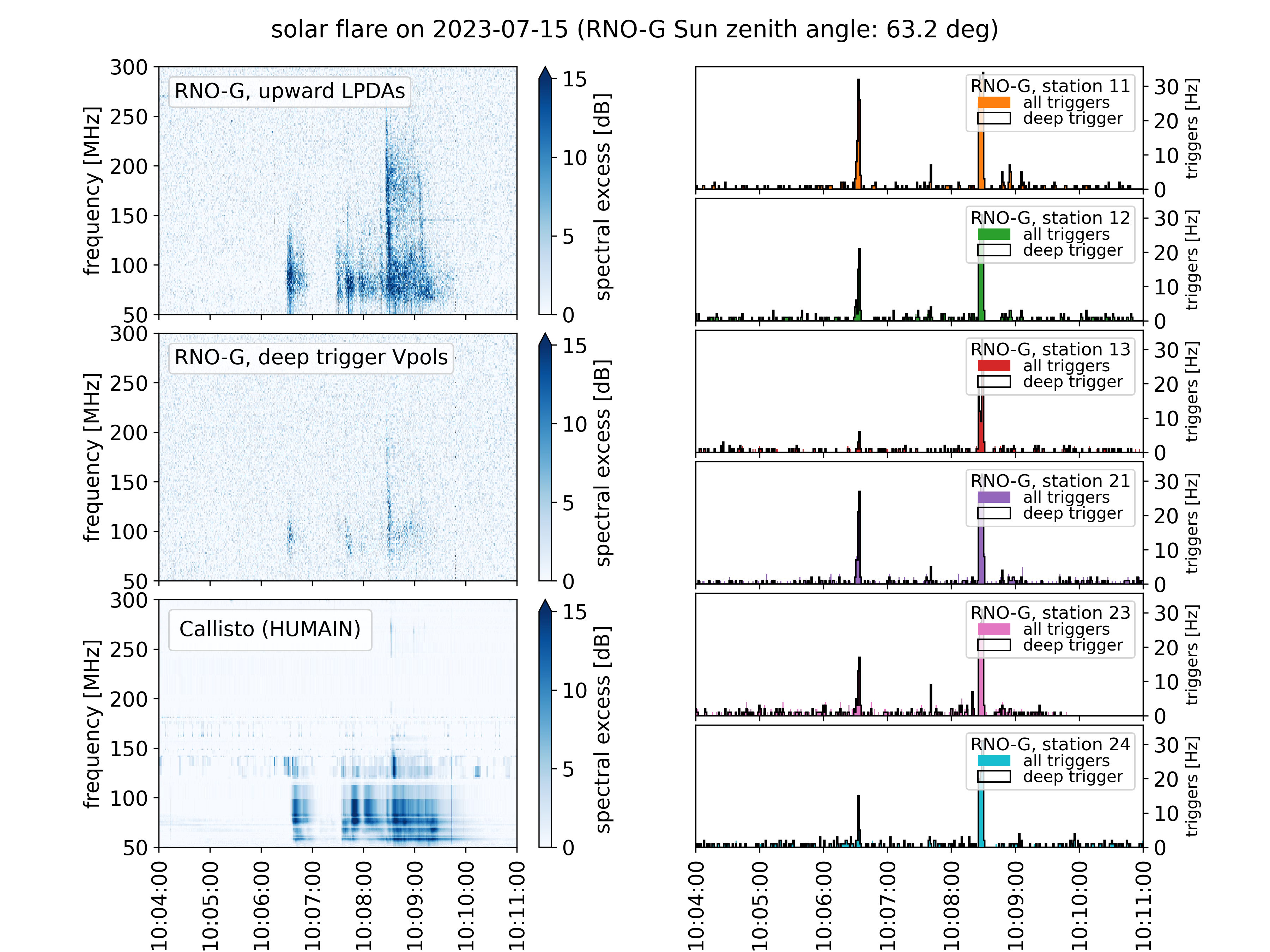}
\caption{A solar flare as observed in RNO-G on July 15th, 2023. Left: Amplitude spectrum as function of time as observed in the upward facing RNO-G antennas (top), in the triggering phased array channels (middle) and, for reference, in the HUMAIN solar monitor located in Belgium \cite{HUMAIN} (bottom). Right: Trigger rate in the different RNO-G stations. The black lines show the trigger rate of the phased array, the colored areas the surface trigger rates. }
\label{fig:solar} 
\end{figure}

\subsubsection{Observation of solar flares}
The sun is one of the most bright radio sources in the sky when in outburst~\cite{Dulk1985}, and is regularly observed by many radio telescopes around the world. Most of them are dedicated to the single purpose of observing the Sun, with many of them being solar spectrographs. The ARA~\cite{Clark:2019ady} and ARIANNA~\cite{Nelles:2015tch} experiments have previously reported on the observation of solar flares, although with some unresolved issues of timing and reconstruction. Observations with these Antarctic-based radio-neutrino experiments also did not have a concurrent heliograph measurement available for direct comparison. 

RNO-G regularly observes solar flares \cite{RNO-G2024solar}, an example of which is shown in \autoref{fig:solar}. The figure shows unfiltered data from several different channels of RNO-G. For better visibility of the structures induced by the Sun, the long-term average spectrum has been subtracted from the data for every antenna. The thus obtained flat spectrum shows a significant increase for several minutes, with significant fine-structure. Its shape matches the observation of the heliograph HUMAIN located in Belgium~\cite{HUMAIN}. It should be noted that the flare is naturally much more visible in the upward facing antennas. However, the signal is large enough to cause a trigger rate increase also in the deep Vpol antennas forming the low threshold trigger, even after accounting for the limited sensitivity of the deep antennas to downward-going signals. 

The solar radio signature provides RNO-G with an absolute pointing measurement and will be a good external source for calibrating the antenna positions. As shown in~\cite{RNO-G2024solar}, the absolute pointing of the RNO-G stations to the incoming radio signal direction can be calibrated to half a degree, which exceeds the design goals, as the expected angular resolution  for neutrino detection will be dominated by signal parameters other than the absolute signal arrival direction \cite{Plaisier:2023cxz}.

\subsection{In-situ position calibration}
\label{sec:pos_calibration}
 
A comprehensive and diverse calibration dataset is taken both during and after station deployment, consisting of measured antenna depths, GNSS surveying, and in situ antenna ranging. This dataset is used to simultaneously fit the antenna positions in the ice, the index of refraction model of the ice as a function of depth, and the cable delays for each channel, and is complementary to the application of extrinsic radio signals. 

Detailed measurements of antenna depth are recorded by the field team during string deployment, which is followed by a measurement of the locations of the surface components and boreholes using GNSS surveying tools with real-time kinematic positioning precision. A phenomenological index of refraction model is developed by fitting an exponential polynomial function \cite{Oeyen:2023eN} to the density data taken at Summit Station by glaciologists \cite{Meese_1997,Hawley_Morris_McConnell_2008,Banta_2007}. Lab measurements of the cable group-delays are taken for each channel on each station, although the calibration fit accommodates small deviations to the detailed lab measurements arising from temperature variations once fielded.  
For a typical post-deployment calibration fit, the antenna positions are corrected by 20-30cm relative to the surveyed positions and the relative cable delays are modified by $\mathcal{O}$(\qty{100}{ps} from pre-deployment lab measurements.
The full procedure of calibrating antenna locations and the resultant accuracy will be discussed in a forthcoming publication.

\section{Summary and Outlook}

We have reported the station and instrument design of the RNO-G-7, and highlighted the performance in the first 3 seasons at Summit Station in Greenland. The RNO-G stations have demonstrated livetimes that are close to design expectations (50\%) while operating on solar-power, and have commenced collecting scientific datasets to be used for UHE neutrino searches. 

The design and performance of the initial RNO-G instrument is established. 
The noise environment and trigger thresholds at the phased-array antennas on the power string have excellent stability, which meet the design goals of the experiment. While the array has been operating primarily using a simplified coincidence trigger, the newly-deployed beamforming trigger demonstrates a threshold at the 2.9-3.4~SNR level, depending on the off-cone angle of the Askaryan signal. This trigger performance does not reach the optimistic 2.0~SNR threshold design target that was originally outlined in the whitepaper, however, further algorithm development is underway to push closer towards this goal.
Several external signals (anthropogenic transmitters, the Sun, the Galaxy) have been measured with RNO-G in high quality, highlighting the performance of the instrument and the RF suitability of the site near Summit Station. 
The RNO-G-7 installed stations significantly enhanced
the technology readiness for a number of new custom 
instrumental systems, deployed into harsh polar environment for the first time, including the new RF signal chains,
borehole string design, and winter-power operational mode.
RNO-G-7 was built and delivered with production-scale efforts.

RNO-G is the first large-scale implementation of the in-ice radio neutrino technique and important lessons have been learned in the first years. New aspects of the RNO-G instrument architecture are currently under test or exist as conceptual design, including the integration of micro wind-turbines,  to enable science operation over-winter, and new DAQ systems, so that the instrument can operate with even higher reliability and autonomy. For a large-scale array of distributed stations, it is essential to reduce the need of user intervention for station operations.
We are also improving the design of the above-surface structures that house the solar panels and communications antenna, to 
reduce yearly maintenance as we better adapt to the high accumulation and snow-drift rates on the Greenland ice sheet.

Once the 35-station array is completed, the collaboration is considering several expansion scenarios, such as enhancing the air shower capabilities by adding infill stations or further expanding the array. 
RNO-G critically informs the design decisions for the proposed radio array of IceCube-Gen2 \cite{IceCube-Gen2:2020qha}. IceCube-Gen2 is envisioned to be roughly a factor of 10 more sensitive than RNO-G, which further impresses the need for demonstrated mass-production and proved-scalable technologies.

\section{Acknowledgments}
We are thankful to the support staff at Summit Station for making RNO-G possible. We also acknowledge our colleagues from the British Antarctic Survey for building and operations of the BigRAID drill for our project. 
We also thank and acknowledge the late Prof.~Gary Varner for the LAB4D ASIC. 

We would like to acknowledge our home institutions and funding agencies for supporting the RNO-G work; in particular the Belgian Funds for Scientific Research (FRS-FNRS and FWO) and the FWO programme for International Research Infrastructure (IRI), the National Science Foundation (NSF Award IDs 2118315, 2112352, 2111232, 2112352, 2111410, and collaborative awards 2310122 through 2310129), and the IceCube EPSCoR Initiative (Award ID 2019597), the German research foundation (DFG, Grant NE 2031/2-1), the Helmholtz Association (Initiative and Networking Fund, W2/W3 Program), the Swedish Research Council (VR, Grant 2021-05449 and 2021-00158), the Carl Trygers foundation (Grant CTS 21:1367), the University of Chicago Research Computing Center, and the European Union under the European Unions Horizon 2020 research and innovation programme (grant agreements No 805486), as well as ERC grants Pro-RNO-G (No 101115122) and NuRadioOpt (No 101116890).

\providecommand{\href}[2]{#2}\begingroup\raggedright\endgroup

\end{document}